\documentclass[twocolumn,aps,showpacs,prb,tightenlines,amsmath,amssymb]{revtex4}
\usepackage{bm}
\usepackage{graphicx}              
\usepackage{amssymb}               
\usepackage{amsmath}                     
\usepackage{colordvi}
\usepackage{calrsfs} 
\usepackage[thinlines,thiklines]{easybmat}
\newcommand{\bgreek}[1]{\mbox{\boldmath$#1$\unboldmath}}
\begin{document}   

\title{Gauge-invariant theory of optical response to
 THz pulses in $s$-wave and ($s$+$p$)-wave superconducting semiconductor quantum wells}
 
\author{T. Yu}
\email{taoyuphy@mail.ustc.edu.cn}
\author{M. W. Wu}
%\thanks{Author to whom correspondence should be addressed}
\email{mwwu@ustc.edu.cn}
\affiliation{Hefei National Laboratory for Physical Sciences at Microscale, Department of Physics,
and CAS Key Laboratory of Strongly-Coupled Quantum Matter Physics,
University of Science and Technology of China, Hefei, Anhui, 230026,
China}

\date{\today}

\begin{abstract} 
We investigate the optical response to the THz pulses in the $s$-wave and ($s$+$p$)-wave
 superconducting semiconductor quantum wells by using the gauge-invariant
 optical Bloch equations, in which the gauge structure in the superconductivity
 is explicitly retained. By using the gauge transformation, not only can the microscopic
description for the quasiparticle dynamics be realized,
but also the dynamics of the condensate is included, 
with the superfluid velocity and the effective
chemical potential naturally incorporated.
 We reveal that the 
superfluid velocity itself can contribute to the pump of quasiparticles
(pump effect), with its rate of change  
acting as the drive field to drive the quasiparticles (drive
effect). Specifically, the drive effect can contribute to the formation of
the blocking
region for the quasiparticle, which directly suppresses the anomalous
correlation of the Cooper pairs. We find that both the
pump and drive effects contribute to the oscillations of the Higgs mode with
twice the frequency of the optical field. However, it is shown that the contribution from
the drive effect to the excitation of
Higgs mode is dominant as long as the driven superconducting momentum is less
than the Fermi momentum. This is in contrast to the conclusion from the
Liouville or Bloch equations in the literature, in which the drive effect on the
anomalous correlation is
overlooked with only the
pump effect considered. Furthermore, in the gauge-invariant optical Bloch equations,
the charge neutrality condition is {\em consistently} considered based on the two-component model for the
charge, in which the charge imbalance of quasiparticles can cause the
  fluctuation of the effective chemical potential. It is predicted that during the optical
process, the quasiparticle charge imbalance 
  can be induced by both the pump and drive effects, leading to the fluctuation of the chemical
 potential. This fluctuation of the chemical potential is further demonstrated
 to directly lead to 
a relaxation channel for the charge imbalance even with the elastic scattering
due to impurities. This is contrast to the previous understanding that in
the isotropic $s$-wave superconductivity, the impurity scattering cannot cause
any charge-imbalance relaxation.  Furthermore, it is  
revealed that when the momentum scattering is weak (strong), the charge-imbalance relaxation
is enhanced (suppressed) by the momentum scattering. Finally, we predict that in
the ($s$+$p$)-wave
 superconducting (100) quantum wells, with the vector potential parallel to the
 quantum wells, the optical field can cause the total 
 spin polarization of Cooper pairs, oscillating with the frequency of the
 optical field. The direction of the total Cooper-pair spin polarization
 is shown to be parallel to the vector potential.

\end{abstract}
%74.40.Gh	Nonequilibrium superconductivity
%74.25.Gz       Optical properties
%74.25.N-       Response to electromagnetic fields
%73.21.Fg	Quantum wells

\maketitle
\section{Introduction}
In recent decades, the nonequilibrium property of superconductors has attracted much attention
for providing new understandings in
superconductivity\cite{Tinkham_book,Kopnin,Chandrasekhar,Triplet_S_F,Buzdin_S_F,Gray,Larkin,Pethick_review}
 and/or exploring novel phases or regimes.\cite{Leggett_book,Liao,Ming_Gong,Twisting_Anderson,Soliton,Critical}
Among them, the optical response plays an important role in both
linear\cite{Mattis,Nam,Schrieffer,two_fluid,Lee} and nonlinear
regimes.\cite{MgB2_1,MgB2_2,YBCO,BSCCO,voltage,Earliest,Matsunaga_1,Matsunaga_2,Matsunaga_3,Matsunaga_4}
The former has been well established from the understanding of the optical
conductivity in the linear response of the superconducting state, which
sheds light on the determination of the pairing symmetry of the
superconducting order parameter.\cite{Mattis,Nam,Schrieffer,two_fluid} The latter is inspired
by the recently-developed THz technique, whose frequency lies around the
superconducting gap.\cite{MgB2_1,MgB2_2,YBCO,BSCCO,voltage,Earliest,Matsunaga_1,Matsunaga_2,Matsunaga_3,Matsunaga_4}
 With an intense THz optical field, the
superconductor can be even excited to the states far away from the equilibrium, opening
a window to reveal the dynamical properties of both the Bogoliubov quasiparticles
and the condensate.\cite{MgB2_1,MgB2_2,YBCO,BSCCO,voltage,Earliest,Matsunaga_1,Matsunaga_2,Matsunaga_3,Matsunaga_4}

In the linear regime, in the dirty limit at zero
temperature, Mattis and Bardeen\cite{Mattis} revealed that the
optical absorption is realized by breaking the Cooper pairs into the quasi-electron
and quasi-hole when the photon energy is larger than twice the magnitude of the 
superconducting gap.\cite{Nam,Schrieffer} Nevertheless, in the
early-stage work,\cite{Mattis} a physical optical
conductivity is
 established only for a specific gauge with transverse vector potential and
 zero scalar potential.\cite{Mattis,Nam,Schrieffer} 
A gauge-invariant description
with charge conservation for the optical
conductivity tensor is later established by Nambu based on the generalized
Ward's identity,\cite{Nambu_gauge,Goldstone} in which the collective excitation is revealed to cancel the
unphysical longitudinal current.\cite{Schrieffer,Nam,Lee}
Furthermore, Ambegaokar and Kadanoff\cite{Ambegaokar} showed that in the long wave limit, the
collective mode can be actually described as a state in which the {\em superconducting phase} of the
order parameter varies periodically in time and
space.\cite{Ambegaokar,Altland,conjugacy,Lee,Schrieffer,Enz,Griffin}
Actually, without considering the response of the order parameter to the optical
field, the absence of the charge conservation naturally arises because the particle
number is not a conserved quantity in the mean-field description
of the superconductor with a {\em global}
$U(1)$ symmetry spontaneously broken.\cite{Altland,conjugacy,Lee,Schrieffer,Nambu_gauge}

When the photon energy is
far below the superconducting gap, a simple physical
picture for the optical response  
 can be captured based on the two-fluid model, in which the optical conductivity at
  {\em finite} frequency $\omega$ reads\cite{two_fluid,MgB2_1,MgB2_2,YBCO,BSCCO} 
\begin{equation}
\sigma(\omega)=\frac{\rho_ne^2\tau}{m^*}\frac{1}{1+\omega^2\tau^2}
+i\Big(\frac{\rho_ne^2\tau}{m^*}\frac{\omega\tau}{1+\omega^2\tau^2}+\frac{\rho_se^2}{m^*}\frac{1}{\omega}\Big).
\label{optical_conductivity}
\end{equation}
Here, $\rho_n$ and $\rho_s$ denote the normal-fluid and super-fluid densities
in the {\em equilibrium} state, respectively; $m^*$ is the effective mass of the electron; and
$\tau$ represents the momentum relaxation time. Based on
Eq.~(\ref{optical_conductivity}),
the optical absorption can be
well understood from the electric current driven by the optical
field.\cite{Mattis,Nam,Schrieffer,two_fluid} 
In the clean limit, the optical
conductivity is purely imaginary with the phase difference between the
induced current 
and the optical field being exactly $\pi/2$, and hence no optical absorption is
expected. Nevertheless,
in the dirty sample, the real part of the optical
conductivity arises due to the existence of the normal-fluid,
which contributes to the electric current in phase to the optical field, and
hence, the optical
absorption. Thus, in the pump-probe measurement, after strongly excited by the
pump field, the non-equilibrium normal-fluid and
super-fluid densities can be estimated from 
the optical response to the probe field with photon energy far below the superconducting
gap.\cite{MgB2_1,MgB2_2,YBCO,BSCCO}
However, to the best of our knowledge, a microscopic theoretical-description
 for the evolution of  the normal- and super-fluids from the
equilibrium state to the non-equilibrium ones is still lacking.

In the nonlinear regime, in which the superconducting state can be markedly
influenced by the optical field, 
the experimental\cite{Earliest,Matsunaga_1,Matsunaga_2,Matsunaga_3,Matsunaga_4}
 and theoretical\cite{Soliton,Critical,Axt1,Axt2,Anderson1,
   Higgs_e_p,multi_component,Leggett_mode,Xie,Tsuji_e_p,multiband,path_integral}
 studies are still in progress. Very
recently, it was reported 
in several experiments in the film of the conventional superconducting metal that the oscillations
of the Higgs mode, i.e., the fluctuation of the order-parameter magnitude, 
 can be excited by the intense THz
 field.\cite{Earliest,Matsunaga_1,Matsunaga_2,Matsunaga_3,Matsunaga_4} It is revealed that the
 oscillation frequency of the Higgs mode is twice the frequency of the THz 
 field, no matter the photon energy is larger or smaller than twice the
 magnitude of the 
 superconducting gap.\cite{Matsunaga_3,Matsunaga_4} Moreover, a large THz 
 third-harmonic generation was reported when the photon energy 
 is tuned to be resonant with the superconducting gap.\cite{Matsunaga_3,Matsunaga_4} 
 Finally, it was discovered that there exists plateau for the
 Higgs mode after the
 THz pulse in most situations, whose value increases with the increase of the field
 intensity.\cite{Matsunaga_1,Matsunaga_2} These observations indicate that there
 exists 
 strong optical absorption with the quasiparticles considerably excited 
 by the strong optical field.\cite{Matsunaga_1,Matsunaga_2,Matsunaga_3,Matsunaga_4}

These experimental findings have been theoretically clarified based on the
Liouville equation\cite{Axt1,Axt2,
   Higgs_e_p} or the Bloch
 equation\cite{Anderson1,multi_component,Leggett_mode,Xie,Tsuji_e_p,multiband,Matsunaga_3,Matsunaga_4}
 derived in the Anderson
pseudospin representation\cite{Anderson_origin} {\em in the clean limit}.
 Specifically, the optical absorption in the clean limit is naturally 
understood by the nonlinear term proportional to ${\bf A}^2$, with ${\bf A}$
standing for the vector
potential of the optical field. It is shown that this non-linear term
contributes to the precessions between the quasi-electron and
quasi-hole states,\cite{Axt1,Axt2}
 which directly contribute to the excitation of the
quasiparticles (pump effect).\cite{Soliton,Critical,Axt1,Axt2,Anderson1,Higgs_e_p,multi_component,Leggett_mode,Xie,
  Tsuji_e_p,multiband,path_integral} Thus, the optical
absorption is realized in the clean limit due to this pump effect,
 from which the Cooper pairs are 
broken into the quasi-electrons and
 quasi-holes.\cite{MgB2_1,MgB2_2,YBCO,BSCCO,voltage,Earliest,Matsunaga_1,Matsunaga_2,Matsunaga_3,Matsunaga_4}
Furthermore, because the frequency of ${\bf A}^2$ is $2\omega$, the pump effect
 contributes to the oscillation of the Higgs mode with twice the frequency
of the optical field.\cite{Axt1,Axt2,Anderson1,Higgs_e_p,
  multi_component,Leggett_mode,Xie,Tsuji_e_p,multiband} Moreover, it is revealed
that the Higgs mode can be resonant with the optical field when the photon energy
equals to the superconducting gap, which is further shown to contribute to the large third
harmonic generation.\cite{Matsunaga_3,Matsunaga_4,path_integral}

However, there still exist several difficulties inherited in the
Liouville\cite{Axt1,Axt2,Higgs_e_p} or Bloch\cite{Anderson1,
  multi_component,Leggett_mode,Xie,Tsuji_e_p,multiband} equations used in the
literature. Firstly, the
anomalous correlation is calculated between the two electrons with momenta ${\bf
k}$ and $-{\bf k}$, no matter the optical field is slowly or rapidly varied.
This means that it is preconceived that no center-of-mass momentum ${\bf q}$ of the Cooper pairs
can be excited.\cite{Axt1,Axt2,Higgs_e_p,Anderson1,
  multi_component,Leggett_mode,Xie,Tsuji_e_p,multiband}
Nevertheless, in the nonlinear regime, with a strong electric field applied, a large
supercurrent is expected to be induced, which should arise from the center-of-mass
momentum of the
Cooper pairs.
It has been well understood that in the static
 situation, a large ${\bf q}$ contributes to the Doppler shift in the energy
 spectra of the elementary excitation, which can lead to the formation of the blocking region with
 the anomalous correlation of the Cooper pairs significantly
 suppressed.\cite{Tao_2,FF,LO,FFLO_Takada,supercurrent,Yang,Doppler_1,Doppler_2}
 Nevertheless, the induction of the center-of-mass momentum for the Cooper pairs
 and its further influence on the superconducting state are absent in the 
description of the
 Liouville equation or the Bloch equation in the Anderson pseudospin
 representation.\cite{Axt1,Axt2,Anderson1,Higgs_e_p,
  multi_component,Leggett_mode,Xie,Tsuji_e_p,multiband} In fact, in the
Liouville equation, the generalized coordinate, i.e., the momentum ${\bf k}$, is
treated to be time-independent or fixed, whereas 
the velocity field ${\bf v}({\bf k})={\bf k}-(e/c){\bf A}$ located at the generalized
coordinate varies with time.
This is similar to the
Euler description in the fluid mechanics, in contrast to the Lagrangian
description with time-dependent generalized coordinate.\cite{Landau} Thus, the anomalous
correlation is always described between ${\bf k}$ and $-{\bf k}$ in the
 Liouville or Bloch equations used in the literature.\cite{Axt1,Axt2,Anderson1,Higgs_e_p,
  multi_component,Leggett_mode,Xie,Tsuji_e_p,multiband}

Secondly, the scattering effect, which is inevitable in the dirty
superconducting metal,\cite{Anderson1,Higgs_e_p}
cannot be simply included in the Liouville
equation in the presence of the optical field.\cite{Jianhua_Liouville} 
Moreover, a simple inclusion of the elastic scattering with the Boltzmann
 description\cite{Kopnin,Tinkham_book} in the
 Liouville equation does not influence the calculated results, because the pump
 effect is isotropic in the momentum
 space.\cite{Axt1,Axt2,Anderson1,Higgs_e_p,
   multi_component,Leggett_mode,Xie,Tsuji_e_p,multiband,path_integral}
 However, this is un-physical because the normal-fluid can still be
 scattered. 
Finally, the {\em gauge invariance}\cite{Nambu_gauge,Altland,Schrieffer} in the Liouville or
the Bloch equations used in
the literature is not clearly addressed.\cite{Axt1,Axt2,Anderson1,Higgs_e_p,
  multi_component,Leggett_mode,Xie,Tsuji_e_p,multiband} 
On one hand, two quantities in the vector potential, scalar potential and
superconducting phase are simultaneously taken to be
zero.\cite{Nambu_gauge,Altland,Schrieffer} Specifically, with the vector
potential chosen, the resulted physical current is shown to be proportional to
${\bf A}$, which is {\em not} a gauge-invariant physical quantity unless a transverse 
gauge for ${\bf A}$ is further restricted.\cite{Schrieffer,Mattis} 
 On the other
hand, from different choices of gauge, different forms of the equation can be
expected. Specifically, with only the scalar
potential, the ${\bf A}^2$-term vanishes and the electric field contributes to
the drive field; whereas with only the
superconducting phase ${\bf q}\cdot{\bf r}$, its rate of change can also contribute to a drive
field.\cite{Nambu_gauge,Altland,Schrieffer,Stephen}

In fact, as pointed out by Nambu,\cite{Nambu_gauge} 
 the absence of the gauge invariance in the theoretical
 description is equivalent to the breaking of the
 charge conservation.\cite{Altland,conjugacy,Lee,Schrieffer}
By restoring the gauge invariance, in the linear regime, 
Nambu revealed a collective
excitation stimulated in the optical
process,\cite{Nambu_gauge} which was further shown by Ambegaokar and
  Kadanoff\cite{Ambegaokar} to be described by a state with the
  period variations in time and space for the superconducting phase in the
  long-wave limit.\cite{Ambegaokar,Altland,conjugacy,Lee,Schrieffer,Enz,Griffin} 
 The temporal and
 spacial variations of the superconducting
 phase can further contribute to the effective chemical potential
and superconducting
velocity.\cite{Ambegaokar,Altland,conjugacy,Lee,Schrieffer,Enz,Griffin}
Then, it is
inspired by this
scheme\cite{Altland,conjugacy,Lee,Schrieffer,Nambu_gauge}
 that with the gauge invariance retained in the kinetic equation, the
collective excitation can also arise naturally.\cite{path_integral}
Specifically, by noting that in the
mean-field description based on the Bogoliubov-de Gennes (BdG)
  Hamiltonian, only the
  dynamics of the quasiparticle is considered.
  It has been suggested that the ``condensate'' can respond to the dynamics
  of the quasiparticles from the consideration of the gauge structure in
  superconductor,\cite{Nambu_gauge}
 with the charge
  conservation restored by the fluctuation of
 the chemical
 potential.\cite{Tinkham1,Tinkham2,Tinkham_condensate,
   Takahashi_SHE1,Takahashi_SHE2,Hirashima_SHE,Takahashi_SHE3,Takahashi_SHE_exp,Hershfield,Spivak,kinetic_book}

 One way to understand the interplay between the particle charge and chemical potential
is based on the two-component model for the
  charge.\cite{Tinkham1,Tinkham2,Takahashi_SHE2,Tinkham_book,Pethick_review,Larkin,Gray,Tinkham_condensate}
 In the two-component model, the electron charge is treated to be carried by the quasiparticle and
  condensate, respectively. This can be easily seen in
  the electrical injection process. In that process, the injection of one electron with charge $e$
 into the {\em conventional} superconductor can add a quasiparticle with
charge $e(u_{\bf k}^2-v_{\bf k}^2)$ and one
Cooper pair with charge $2e v_{\bf k}^2$,
respectively. Here, $u_{\bf k}$ and $v_{\bf k}$ comes from the Bogoliubov
transformation with
$u_{\bf k}^2+v_{\bf k}^2=1$, indicating the charge conservation in the
electrical injection process.\cite{Tinkham1,Tinkham2,Takahashi_SHE2} Thus, the
fluctuation of the quasiparticle charge is associated with the fluctuation of the condensate
density.\cite{Tinkham1,Tinkham2,Takahashi_SHE2,Tinkham_book,Pethick_review,Larkin,Gray,Tinkham_condensate}
This is consistent with the conjugacy
relationship between the particle number and superconducting
phase.\cite{conjugacy,Nambu_gauge}

 Furthermore, in the dynamical process, the charges for the quasiparticle and condensate can both 
 be deviated from their equilibrium values. This is referred to as the
 charge imbalance,\cite{Tinkham1,Tinkham2,Takahashi_SHE2,Tinkham_book,
   Pethick_review,Larkin,Gray,Clarke_first,Clarke_thermal}
 which has been measured for  both the quasiparticle\cite{Tinkham1,Tinkham2,
   Pethick_review,Clarke_first,Clarke_thermal} and
 condensate.\cite{voltage}
For the quasiparticle, due to the momentum-dependence of the charge,
 its non-equilibrium distribution can
lead to the charge imbalance, whose creation and relaxation are intensively studied in the
electrical 
experiment.\cite{Tinkham_book,Tinkham1,Tinkham2,Takahashi_SHE2,Clarke_first,Clarke_thermal} 
It is so far widely believed that for the isotropic
$s$-wave superconductor, the {\em elastic} scattering due to the impurity
 cannot cause the relaxation of
the charge imbalance.\cite{Tinkham_book,Tinkham1,Tinkham2,Pethick_review}
 This is because there exists the coherence 
factor $(u_{\bf k}u_{{\bf k}'}-v_{\bf k}v_{{\bf k}'})$ in the scattering
potential, where ${\bf k}$ and ${\bf k}'$ are the initial and final momenta
during the scattering, 
 due to which the elastic scattering cannot exchange the electron-like and
hole-like quasiparticles.\cite{Tinkham_book,Tinkham1,Tinkham2,Pethick_review}
 However, in that relaxation process, the
condensate is assumed to be in its equilibrium state, meaning that the
charge conservation or neutrality
is not explicitly considered in the literature. Moreover, the correlation between the
quasi-electron and quasi-hole is often
neglected.\cite{Tinkham2,Takahashi_SHE2,Tinkham_book,Pethick_review,Larkin,Gray}
Thus, it is essential to check the influence of the condensate on the
charge-imbalance relaxation in the framework of charge neutrality.
Furthermore, although the charge imbalance including its creation and
relaxation is
intensively
studied in the electrical
experiment,\cite{Tinkham1,Tinkham2,Takahashi_SHE2,Tinkham_book,
  Pethick_review,Larkin,Gray,Clarke_first,Clarke_thermal}
it has yet been well investigated in the optical process.\cite{voltage}

So far we have addressed the experimental and theoretical investigations on the
  optical response in the conventional superconductivity, in which the Cooper
  pairs do not carry any net spin. As the optical method is
  often used to create and manipulate the  electron spins 
 in semiconductors\cite{Jinluo,Jianhua_Liouville,Ivchenko,wureview}
 or topological insulator\cite{topological_0} by inducing an 
 effective spin-orbit coupling (SOC), it is intriguing to consider the possibility
  of the creation and manipulation for the Cooper-pair spin polarization. This is
  possible in the triplet superconductivity
 as the triplet Cooper pairs can carry net
 spin polarization.\cite{Sigrist,Supercurrents,Superconducting_spintronics,Eschrig_supercurrent,Linder1,Linder2}
It is noticed that the inclusion of the optical field can break the time-reversal
symmetry. Previous works have shown that the breaking of
 the time-reversal symmetry by the Zeeman field\cite{Linder_SFS} or the
 supercurrent\cite{Tkachov} can induce the Cooper-pair spin polarization.
 In the former situation, for the conventional $s$-wave
superconductor in proximity to a ferromagnet, the triplet Cooper
pairing can be induced in the superconductor-ferromagnet interface.\cite{Buzdin_S_F,Triplet_S_F,Linder1,Tokatly_PRLB,
exp_realization1,exp_realization2,Berezinskii} Then Jacobsen {\em et al.} showed
that in the
superconductor-ferromagnet-superconductor
Josephson junction, when there exists the SOC in the superconductor-ferromagnet interface,
 the net spin polarization of triplet Cooper pairs can be
created, and a superconducting spin flow with
spin-flip immunity can be realized.\cite{Linder_SFS} In the latter situation,
Tkachov pointed 
out that in noncentrosymmetric superconductors, a {\em nonunitary}
triplet pairing\cite{Leggett,Leggett_book,Sigrist} can be induced by the supercurrent,
 which contributes to the spin polarization of triplet
Cooper pairs and can be detected by the magnetoelectric Andreev effect.\cite{Tkachov} 
It is further noted that the optical field can also induce a supercurrent
 in noncentrosymmetric superconductivity, which is expected to dynamically
generate the Cooper-pair spin polarization.

Recently, the proximity-induced
superconductivity has been realized in InAs\cite{InAs_1,InAs_2,Doppler_2} and
GaAs\cite{GaAs_1,GaAs_2,GaAs_3} heterostructures. Thus, based on the well-developed techniques in
semiconductor optics,\cite{Haug,Axt_optics,Rossi_optics} the superconducting
semiconductor quantum wells (QWs) can provide an ideal platform to study the optical
response of superconductivity. Compared to the film of
the superconducting metal, the QWs can be synthesized to be extremely
clean. Furthermore, the
material parameters in the QWs, e.g., the electron density, the strength of the
SOC and the
interaction strengths including the Coulomb, electron-phonon and
electron-impurity interactions, can be easily tuned.
Moreover, in the QWs, the simple Fermi surface and exactly-known
interaction forms can significantly reduce the difficulties
in the comparison between the theory and experiment. Finally, the predictions
revealed in the superconducting QWs can still
shed light on the optical response in the superconducting metal even with 
complex Fermi surfaces.

In the present work, we investigate the optical response to the THz pulses in both the $s$-wave and
($s$+$p$)-wave superconducting semiconductor QWs. 
The gauge-invariant optical Bloch equations are set up via the gauge-invariant
nonequilibrium Green
function approach,\cite{Tao_2,Haug,Levanda,GKB,BoYe} in which the
gauge-invariant Green function with the Wilson line\cite{Wilson_line,Haug,Levanda}
is constructed by using the gauge structure revealed by
Nambu.\cite{Nambu_gauge} In the optical Bloch equations, the structure can
  be easily captured by a special
  gauge, in which the superconducting phase is chosen to be zero
  among the vector potential, scalar potential and
  superconducting phase.  This gauge is referred to as the ${\bf p}_s$-gauge here,
  with ${\bf p}_s$ being the superfluid momentum driven by the optical
  field. It is noted that this superfluid momentum directly contributes to the center-of-mass
  momentum of Cooper pairs. Furthermore, in the ${\bf p}_s$-gauge,
 not only can the microscopic
description for the quasiparticle dynamics be realized,
but also the dynamics of the condensate is included,
with the superconducting velocity and the effective
chemical potential naturally incorporated. 
Then in the derived gauge-invariant optical Bloch equations, this 
superconducting velocity $\propto {\bf p}_s^2$ is shown to directly contribute to the pump of the
quasiparticles (pump effect), whose rate of
change $\partial _t{\bf p}_s$ induces a drive field to drive the quasiparticle (drive effect).
We find that both the
pump and drive effects contribute to the oscillation of the Higgs mode with
twice the frequency of the optical field. However, it is shown that the contribution from
the drive effect to the excitation of
Higgs mode is dominant as long as the superconducting momentum ${\bf p}_s$ is smaller
than the Fermi momentum $k_F$, thanks to the efficient suppression of the pump effect
by the Pauli blocking. This is in sharp contrast to the results from
the Liouville\cite{Axt1,Axt2,Higgs_e_p} or
Bloch\cite{Anderson1,multi_component,Leggett_mode,Xie,Tsuji_e_p,multiband}
 equations in the
literature, where only the pump effect is considered and the effects of the
center-of-mass momentum on the superconducting state are overlooked. 
 The influence of the electron-impurity scattering is also addressed,
 which is shown to further suppress the Cooper pairing on the basis of the drive
 effect.

\begin{figure}[ht]
  {\includegraphics[width=9.5cm]{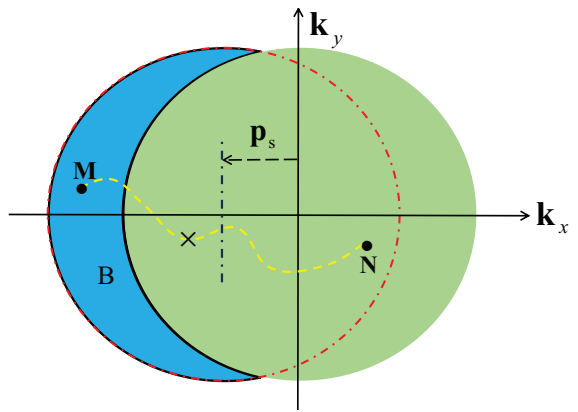}}
  \caption{(Color online) Schematic of the electron drift states in
    response to the
    optical field, with the Fermi surface labeled by the red
    chain curve. Here, the superconducting
    momentum ${\bf p}_s=p_s\hat{\bf x}$ with $p_s<0$. With the drift of the
  electron states, a blue region labeled by ``B'' arises, in which the electrons
  deviate from their equilibrium states. Actually, these electrons are directly excited to
be the quasiparticles, whose population can be close to
one.\cite{FF,LO,FFLO_Takada,supercurrent,Tao_2,Yang}
 By using the
terminology in the FFLO state,\cite{FF,LO,FFLO_Takada} this blue region populated by the quasiparticles
is referred to as the blocking region. Furthermore, due to the
induction of the center-of-mass momentum for the Cooper pairs by the
applied optical field, the two electrons with momenta
${\bf k}+{\bf p}_s$ and $-{\bf k}+{\bf p}_s$ are paired together. 
Nevertheless, once the  electrons are excited in the blocking region, they no longer
participate in the Cooper
pairing.\cite{FF,LO,FFLO_Takada,supercurrent,Tao_2,Yang}
 For instance, the electron labeled by ``N''
cannot pair with its corresponding one labeled by ``M'' in the blocking region,
 which has been excited to be the
quasiparticle. Accordingly, the anomalous correlation is directly suppressed due to the
drift of the electron states.}
  \label{figyw1}
\end{figure}

 The physical picture for the suppression of the anomalous correlation of
 Cooper pairs by the
 optical field can be understood as follows.
 Thanks to the drive of
 the optical field, the electron states are drifted, obtaining exactly the center-of-mass momentum
 ${\bf p}_s$ in the impurity-free situation. The drift states of electrons are schematically
 presented in Fig.~\ref{figyw1} with the Fermi surface labeled by the red
 chain curve. 
In Fig.~\ref{figyw1}, without loss of generality, the superconducting
    momentum is taken to be along the $\hat{\bf x}$-direction,
 i.e., ${\bf p}_s=p_s\hat{\bf x}$, with $p_s<0$. It can be seen that with the drift of the
  electron states, a blue region labeled by ``B'' arises, in which the electrons
  deviate from their equilibrium states. Actually, these electrons are directly excited to
be the quasiparticles, whose population can be close to
one.\cite{FF,LO,FFLO_Takada,supercurrent,Tao_2,Yang}
 By using the
terminology in the Fulde-Ferrell-Larkin-Ovchinnikov
 (FFLO) state,\cite{FF,LO,FFLO_Takada} this blue region populated by the quasiparticles
is referred to as the blocking region.

 Furthermore, it is noted that the applied 
optical field breaks the time-reversal symmetry. Thus, the paired electrons
do not necessarily come from two time-reversal partners with momenta ${\bf
  k}$ and $-{\bf k}$.
 On the contrary, due to the
induction of the center-of-mass momentum for the Cooper pairs by the
applied optical field, the two electrons with momenta
${\bf k}+{\bf p}_s$ and $-{\bf k}+{\bf p}_s$ are paired together.
Nevertheless, once the  electrons are excited in the blocking region, they no longer
participate in the Cooper
pairing.\cite{FF,LO,FFLO_Takada,supercurrent,Tao_2,Yang}
 One typical example is shown in Fig.~\ref{figyw1}, in which the electron labeled by ``N''
cannot pair with its old partner labeled by ``M'' in the blocking region,
 which has been excited to be the
quasiparticle.
 Consequently, the anomalous correlation in the blocking region is
significantly suppressed, directly leading to the suppression of the magnitude of the
order parameter.\cite{FFLO_Takada,Tao_2,Yang} This is responsible for the oscillation of the Higgs
mode. Nevertheless, at high frequency, this oscillation is suppressed due to the
suppression of the drift effect and hence the range of the blocking region.
 This picture is consistent with the the {\em static} case when the center-of-mass
momentum of the Cooper pairs emerges due to either the spontaneous
symmetry-breaking\cite{FF,LO,FFLO_Takada} or the
supercurrent.\cite{supercurrent,Tao_2,Yang}

In the derived optical Bloch equations,
the charge neutrality condition is consistently considered based on the two-component model for the
charge, in which the induction of the charge imbalance of quasiparticles can cause the
  fluctuation of the condensate chemical
  potential.\cite{Tinkham1,Tinkham2,Takahashi_SHE2,Tinkham_book,Pethick_review,Larkin,Gray}
 We predict that during the optical
process, the charge imbalance 
  can be created by both the pump and drive effects, with the former
    arising from the AC Stark effect and the latter coming from the breaking of
    Cooper pairs by the electrical field. The induction of the
    charge imbalance directly leads to the fluctuation of the chemical
 potential.  This fluctuation is further found to directly provide
a relaxation channel for the charge imbalance even with the elastic scattering
due to impurities. This is in contrast to the previous understanding that in
the isotropic $s$-wave superconductivity, the impurity scattering cannot cause
any charge-imbalance relaxation.\cite{Tinkham1,Tinkham2,Pethick_review}
Specifically, we reveal that when the momentum scattering is weak
(strong), the charge-imbalance relaxation
is enhanced (suppressed) by the momentum scattering.

\begin{figure}[ht]
  {\includegraphics[width=9.8cm]{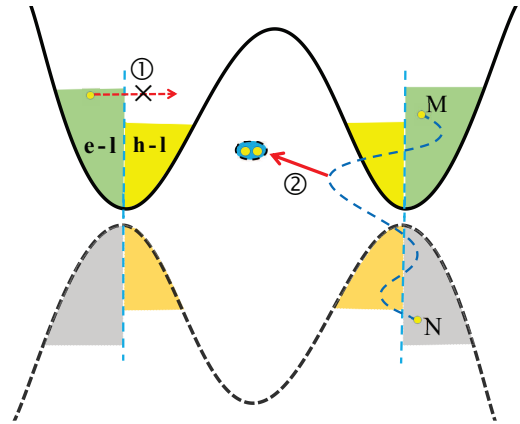}}
  \caption{(Color online) Schematic of the charge-imbalance relaxation
    channels.
    The upper and lower bands, plotted by the black solid and dashed curves, represent
    the quasi-electron and quasi-hole
    bands, respectively. In the quasi-electron (quasi-hole) band, the green
    (gray) and yellow (orange) regions denote the
  electron- (hole-) and hole-like (electron-like) quasi-electrons
  (quasi-holes), respectively.
  One sees that the 
  quasi-electron number in the 
  electron-like branch is larger than the one in the hole-like branch.
  In this situation, 
  the charge imbalance is created with net negative charges. The two
  charge-imbalance relaxation channels labeled by ``{\textcircled 1}'' and
  ``{\textcircled 2}'' can be understood as follows.
  Process {\textcircled 1} has been addressed in the previous works,
 representing the direct scattering of quasiparticles between
 the electron- and hole-like branches, which is actually forbidden in the elastic
 scattering process in the isotropic $s$-wave superconductor.\cite{Tinkham1,Tinkham2,Pethick_review}
 In process {\textcircled 2}, the quasi-electron and quasi-hole, labeled by ``M'' and ``N'', become
 correlated due to the fluctuation of the effective chemical potential, which then
 annihilate into one
 Cooper pair due to the momentum scattering. Here, one notes that the momenta of
   the correlated quasi-electron (``M'') and quasi-hole (``N'') are the same, in
 consistent with the Bogoliubov transformation [refer to
   Eq.~(\ref{modified_B}) in the main text].
 Thus, the annihilation of 
 extra quasiparticles directly leads to the charge-imbalance relaxation.}
\label{figyw2}
\end{figure}

We demonstrate that the
fluctuation of the condensate chemical potential can {\em first} induce the
quasiparticle correlation
between the quasi-electron and quasi-hole, which {\em then} provides the
charge-imbalance relaxation channel for the quasiparticle populations {\em in the
  presence of the elastic momentum scattering}. In
  the previous works, it was revealed that in the presence of the impurities, 
  the charge-imbalance relaxation is induced by the direct scattering of
  quasiparticles between the electron- and hole-like
  branches,\cite{Tinkham1,Tinkham2,Pethick_review} during which the
  quasiparticle number is conserved.  
  Nevertheless, this is demonstrated to be forbidden in the
  isotropic $s$-wave superconductors.\cite{Tinkham1,Tinkham2,Pethick_review} 
  Differing from this charge-imbalance relaxation
  channel,\cite{Tinkham1,Tinkham2,Pethick_review} in this work, the
  charge-imbalance relaxation is actually caused by the direct
  annihilation of the quasiparticles in the quasi-electron and quasi-hole bands,
  in which the
  quasiparticle-number conservation is broken. 
  These two charge-imbalance relaxation channels are schematically
  shown in Fig.~\ref{figyw2}, labeled by ``{\textcircled 1}'' and
  ``{\textcircled 2}'', respectively.
 Specifically, process {\textcircled 1} represents the direct scattering of quasiparticles between
    the electron- and hole-like branches. Whereas 
in process {\textcircled 2}, the quasi-electron and quasi-hole, labeled by ``M'' and ``N'', become
 correlated due to the fluctuation of the effective chemical potential, which then
 annihilate into one
 Cooper pair due to the momentum scattering.

 Actually, it is overlooked in the previous
 studies\cite{Tinkham1,Tinkham2,Pethick_review}
 that the non-equilibrium effective chemical
 potential itself can induce the precession between the quasi-electron
 and quasi-hole states and hence the quasiparticle correlation.\cite{Tinkham1,Tinkham2,Pethick_review}
 The quasiparticle correlation is crucial to induce the
 quasiparticle-number fluctuation. As addressed in our previous
 work,\cite{Tao_2} the induction of the
 quasiparticle correlation is related to the process of the condensation with two quasiparticles
 binding into one Cooper pair in the condensate, or vice versa.\cite{Bardeen,Tinkham2,Josephson_B}
 These processes can directly cause {\em the annihilation of the extra
   quasiparticles}
 in the quasi-electron or quasi-hole bands, inducing the charge-imbalance relaxation for the 
 quasiparticles. Meanwhile, with the condensation or breaking of the Cooper pairs in
 the condensate, the fluctuation of the effective chemical
 potential is also induced.
 If only the induction of the quasiparticle correlation was not influenced
 by the momentum scattering, the charge-imbalance relaxation 
 rate would be proportional to the electron-impurity scattering
 strength. Nevertheless, it is further revealed
 that the induction of the quasiparticle correlation can be suppressed by the impurity scattering.
 Thus, the competition
 between the relaxation channels due to the
  quasiparticle correlation and population leads to the non-monotonic dependence
  on the momentum scattering for the charge-imbalance relaxation.

 Finally, we predict that in
the ($s$+$p$)-wave 
 superconducting InSb (100) QWs, with the vector potential being along
 the $\hat{\bf x}$-direction, the optical field can cause the
 spin polarization of Cooper pairs, which is also along the $\hat{\bf x}$-direction
 and oscillates with the frequency of the optical field. 
Specifically, in our previous work, it has been revealed that in InSb (100) QWs in proximity to an $s$-wave
superconductor, due to the Rashba-like SOC, there exists $p$-wave triplet Cooper
correlation in $(p_x\pm ip_y)$-type,\cite{Tao_1} represented by $[{\bf
  l}({\bf k})\cdot{\bgreek \sigma}]i\sigma_y$ with ${\bgreek
  \sigma}=(\sigma_x,\sigma_y,\sigma_z)$
 denoting the Pauli matrices.
 In the equilibrium state, the
${\bf l}$-vector of the triplet Cooper correlation is parallel to the effective
magnetic field ${\bgreek \Omega}({\bf k})$ due to the SOC in the momentum space.
Here, the ${\bf l}$-vector is
  defined from
\begin{equation}
\big[{\bf l}({\bf k})\cdot {\bgreek \sigma}\big]i\sigma_y=\left(
\begin{array}{cccc}
F_{\uparrow\uparrow}({\bf k}) & F_{\uparrow\downarrow}({\bf k})\\
F^*_{\uparrow\downarrow}({\bf k})& F_{\downarrow\downarrow}({\bf k})
\end{array}
\right), 
\label{definition_first}
\end{equation}
with $F({\bf k})$ representing the anomalous correlations of {\em triplet} Cooper
pairs.\cite{Leggett_book}

Actually, the anomalous correlations $F({\bf k})$, calculated by the optical
Bloch equations in this work, are just the Fourier components of the wavefunction of
triplet Cooper pairs in the spatial space.\cite{Leggett_book} By further considering the spin
space, the wavefunction of the triplet Cooper pairs is expressed as\cite{Leggett_book}
\begin{eqnarray}
\nonumber
F_t({\bf r})&=&F_{\uparrow\uparrow}({\bf r})|\uparrow_1\rangle|\uparrow_2\rangle
+F_{\downarrow\downarrow}({\bf r})|\downarrow_1\rangle|\downarrow_2\rangle\\
\mbox{}&+&F_{\uparrow\downarrow}({\bf r})(1/\sqrt{2})\big(|\uparrow_1\rangle|
\downarrow_2\rangle+|\downarrow_1\rangle|\uparrow_2\rangle\big).
\label{pair_function}
\end{eqnarray} 
Here, $F_{\uparrow\uparrow}({\bf r})$, $F_{\downarrow\downarrow}({\bf r})$
and $F_{\uparrow\downarrow}({\bf r})$ denote the wavefuctions of the triplet
Cooper pairs with total spin $S_z=1$, $-1$ and $0$,\cite{Leggett_book} respectively, with ${\bf r}$
being the relative coordinate for the two electrons (labeled by ``1'' and ``2'') in
the Cooper pairs. 
Thus,
with the $\hat{\bf z}$-direction of
the spin operator chosen to be perpendicular to the QWs, in the equilibrium state of
superconducting InSb (100) QWs, $F_{\uparrow\downarrow}({\bf r})=0$
 and $|F_{\uparrow\uparrow}({\bf r})|=|F_{\downarrow\downarrow}({\bf
   r})|$.\cite{Tao_1} From the Cooper-pair wavefunction, the total spin polarization
 of Cooper pairs is determined by
 \begin{equation}
 {\bf P}_{\rm C}=\int d{\bf r}F_t^*({\bf r})\hat{\bf
   S}F_t({\bf r})\propto \sum_{\bf k} i{\bf l}({\bf k})\times {\bf l}^*({\bf
    k}),
 \label{polarization_definition}
 \end{equation}
 with $\hat{\bf
   S}\equiv \hat{\bf s}_1+\hat{\bf s}_2$ being the total spin operator by the
 sum of the spin operators $\hat{\bf s}_1$ and $\hat{\bf s}_2$ of two
 electrons.

When the optical
field with the vector potential along the $\hat{\bf x}$-direction is applied to the
  superconducting system, the superconducting velocity is induced, which is
  shown to contribute to an effective SOC along the $\hat{\bf x}$-direction. 
  This effective SOC can cause the
  precession of the ${\bf l}$-vectors, with a component perpendicular to ${\bgreek
    \Omega}({\bf k})$ induced. Thus, with the Cooper-pair spin vector
  defined as
  ${\bf n}({\bf k})=i{\bf l}({\bf k})\times {\bf l}^*({\bf
    k})$,\cite{Leggett,Sigrist,Tkachov,non_unitary} whose momentum integral 
  contributes to the total Cooper-pair spin polarization ${\bf P}_{\rm C}$ [refer to
    Eq.~(\ref{polarization_definition})], the
  $\hat{\bf x}$-component of ${\bf P}_{\rm C}$ can be induced.
  Specifically, the $\hat{\bf x}$-component of the Cooper-pair spin
 polarization is ${\bf P}_{\rm C}^x=(1/\sqrt{2})\int d{\bf r}\big\{F_{\uparrow\downarrow}^*({\bf
   r})\big[F_{\uparrow\uparrow}({\bf r})+F_{\downarrow\downarrow}({\bf
   r})\big]+{\rm h.c.}\big\}$. Accordingly, one finds
that the excitation of the $\hat{\bf x}$-component of the Cooper-pair spin
 polarization is the reflection of the optical-induction of the triplet Cooper-pair
 wavefunction $F_{\uparrow\downarrow}({\bf
   r})$ with $S_z=0$. Actually, the Fourier component of $F_{\uparrow\downarrow}({\bf
   r})$ is exactly ${\bf l}_z({\bf k})$.\cite{Leggett,Sigrist,Tkachov,non_unitary}
Furthermore, we reveal that the Cooper-pair spin polarization is proportional
  to the superconducting velocity, which oscillates with the frequency of the optical field.

This paper is organized as follows. We first focus on the $s$-wave
superconducting semiconductor QWs in Sec.~\ref{s_wave}, whose
framework is then generalized to the ($s$+$p$)-wave one in (100) QWs in
Sec.~\ref{p_wave}.
Specifically, for the $s$-wave [($s$+$p$)-wave]
superconducting QWs, we present the Hamiltonian in Sec.~\ref{Hamiltonian_s}
(Sec.~\ref{Hamiltonian_p}); then 
in Sec.~\ref{KSBE} (Sec.~\ref{KSBE_p}), the optical Bloch equations are derived via the
gauge-invariant non-equilibrium Green function approach; the numerical results are presented in
Sec.~\ref{Numerical_s} (Sec.~\ref{Numerical_p}).
We conclude and discuss in Sec.~\ref{summary}.

\section{$s$-wave superconducting QW$s$}
\label{s_wave}
In this section, we investigate the optical response to the THz pulses in the $s$-wave
superconducting QWs,
which can be realized in the GaAs QWs in proximity to an $s$-wave superconductor
with negligible SOC. We
first present the Hamiltonian, in which the gauge structure is emphasized (Sec.~\ref{Hamiltonian_s}). Then
the optical Bloch equations via the nonequilibrium
Green function method with the generalized Kadanoff-Baym (GKB)
ansatz are set up, in which the gauge invariance is retained explicitly by using the
gauge-invariant Green function
(Sec.~{\ref{KSBE}}).\cite{Tao_2,Haug,Levanda,GKB,BoYe}
 Finally, we numerically calculate the
optical response by solving the optical Bloch equations including the THz-field--induced
oscillations of the Higgs mode and THz-field--induced charge
imbalance, in which a novel charge-imbalance relaxation channel due to
 the elastic momentum scattering is revealed (Sec.~{\ref{Numerical_s}}).

\subsection{Hamiltonian and Gauge Structure}
\label{Hamiltonian_s}
In the $s$-wave superconducting QWs with negligible SOC, the Hamiltonian is composed by the free  
 BdG Hamiltonian $H_0$ and the interaction
Hamiltonian including the electron-electron Coulomb, electron-phonon and
electron-impurity interactions $H_{\rm ee}$, $H_{\rm ep}$ and $H_{\rm ei}$.   
Specifically, $H_0$ is written as ($\hbar \equiv 1$ throughout this paper)
\begin{equation}
H_0=\int \frac{d{\bf r}}{2}\Psi^{\dagger}\left(
\begin{array}{cccc}
\zeta_{{\bf k}}^-(x)+e\phi(x) & |\Delta|e^{i\zeta(x)}\\
|\Delta|e^{-i\zeta(x)}&-\zeta_{{\bf k}}^+(x)-e\phi(x) 
\end{array}
\right)\Psi,
\label{BdG_s1}
\end{equation}
in which $\zeta_{{\bf k}}^{\pm}(x)=\big[{\bf k}\pm\frac{e}{c}{\bf
  A}(x)\big]^2/(2m^*)-\mu$ with $x\equiv(t,{\bf r})$ being the time-space
point, ${\bf A}(x)$ denoting the vector potential and $\mu$ representing the chemical potential of the system; 
$\Psi(x)=(\psi_{\uparrow}(x),\psi^{\dagger}_{\downarrow}(x))^T$ is the particle field
operator in the Nambu space; $\phi(x)$ denotes the scalar potential; $\Delta$ and
$\zeta(x)$ stand for the $s$-wave order parameter and 
the superconducting phase. The electron-electron, electron-phonon and electron-impurity interactions are written as 
\begin{eqnarray}
\hspace{-0.8cm}&&H_{\rm ee}=\int \frac{d{\bf r}d{\bf r}'}{2}U({\bf r}-{\bf
  r}')\big[\Psi^{\dagger}({\bf r})\tau_3\Psi({\bf r})\big]
\big[\Psi^{\dagger}({\bf r}')\tau_3\Psi({\bf r}')\big],\\
\hspace{-0.8cm}&&H_{\rm ep}=\frac{1}{2}\int d{\bf r}d{\bf r}'g^{\lambda}({\bf r}-{\bf
  r}')\Psi^{\dagger}({\bf r})\tau_3\Psi({\bf r})\chi({\bf r}'),\\
\hspace{-0.8cm}&&H_{\rm ei}=\frac{1}{2}\int d{\bf r}\Psi^{\dagger}({\bf
  r})V({\bf r})\tau_3\Psi({\bf r}),
\end{eqnarray}
respectively. 
Here, ${\bgreek \tau}\equiv (\tau_1,\tau_2,\tau_3)$ represent the Pauli matrices
in the Nambu space; $U({\bf r})$ and $V({\bf r})$ denote the
screened Coulomb potentials whose
expressions have been derived in Ref.~\onlinecite{Tao_1}; $\chi({\bf
  r})$ is the phonon field operator; and $g^{\lambda}({\bf r}-{\bf
  r}')$ stand for the electron-phonon interactions due to the deformation potential in the LA branch
and piezoelectric coupling including LA and TA branches, with 
  $\lambda$ denoting the corresponding phonon branch.\cite{e_p_formula1,e_p_formula2} Their Fourier
components $g^{\lambda}({\bf p})$
are explicitly given in Refs.~\onlinecite{e_p_formula1,e_p_formula2}.

The gauge structure in the $s$-wave
superconductivity was first revealed by Nambu.\cite{Nambu_gauge,Stephen,Altland}
By performing the gauge transformation, i.e., 
\begin{equation}
  \Psi(x)\rightarrow
  e^{i\tau_3\Lambda(x)/2}\Psi(x),
\label{gauge_transformation}
\end{equation}
the gauge invariance of the BdG Hamiltonian [Eq.~(\ref{BdG_s1})] requires 
the vector potential,
scalar potential and superconducting phase transforming as\cite{Nambu_gauge,Stephen,Altland} 
\begin{eqnarray}
\label{vector}
&&{\bf A}(x)\rightarrow {\bf A}(x)+(c/2e)\nabla\Lambda(x),\\
\label{scalar}
&&\phi(x)\rightarrow \phi(x)-(1/2e)\partial_t\Lambda(x),\\
&&\zeta(x)\rightarrow \zeta(x)+\Lambda(x).
\label{phase}
\end{eqnarray}
From Eqs.~(\ref{vector}-\ref{phase}), one can construct
the gauge-invariant physical quantities\cite{Nambu_gauge,Stephen,Altland}
\begin{eqnarray}
\label{superconducting_momentum}
&&{\bf p}_s(x)=(1/2)\nabla\zeta(x)-(e/c){\bf A}(x),\\
&&\mu_{\rm eff}(x)=(1/2)\partial_t\zeta(x)+e\phi(x),
\label{effective potential}
\end{eqnarray}
which represent the superconducting momentum and effective chemical
potential. It is noted that the above two gauge-invariant quantities are related
by the acceleration relation\cite{Nambu_gauge,Stephen,Altland} 
\begin{equation}
\partial_t{\bf p}_s=\nabla\mu_{\rm eff}+e{\bf E},
\label{EOM}
\end{equation}
 which is valid under any
  circumstances. Thus, with an optical field applied to the superconducting
  system, Eq.~(\ref{EOM}) shows that in the homogeneous limit, a
  time-dependent superconducting momentum can be induced, which is always a
  transverse physical quantity in the presence of the optical field.\cite{Mattis,Nambu_gauge}

\subsection{Optical Bloch Equations}
\label{KSBE}
In this section, we derive the optical Bloch equations 
  in the $s$-wave superconducting QWs via the nonequilibrium Green
function method with the GKB ansatz.\cite{Haug,wureview,GKB,Tao_2} 
 From Sec.~\ref{Hamiltonian_s}, one notices that there exists a 
nontrivial gauge structure in the BdG Hamiltonian. To account for this gauge
structure, the gauge-invariant Green function is used to obtain the
gauge-invariant kinetic equations.\cite{Haug,BoYe,Levanda}  

\subsubsection{Gauge-invariant Green function}
\label{Derivation}
The optical Bloch equations can be
  constructed from the ``lesser'' Green function $G_{12}^<\equiv
  i\langle\Psi_2^{\dagger}\Psi_1\rangle$, in which $1\equiv x_1=(t_1,{\bf r}_1)$ represents the
time-space point and $\langle \cdots\rangle$ denotes the ensemble
average.\cite{Haug,wureview,Tao_2}
 With the gauge transformation in Eq.~(\ref{gauge_transformation}), the
 ``lesser'' Green
  function transforms as $G_{12}^<\rightarrow
  e^{i\tau_3\Lambda(x_1)/2}G^<_{12}e^{-i\tau_3\Lambda(x_2)/2}$. As in the
  kinetic equations {\em in the quasiparticle approximation},\cite{Haug} only the
  center-of-mass coordinates are retained, the gauge structure cannot be easily 
  realized in the kinetic equations constructed from
  $G^<_{12}$.\cite{Levanda,Haug}
 Nevertheless, the gauge invariance can be retained by
  introducing the Wilson
  line to construct the gauge-invariant Green function,\cite{Levanda,Haug,Wilson_line} which is constructed as
\begin{equation}
\tilde{G}^<_{12}=Pe^{-ie\int_{x_1}^RA_jdx^j\tau_3}G^<_{12}e^{-ie\int_{R}^{x_2}A_jdx^j\tau_3}.
\label{Wilson}
\end{equation}
In Eq.~(\ref{Wilson}), $A_jdx^j\equiv\phi dt-(1/c){\bf A}\cdot d{\bf r}$, $R\equiv({\bf R},T)=\big(({\bf
      r}_1+{\bf r}_2)/2,(t_1+t_2)/2\big)$ are the center-of-mass
  coordinates, and ``$P$'' indicates that the line integral is path-dependent. Then
by the gauge transformation in Eq.~(\ref{gauge_transformation}), the
gauge-invariant Green function is transformed as
$\tilde{G}_{12}^<\rightarrow
  e^{i\tau_3\Lambda(R)/2}\tilde{G}^<_{12}e^{-i\tau_3\Lambda(R)/2}$, in which the
  transformed phase only depend on the center-of-mass coordinates.

Finally, by choosing the path to be the straight line connecting $x_1$ and $x_2$,\cite{Levanda,Haug} the
 gauge-invariant Green function reads
\begin{eqnarray}
\nonumber
\hspace{-0.5cm}&&\tilde{G}^<_{12}=\exp\Big[ie\int_0^{\frac{1}{2}}d\lambda
A_j(T+\lambda \tau,{\bf R}+\lambda{\bf r})x^j\tau_3\Big]\\
\hspace{-0.5cm}&&
\mbox{}\times
G^<_{12}\exp\Big[ie\int_{-\frac{1}{2}}^0d\lambda
A_j(T+\lambda \tau,{\bf R}+\lambda{\bf r})x^j\tau_3\Big],
\end{eqnarray}
in which $x=(\tau,{\bf r})=(t_1-t_2,{\bf r}_1-{\bf r}_2)$ are the relative coordinates.

\subsubsection{Derivation on the optical Bloch equations}
In this part, we derive the optical Bloch equations in the 
$s$-wave superconducting QWs, with special attention paid to the gauge structure.
 Accordingly, we do not specify any gauge in the beginning of 
the derivation, and finally choose a special gauge for the convenience of
physical analysis and numerical calculation. Thus, in the derived equations, there
exist ${\bf A}({\bf r},t)$, $\phi({\bf r},t)$ and $\zeta({\bf r},t)$, which
are not physical quantities.

We begin from the two Dyson equations,\cite{Haug,wureview,Tao_2} 
\begin{eqnarray}
\label{Dyson1}
&&\hspace{-1.2cm}i\partial_{t_1} G_{12}^<-H_{{\bf k}_1}G_{12}^<
=\int d3(\Sigma_{13}^RG_{32}^<+\Sigma_{13}^<G_{32}^A),\\
&&\hspace{-1.2cm}-i\partial_{t_2}
G_{12}^<-G_{12}^<\stackrel{\leftarrow}{H}_{{\bf k}_2}
=-\int d3(G_{13}^R\Sigma_{32}^<+G_{13}^<\Sigma_{32}^A),
\label{Dyson2}
\end{eqnarray}
in which ``$R$'' and ``$A$'' label the retarded and advanced Green
functions, and $\Sigma$ are the self-energies contributed
 by the electron-electron and electron-impurity
 interactions.\cite{Haug,wureview,Tao_2}
 In Eqs.~(\ref{Dyson1}) and (\ref{Dyson2}), 
\begin{equation}
H_{{\bf k}_1}=\left(
\begin{array}{cccc}
\frac{({\bf k}_1-\frac{e}{c}{\bf A}_1)^2}{2m^*}-\mu+e\phi_1 & |\Delta|e^{i\zeta_1}\\
|\Delta|e^{-i\zeta_1}&-\frac{({\bf k}_1+\frac{e}{c}{\bf A}_1)^2}{2m^*}+\mu-e\phi_1 
\end{array}
\right),
\end{equation}
and
\begin{equation}
H_{{\bf k}_2}=
\left(
\begin{array}{cccc}
\frac{({\bf k}_2+\frac{e}{c}{\bf A}_2)^2}{2m^*}-\mu+e\phi_2 & |\Delta|e^{i\zeta_2}\\
|\Delta|e^{-i\zeta_2}&-\frac{({\bf k}_2-\frac{e}{c}{\bf A}_2)^2}{2m^*}+\mu-e\phi_2 
\end{array}
\right).
\end{equation}
We first present the derivation of the free terms in the kinetic equations including the coherent, pump,
drive and diffusion terms, in which the gauge-invariant scheme is
used. Specifically, 
from the left-hand side of Eqs.~(\ref{Dyson1}) and (\ref{Dyson2}),
one obtains the equations for the gauge-invariant Green function
$\tilde{G}_{12}^<$. Then by
using the gradient expansion, the
kinetic equations are derived from the Fourier component of the gauge-invariant
Green function $\tilde{G}({\bf
    k},\omega;{\bf R},T)=\int d{\bf r}d{\tau}e^{i\omega\tau-i{\bf
      k}\cdot{\bf r}}\tilde{G}^<_{12}$. Finally, after the
  integration over the frequency, one obtains the optical Bloch equations for
  the $2\times 2$ density matrix in the Nambu space 
\begin{equation} 
\tilde{\rho}_{\bf k}({\bf R},T)=\int \frac{d\omega}{2\pi}\tilde{G}({\bf
    k},\omega;{\bf R},T), 
\label{integration}
\end{equation}
whose diagonal terms represent the distributions of electron and hole, and
off-diagonal terms denote the anomalous correlations. 
Finally, the optical kinetic equations are written as 
\begin{widetext}
\begin{eqnarray}
\nonumber
\hspace{-0.85cm}&&\frac{\partial \tilde{\rho}_{\bf k}}{\partial T}+
i\Big[\Big(\frac{{\bf k}^2}{2m^*}-\mu+e\phi\Big)\tau_3,\tilde{\rho}_{\bf k}\Big]
+i\Big[\left(
\begin{array}{cc}
0&|\Delta|e^{i\zeta(R)} \\
|\Delta|e^{-i\zeta(R)}&0
\end{array}
\right),\tilde{\rho}_{{\bf k}}\Big]+
i\Big[\frac{1}{2m^*}\Big(\frac{e}{c}{\bf
  A}\Big)^2\tau_3,\tilde{\rho}_{\bf k}\Big]
+\frac{1}{2}
\Big\{e{\bf E}\tau_3,\frac{\partial \tilde{\rho}_{\bf k}}{\partial {\bf
    k}}\Big\}\\
\hspace{-0.85cm}&&\mbox{}-i\Big[\frac{1}{8m^*}\tau_3,\frac{\partial^2\tilde{\rho}_{\bf
    k}}{\partial {\bf R}^2}\Big]+\frac{1}{2}\Big\{\frac{{\bf
      k}}{m^*}\tau_3,\frac{\partial
 \tilde{\rho}_{\bf k}}{\partial {\bf
   R}}\Big\}+\Big[\frac{e{\bf A}}{2m^*c}\tau_3,\frac{\partial \tilde{\rho}_{\bf k}}{\partial
{\bf R}}\tau_3\Big]+\Big[\frac{e}{4m^*c}\nabla\cdot{\bf A}\tau_3,
  \tilde{\rho}_{\bf k}\tau_3\Big]
=\frac{\partial \tilde{\rho}_{\bf
    k}}{\partial t}\Big|_{\rm HF}+\frac{\partial \tilde{\rho}_{\bf k}}{\partial t}\Big|_{\rm scat},
\label{KSBE_full}
\end{eqnarray}
\end{widetext}
with ${\bf E}=-\nabla_{\bf R}\phi-(1/c)\partial_T{\bf A}$. Here, $[{\rm A},{\rm
  B}]={\rm AB}-{\rm BA}$ and
$\{{\rm A},{\rm B}\}={\rm AB}+{\rm BA}$ represent the commutator and
anti-commutator, respectively. It is noted that in the equation, the
  gradient expansion has been performed to the second order in ${\bf R}$,
  i.e., the sixth term on the left-hand side in Eq.~(\ref{KSBE_full}), to
  retain the gauge-invariance structure in the optical kinetic equations.

In Eq.~(\ref{KSBE_full}), on the left-hand side, the second and third terms
represent the coherent terms contributed by the kinetic energy and the order
parameter, respectively;
the fourth term describes the pump
term, as addressed in the Liouville equation in the literature;\cite{Soliton,Critical,Axt1,Axt2,Anderson1,
  Higgs_e_p,multi_component,Leggett_mode,Xie,Tsuji_e_p,multiband,path_integral} the fifth term is the
drive term, which can directly induce
 the center-of-mass momentum of the Cooper pairs [Eq.~(\ref{EOM})];\cite{Altland,Nambu_gauge,Schrieffer}
the diffusion terms are contributed by the sixth to the ninth
terms.
On the right-hand side of the equation, $\partial_t \tilde{\rho}_{\bf
  k}|_{\rm HF}$ and $\partial_t \tilde{\rho}_{\bf k}|_{\rm scat}$ represent the
Hartree-Fock (HF) term contributed by the Coulomb interaction and scattering term due
to the electron-impurity and electron-phonon interactions,
which are derived from the right-hand side of Eqs.~(\ref{Dyson1}) and
(\ref{Dyson2}). The gauge-invariant versions of the scattering terms are 
complex.\cite{Haug,BoYe,Levanda} Nevertheless, these terms can
be approximated by the ones without gauge-invariant treatments as long as the applied
field is not very strong with {\em the driven center-of-mass momentum of the system
being much smaller than the Fermi momentum $k_{\rm F}$}.\cite{Haug,BoYe,Stephen}
 In this situation, the energy spectra is not significantly disturbed. 
The gauge structure of Eq.~(\ref{KSBE_full}) is then checked by the gauge
transformation $\tilde{\rho}_{\bf k}\rightarrow
e^{i\tau_3\Lambda(R)/2}\tilde{\rho}_{\bf k}e^{-i\tau_3\Lambda(R)/2}$. The same
gauge structures as Eqs.~(\ref{vector}), (\ref{scalar}) and (\ref{phase}) are
obtained for the vector potential, scalar potential and superconducting phase.

For the convenience of the physical analysis and numerical calculation,
 a specific gauge is chosen.
 It is noted that
generally one
cannot choose two quantities in the vector potential, scalar potential and
superconducting phase to be zero. Nevertheless, in the
 Liouville and Bloch equations used in the literature, both the scalar potential and
superconducting phase are taken to be zero.\cite{Soliton,Critical,Axt1,Axt2,Anderson1,
Higgs_e_p,multi_component,Leggett_mode,Xie,Tsuji_e_p,multiband,path_integral}
 Here, we choose a special gauge referred to as the ${\bf p}_s$-gauge,
 in which the superconducting phase $\zeta$ is
 zero.\cite{Kerson_Huang,Altland}
 This can be realized by
the gauge transformation $\tilde{\rho}_{\bf k}\rightarrow
  e^{-i\tau_3\zeta(R)/2}\tilde{\rho}_{\bf k}e^{i\tau_3\zeta(R)/2}\equiv
  \rho_{\bf k}$ in Eq.~(\ref{KSBE_full}). Then by using the definition of the
  superconducting momentum [Eq.~(\ref{superconducting_momentum})]
  and effective chemical potential [Eq.~\ref{effective potential}],
 the optical Bloch equations become 
\begin{eqnarray}
\nonumber
\hspace{-0.5cm}&&\frac{\partial \rho_{\bf k}}{\partial T}+
i\Big[\Big(\frac{{\bf k}^2}{2m^*}-\Phi\Big)\tau_3,\rho_{\bf k}\Big]+i\Big[\left(
\begin{array}{cc}
0&|\Delta| \\
|\Delta|&0
\end{array}
\right),\rho_{{\bf k}}\Big]\\
\nonumber
\hspace{-0.5cm}&&\mbox{}+
i\Big[\frac{{\bf
    p}_s^2}{2m^*}\tau_3,\rho_{\bf k}\Big]+\frac{1}{2}
\Big\{\Big(\frac{\partial {\bf p}_s}{\partial T}-\nabla_{\bf R}\mu_{\rm eff}\Big)\tau_3,
\frac{\partial \rho_{\bf k}}{\partial {\bf k}}\Big\}\\
\nonumber
\hspace{-0.5cm}&&\mbox{}+\frac{1}{2}\Big\{\frac{{\bf
      k}}{m^*}\tau_3,\frac{\partial
 \rho_{\bf k}}{\partial {\bf
   R}}\Big\}-i\Big[\frac{\tau_3}{8m^*},\frac{\partial^2\rho_{\bf
    k}}{\partial {\bf R}^2}\Big]-\Big[\frac{{\bf p}_s}{2m^*}\tau_3,\frac{\partial \rho_{\bf k}}{\partial
{\bf R}}\tau_3\Big]\\
\hspace{-0.5cm}&&\mbox{}-\Big[\frac{1}{4m^*}\nabla_{\bf R}\cdot{\bf
    p}_s\tau_3,\rho_{\bf k}\tau_3\Big]
=\frac{\partial\rho_{\bf
    k}}{\partial t}\Big|_{\rm HF}+\frac{\partial\rho_{\bf k}}{\partial t}\Big|_{\rm scat},
\label{zeta_gauge}
\end{eqnarray}
where $\Phi=\mu-\mu_{\rm eff}$ is the total chemical potential in the system
including the contribution from the rate of change of the superconducting phase.

 It is noted that
in Eq.~(\ref{zeta_gauge}), 
the electric force $e{\bf E}$ is replaced by $\partial_T {\bf p}_s-\nabla_{\bf R}\mu_{\rm eff}$
according to the acceleration relation [Eq.~(\ref{EOM})]. Accordingly, in Eq.~(\ref{zeta_gauge}), only the gauge
invariant physical quantities ${\bf p}_s$ and $\mu_{\rm eff}$
appear. In fact, in the gauge-invariant framework, from any specific gauge at the
beginning of the derivation, one can obtain Eq.~(\ref{zeta_gauge}) with the
existence of both the pump
and drive terms.\cite{Haug} Moreover, in Eq.~(\ref{zeta_gauge}),
with ${\bf p}_s$ and $\mu_{\rm eff}$ describing the kinetics of the condensate,
Eq.~(\ref{zeta_gauge}) not only describes the dynamics of the quasiparticle, bot
also includes the influence of the condensate. This is consistent with the
  two-component description for the charge, in which there exists interplay between the quasiparticle and
  condensate.\cite{Tinkham1,Tinkham2,Takahashi_SHE2,Tinkham_book,Pethick_review,Larkin,Gray}

 When considering the optical excitation by the THz pulses in the
 superconductor, Eq.~(\ref{zeta_gauge}) can be significantly simplified.
Often the spacial dependence in the optical field can be neglected, and hence
Eq.~(\ref{zeta_gauge}) can be solved in the homogeneous limit. 
Specifically, with $\Phi$, ${\bf p}_s$ and
$\rho_{\bf k}$ being independent on ${\bf R}$, the optical Bloch equations
[Eq.~(\ref{zeta_gauge})] are reduced to 
\begin{eqnarray}
\nonumber
\hspace{-0.4cm}&&\frac{\partial \rho_{\bf k}}{\partial T}+
i\Big[\Big(\frac{{\bf k}^2}{2m^*}-\Phi\Big)\tau_3,\rho_{\bf k}\Big]
+i\Big[\left(
\begin{array}{cc}
0&|\Delta| \\
|\Delta|&0
\end{array}
\right),\rho_{{\bf k}}\Big]\\
\nonumber
\hspace{-0.4cm}&&+
i\Big[\frac{{\bf
    p}_s^2}{2m^*}\tau_3,\rho_{\bf k}\Big]
+\frac{1}{2}
\Big\{\frac{\partial {\bf p}_s}{\partial T}\tau_3,\frac{\partial \rho_{\bf k}}{\partial {\bf
    k}}\Big\}=\frac{\partial\rho_{\bf
    k}}{\partial t}\Big|_{\rm HF}+\frac{\partial \rho_{\bf k}}{\partial t}\Big|_{\rm scat}.\\
\hspace{-0.4cm}&&
\label{homogeneous}
\end{eqnarray}
It is addressed that Eq.~(\ref{homogeneous}) is different from the Liouville\cite{Axt1,Axt2,
   Higgs_e_p} or
 Bloch\cite{Anderson1,multi_component,Leggett_mode,Xie,Tsuji_e_p,multiband,Matsunaga_3,Matsunaga_4}
 equations used in
  the literature in several aspects. Firstly, the momenta of the two electrons
  participating 
  in the anomalous correlation are no longer ${\bf k}$ and $-{\bf k}$ during the
  evolution. This is because in the optical kinetic equation here, similar to the
  Boltzmann equation,\cite{Haug,Kadanoff_Baym,Hershfield,Spivak,kinetic_book,Kopnin} the Lagrangian
  description is used, in which the generalized coordinate evolves with
  time.\cite{Landau} Thus, with the anomalous correlation represented by
  $\langle c_{{\bf k}(T)}c_{{\bf k}'(T)}\rangle$ in which $c_{{\bf k}}$ is the
  annihilation operator of the electron, the center-of-mass momentum of the
  Cooper pairs ${\bf p}_s=[{\bf k}(T)+{\bf k}'(T)]/2$. Then, with $\partial_T {\bf
    k}(T)=\partial_T {\bf
    k}'(T)=e{\bf E}$, the acceleration relation in the homogeneous limit [Eq.~(\ref{EOM})] can be
  directly recovered. One sees that it is natural to include the contribution of the
  center-of-mass momentum in the anomalous correlation in our
  description. Secondly, in the homogeneous limit, with ${\bf p}_s$ and $\partial_T{\bf p}_s$ being
  transverse in the presence of the optical field [Eq.~(\ref{EOM})], the obtained electrical
  current is perpendicular to the propagation direction of the optical field.
  Moreover, the obtained physical quantities are naturally 
  gauge-invariant due to the gauge invariance in ${\bf p}_s$ and $\partial_T{\bf
    p}_s$. Furthermore, the effective chemical potential naturally arises
  from the gauge-invariant treatment in the derivation,
   which corresponds to the collective excitation, evolving
   with time in the homogeneous limit.\cite{Nambu_gauge,Ambegaokar,Enz,Griffin} 
  Finally, the scattering term can be simply included in
  our description which is similar to its setup in the Boltzmann
  equation,\cite{Haug,Kadanoff_Baym,Hershfield,Spivak,kinetic_book,Kopnin}
 with the details addressed as follows.

In Eq.~(\ref{homogeneous}), $\partial_t \rho_{\bf
    k}|_{\rm HF}$ and $\partial_t \rho_{\bf k}|_{\rm scat}$ are derived in
  the GKB ansatz.\cite{Tao_1,Tao_2} For the HF term, it is written as 
\begin{equation}
\partial_t \rho_{\bf
    k}|_{\rm HF}=i\sum_{{\bf k}'}\Big[U_{{\bf k}-{\bf
    k}'}\tau_3(\rho_{{\bf k}'}-\rho^0_{{\bf
      k}'})\tau_3,\rho_{\bf k}\Big].
\label{HF}
\end{equation}
In Eq.~(\ref{HF}), it is assumed that the renormalization energy due to the Coulomb
interaction has been included in the free BdG Hamiltonian [Eq.~(\ref{BdG_s1})], and
hence the density matrix in the equilibrium state $\rho^0_{{\bf
    k}}$ appears in the HF self-energy.
 Accordingly, the fluctuation
of the order parameter is represented by 
\begin{equation}
  \delta\Delta({\bf k})=\sum_{{\bf k}'}U_{{\bf k}-{\bf
      k}'}(\rho_{{\bf k}',12}-\rho^0_{{\bf
      k},12}),
\label{Higgs_formula}
\end{equation}
which can be treated as the Higgs mode when the phase fluctuation
 can be neglected.\cite{Soliton,Critical,Axt1,Axt2,Anderson1,
Higgs_e_p,multi_component,Leggett_mode,Xie,Tsuji_e_p,multiband,path_integral}

For the scattering terms, both the electron-impurity and electron-phonon
interactions are considered, which are written as  
\begin{eqnarray}
\nonumber
\hspace{-0.5cm}&&\partial_t \rho_{\bf k}|_{\rm ei}=-\pi n_i\sum_{{\bf k}'}\sum_{\eta_1\eta_2=\pm}|V_{{\bf
    k}-{\bf k}'}|^2\delta(E_{{\bf k}'\eta_1}-E_{{\bf
    k}\eta_2})\\
\label{scat}
\hspace{-0.5cm}&&\mbox{}\times\big[\tau_3\Gamma_{{\bf k}'\eta_1}\tau_3\Gamma_{{\bf
    k}\eta_2}\rho_{\bf k}-\tau_3\rho_{{\bf k}'}\Gamma_{{\bf
    k}'\eta_1}\tau_3\Gamma_{{\bf k}\eta_2}+{\rm H.c.}\big],\\
\nonumber
\hspace{-0.5cm}&&\partial_t \rho_{\bf k}|_{\rm ep}=-\pi \sum_{{\bf k}'k_z}
\sum_{\eta_1\eta_2=\pm}|g^{\lambda}_{{\bf
    k}-{\bf k}',k_z}|^2\delta(E_{{\bf k}'\eta_1}-E_{{\bf
    k}\eta_2}+\omega_{{\bf k}-{\bf k}'}^{\lambda})\\
\nonumber
\hspace{-0.5cm}&&\mbox{}\times(1+n_{{\bf k}-{\bf k}'})
\big[\tau_3\rho_{{\bf k}'}^>\Gamma_{{\bf k}'\eta_1}\tau_3\Gamma_{{\bf
    k}\eta_2}\rho^<_{\bf k}-\tau_3\rho_{{\bf
      k}'}^<\Gamma_{{\bf k}'\eta_1}\tau_3\Gamma_{{\bf k}\eta_2}\rho_{{\bf
      k}}^>\\
\hspace{-0.5cm}&&\mbox{}+{\rm
    H.c.}\big]+\big[\omega^{\lambda}_{{\bf k}-{\bf k}'}\rightarrow-\omega^{\lambda}_{{\bf k}-{\bf k}'};
  (1+n_{{\bf k}-{\bf k}'})\rightarrow n_{{\bf k}-{\bf k}'}\big].
\label{scat_ep}
\end{eqnarray}
In Eq.~(\ref{scat}), $n_i$ is the
impurity density; $E_{{\bf
    k}\pm}=\pm E_{\bf k}$ in which $E_{\bf
k}=\sqrt{\zeta_{\bf k}^2+|\Delta|^2}$ with $\zeta_{\bf k}\equiv \varepsilon_{\bf k}-\mu={\bf
k}^2/(2m^*)-\mu$;
 $\Gamma_{{\bf k}\pm}=1/2\pm (1/2){\mathcal U}_{\bf
  k}^{\dagger}\tau_3{\mathcal U}_{\bf k}$ represent the projection
operators. Here, 
\begin{equation}
{\mathcal U}_{\bf k}=\left(
\begin{array}{cc}
u_{\bf k} & v_{\bf k}\\
-v_{\bf k} & u_{\bf k} 
\end{array}
\right)
\label{unitary}
\end{equation}
is the unitary transformation matrix from the particle space to the
quasiparticle one with $u_{\bf k}=\sqrt{1/2+\zeta_{\bf k}/(2E_{\bf k})}$ and $v_{\bf
k}=\sqrt{1/2-\zeta_{\bf k}/(2E_{\bf k})}$.
In Eq.~(\ref{scat_ep}),
  $\omega^{\lambda}_{{\bf k}}$ is the $\lambda$-branch--phonon energy with
  momentum ${\bf k}$;
 $n_{{\bf k}}$
represents the phonon distribution function; $\rho_{\bf
  k}^{\stackrel{>}{<}}\equiv \rho_{\bf k}+1/2\pm 1/2$.

Finally, we point out that the structures of the pump, drive and scattering terms in
Eq.~(\ref{homogeneous}) can be analyzed more clearly in the quasiparticle space,
in which the optical Bloch equations are set up by the Bogoliubov transformation
$\rho_{\bf k}^h={\mathcal U}_{\bf k}\rho_{\bf k}{\mathcal U}_{\bf k}^{\dagger}$.
These detailed analysis are presented in Appendix~\ref{AA}.

\subsubsection{Charge neutrality condition}
\label{charge_picure}

Equation~(\ref{KSBE_full}) provides the microscopic description
for the quasiparticle dynamics. Moreover, in the ${\bf p}_s$-gauge,
both the superfluid
momentum ${\bf p}_s$ and the effective chemical potential $\mu_{\rm eff}$ which
are associated with the dynamics of the
condensate, appear in Eq.~(\ref{KSBE_full}), although ${\bf p}_s$ and $\mu_{\rm eff}$
still needs to be determined. Thus, the two-component picture naturally
arises in our description, in which there exists the interplay between the
quasiparticle and
condensate.\cite{Tinkham1,Tinkham2,Takahashi_SHE2,Tinkham_book,
  Pethick_review,Larkin,Gray} Actually, this can be directly seen from the modified Bogoliubov
transformation in which the creation and annihilation of the Cooper-pair
operators $S$ and $S^{\dagger}$ are added,\cite{Bardeen,Tinkham2,Josephson_B}
\begin{equation}
\left(
\begin{array}{c}
c_{{\bf k}\uparrow} \\
\hat{S}c_{-{\bf k}\downarrow}^{\dagger}
\end{array}
\right)=
{\mathcal U}_{\bf k}
\left(
\begin{array}{c}
\alpha_{{\bf k}\uparrow} \\
\beta_{{\bf k}\downarrow}^{\dagger}
\end{array}
\right).
\label{modified_B}
\end{equation}
Here,  $\alpha_{{\bf k}\uparrow}^{\dagger}$ ($\beta_{{\bf
    k}\downarrow}^{\dagger}$) is the creation operator for the quasi-electron (quasi-hole).
 From Eq.~(\ref{modified_B}), one has  
$\alpha_{{\bf k}\uparrow}^{\dagger}=u_{\bf k}c_{{\bf k}\uparrow}^{\dagger}-v_{\bf
  k}\hat{S}^{\dagger}c_{-{\bf k}\downarrow}$ and $\beta_{{\bf k}\downarrow}^{\dagger}=v_{\bf k}c_{{\bf k}\uparrow}+u_{\bf
  k}\hat{S}c_{-{\bf k}\downarrow}^{\dagger}$. By noting that $\hat{S}$ annihilates one 
Cooper pair with charge $2e$, one obtains that $\alpha_{{\bf k}\uparrow}^{\dagger}$ ($\beta_{{\bf
    k}\downarrow}^{\dagger}$)
corresponds to create a quasi-electron (quasi-hole) with charge
  $e$ ($-e$).
 Furthermore, one observes that the creation of one 
quasi-electron and one quasi-hole is associated with the creation and
annihilation of the Cooper pair with probability
  $v_{\bf k}^2$ and $u_{\bf k}^2$, respectively. Thus, the net
  creation of the Cooper pair is $v_{\bf k}^2-u_{\bf k}^2$, which is positive (negative)
  when $|{\bf k}|<k_F$ ($|{\bf k}|>k_F$). Accordingly, when $|{\bf
    k}|<k_F$, both quasiparticles and
  Cooper pairs are created; whereas when $|{\bf k}|>k_F$, the quasiparticles are created by breaking 
  Cooper pairs.

The above physical picture suggests that in the dynamical process, to maintain the charge
neutrality or charge conservation, the Cooper pair condensate has to  
respond to the dynamics of the 
 quasiparticles.\cite{Takahashi_SHE1,Takahashi_SHE2,Hirashima_SHE,
Takahashi_SHE3,Takahashi_SHE_exp,Hershfield,Spivak,kinetic_book}
That is to say, in the dynamical process, once the charge
imbalance for the quasiparticle is created, the chemical potential of the
condensate reacts to screen the extra charge due to the charge
imbalance. Hence it
is suggested that in Eq.~(\ref{homogeneous}), the effective chemical potential 
$\mu_{\rm eff}$ is determined from the charge
 neutrality condition, which actually has been used
 in the dynamical problem in superconductivity.\cite{Takahashi_SHE1,Takahashi_SHE2,Hirashima_SHE,
Takahashi_SHE3,Takahashi_SHE_exp,Hershfield,Spivak,kinetic_book}
Specifically, in the quasiparticle space, the particle number with momentum
${\bf k}$ is expressed as 
\begin{equation}
n_{\bf k}=2v_{\bf k}^2+\frac{\zeta_{\bf k}}{E_{\bf k}}\big[\rho_{11}^{h}({\bf
  k})+\rho_{11}^{h}(-{\bf k})\big]-\frac{\Delta}{E_{\bf
    k}}\Big[\rho_{12}^{h}({\bf k})+\rho_{21}^h({\bf k})\Big],
\label{Tinkham_delta}
\end{equation}
with $v_{\bf k}^2$ treated as the distribution function of the
condensate.\cite{Takahashi_SHE1,Takahashi_SHE2,Hirashima_SHE,
Takahashi_SHE3,Takahashi_SHE_exp,Hershfield,Spivak,kinetic_book} When the system is {\em near 
zero temperature and the equilibrium state}, to keep charge neutrality, 
the chemical potential for the condensate is suggested to be 
  varied $\mu\rightarrow \Phi$.\cite{Tinkham_condensate,Hershfield,Spivak,kinetic_book} Then the time
  evolution of the effective
  chemical potential can be obtained by solving the self-consistent equation
  with the quasiparticle density matrix obtained 
  from Eq.~(\ref{KSBE_full}),\cite{Tinkham_condensate,Hershfield,Spivak,kinetic_book}
\begin{eqnarray}
\nonumber
\hspace{-0.6cm}&&\sum_{\bf k}n_{\bf k}\equiv n_0=
\sum_{\bf k}\Big[1-\frac{\varepsilon_{\bf k}-\Phi}{\sqrt{(\varepsilon_{\bf
      k}-\Phi)^2+\Delta^2}}\\
\hspace{-0.6cm}&&\mbox{}+\frac{\zeta_{\bf k}}{E_{\bf k}}\big[\rho_{11}^{h}({\bf
  k})+\rho_{11}^{h}(-{\bf k})\big]-\frac{\Delta}{E_{\bf
    k}}\Big[\rho_{12}^{h}({\bf k})+\rho_{21}^h({\bf k})\Big].
\label{self_consistent_s}
\end{eqnarray}
Here, $n_0$ is the total electron density. From
  Eq.~(\ref{self_consistent_s}), it can be seen that not only the non-equilibrium quasi-electron and
quasi-hole distributions but also the
correlation between quasi-electron and quasi-hole states contribute to the charge imbalance.

The superfluid momentum ${\bf p}_s$ can be obtained from Eq.~(\ref{EOM}) in the
homogeneous limit with the electrical field in the optical pulse known. With the propagation direction of
  the optical field assumed to be  
perpendicular to the QWs, i.e., the $\hat{\bf z}$-direction, the direction of the
  electrical field is taken 
  to be along the $\hat{\bf
    x}$-direction without loss of generality. Thus, 
\begin{eqnarray}
\label{superfluid_momentum}
\hspace{-1cm}&&{\bf p}_s=(e/\omega)E_{0}\hat{\bf x}\sin(\omega
t)\exp[{-t^2/(2\sigma_t^2)}],\\
\hspace{-1cm}&&\partial_t{\bf p}_s\approx eE_{0}\hat{\bf x}\cos(\omega
t)\exp[{-t^2/(2\sigma_t^2)}].
\label{rate_of_change}
\end{eqnarray}
Here, $E_0$ is the strength of the {\em effective} electrical field in the
  superconductor\cite{Ambegaokar}
and $\sigma_t$ represents the
duration time of the optical pulse. In the numerical calculation,
$-2.5\sigma_t\leq t\leq 5\sigma_t$.

Finally, we address that Eqs.~(\ref{homogeneous}),
(\ref{self_consistent_s}-\ref{rate_of_change})
provide the consistent equations to solve the optical
response to the THz pulses. Here, the condensate is assumed to react to the
quasiparticles {\em simultaneously} due to the charge
neutrality.\cite{Bardeen,Tinkham2,Josephson_B}
 In our previous work in the study of the quasiparticle spin
dynamics {\em with small spin imbalances},
it is assumed that the condensation rate is {\em slower} than the spin relaxation one
and hence the framework with the quasiparticle-number conservation is used.\cite{Tao_2}
Therefore, different assumptions for the condensate dynamics can lead to
different schemes. Nevertheless, for the problem near the equilibrium, the induced change imbalance is
expected to be small and these two schemes can even give similar physical results.

\subsection{Numerical Results}
\label{Numerical_s}
In this subsection, we present the numerical results by solving the
optical Bloch equations [Eqs.~(\ref{homogeneous}),
  (\ref{self_consistent_s}-\ref{rate_of_change})]
in a specific material GaAs QW in
proximity to an $s$-wave superconductor. 
All parameters used in our computation are
listed in Table~\ref{material_s}.\cite{material_book}
\begin{table}[h]
\caption{Parameters used in the computation for GaAs QWs in
proximity to an $s$-wave superconductor.\cite{material_book}}
 \label{material_s} 
\begin{tabular}{ll|ll}
    \hline
    \hline    
    $m^*/m_0$&\;\;\;$0.067$&\;$a~({\rm nm})$&\;\;\;$8$\\[4pt]
    $\kappa_0$&\;\;\;$12.9$ &\;$n_0$~(cm$^{-2}$)&\;\;\;$5\times 10^{11}$\\[4pt]
    $\sigma_t~({\rm ps})$&\;\;\;$4$&\;$T_e~({\rm K})$&\;\;\;$2$ \\[4pt]
    $d~({\rm g/cm^3})$&\;\;\;$5.31$&\;$v_{sl}~({\rm m/s})$&\;\;\;$5290$ \\[4pt]
    $\Xi~({\rm eV})$&\;\;\;$8.5$&\;$v_{st}~({\rm m/s})$&\;\;\;$2480$\\[4pt]
    $e_{\rm 14}~({\rm 10^9~V/m})$&\;\;\;$1.41$ &\\[4pt]
    \hline
    \hline
\end{tabular}
\end{table}

In Table~\ref{material_s}, for the material parameters, $\kappa_0$ stands for the relative dielectric
constant; $a$ denotes the well width; and $d$ is the mass density of the crystal.
 For the parameters associated with the electron-phonon interaction,
   $\Xi$ denotes the
  deformation potential; $e_{14}$ represents the piezoelectric constant; $v_{sl}$ and $v_{st}$
  are the velocities of LA and TA phonons,
  respectively.\cite{e_p_formula1,e_p_formula2}
 Finally, $T_e$ is the environment temperature.

 With these parameters, we
   directly estimate the contribution of the electron-AC-phonon interaction in
   the scattering term at $T_e=2$~K, compared to the one of the
   electron-impurity interaction with the typical impurity density
   $\tilde{n}_i=0.1n_0$.
   In Eq.~(\ref{scat_ep}), at low temperature, $n_{\bf
  k}\approx 0$. Thus, the electron-AC-phonon interaction is approximately determined by its
   strength $\sum_{k_z}|g^{\lambda}_{{\bf k}-{\bf k}',k_z}|^2$.
   We explicitly calculate the electron--AC-phonon interaction strength 
   $\sum_{k_z}|g^{\lambda}_{{\bf k}-{\bf k}',k_z}|^2$ due to the deformation potential in the LA branch
and piezoelectric coupling including LA and TA branches, which are found to be
about three orders of magnitude smaller than $\tilde{n}_i|V_{{\bf k}-{\bf k}'}|^2$. Thus,
the electron-AC-phonon interaction is negligible in our computation.

\subsubsection{Excitations of Higgs mode}
\label{Higgs_s}
Recently, it was reported 
in several experiments in the conventional superconducting metals that the Higgs mode
 can be excited by the intense THz
 field, which oscillates with twice the frequency of the THz 
 field.\cite{Earliest,Matsunaga_1,Matsunaga_2,Matsunaga_3,Matsunaga_4} 
 These experiments also show that there exists plateau for the
 Higgs mode after the
 THz pulse in most situations, whose value increases with the increase of the field
 intensity.\cite{Matsunaga_1,Matsunaga_2} Previously, the oscillation of the
 Higgs mode has been explained by the pump effect from the Anderson
pseudo-spin picture, in which the drive effect on the
 superconducting state is absent.\cite{Soliton,Critical,Axt1,Axt2,
Anderson1,Higgs_e_p,multi_component,Leggett_mode,Xie,Tsuji_e_p,multiband} 
Here, we aim to distinguish the contribution of the pump and drive
effects to the evolution of the Higgs mode in GaAs QW in
proximity to an $s$-wave superconductor.

\paragraph{Different pump regimes}
Before we present the numerical results, we first analyze the behavior
of the pump effect from a simplified model, from which different regimes are divided
according to the pump strength. In the pump term in Eq.~(\ref{homogeneous}), 
$\frac{{\bf
    p}_s^2}{2m^*}=\frac{1}{4m^*}\Big(\frac{e}{\omega_L}\tilde{E}_0\Big)^2(1-\cos2\omega
t)$ with $\tilde{E}_0\equiv E_0\exp[{-t^2/(2\sigma_t^2)}]$ slowly varying with time.
The analytical calculation is simplified for high optical frequency $\omega$,
with which the rotation-wave approximation\cite{Haug} can be applied with $\frac{{\bf p}_s^2}{2m^*}\approx
\frac{1}{4m^*}\Big(\frac{e}{\omega_L}\tilde{E_0}\Big)^2\equiv \eta$. In this situation,
in the free situation without the drive and HF terms, the optical Bloch equations in
the quasiparticle space read [refer to Eq.~(\ref{quasiparticle_s})] 
\begin{equation}
\frac{\partial \rho_{\bf k}^h}{\partial T}+i
\Big[\left(
\begin{array}{cc}
E_{\bf k}+\frac{\zeta_{\bf k}}{E_{\bf k}}\eta&
-\frac{\Delta}{E_{\bf k}}\eta \\
-\frac{\Delta}{E_{\bf k}}\eta&-E_{{\bf k}}-
\frac{\zeta_{\bf k}}{E_{\bf k}}\eta 
\end{array}
\right),\rho_{\bf k}^h\Big]=0.
\label{division}
\end{equation}
With the initial state being the equilibrium distribution, the population for the
quasi-electron is
\begin{equation}
\rho_{{\bf k},11}^h=f^0_{\bf k}+\big[\frac{1}{2}-f^0_{\bf
  k}\big]\Big(\frac{\Delta \eta}{E_{\bf k}\mathcal{E}_{\bf
    k}}\Big)^2\Big(1-\cos2\mathcal{E}_{\bf k}T\Big).
\label{condition_1}
\end{equation}
Here, $f^0_{\bf k}=\{\exp[E_{\bf
  k}/(k_BT_e)]+1\}^{-1}$ 
represents the equilibrium distribution for the quasi-electron with $k_B$
being the Boltzmann constant; 
 $\mathcal{E}_{\bf k}=\sqrt{(\varepsilon_{\bf k}-
\mu+\eta)^2+\Delta^2}$, from which it can be seen that $\eta$ directly contributes to the
AC stark effect in the energy spectrum.\cite{Jinluo,Jianhua_Liouville} 

According to the behavior of $(E_{\bf k}\mathcal{E}_{\bf
    k})^2$, which is further expressed as 
\begin{equation}
(E_{\bf k}\mathcal{E}_{\bf k})^2\equiv F({\bf k})=\big[(\zeta_{\bf
  k}+\eta/2)^2-(\eta^2/4-\Delta^2)\big]^2+\Delta^2\eta^2,
\label{condition_2}
\end{equation}
one can separate different pump regimes. 
When $\eta<2\Delta$, the minimum value of $F({\bf k})$ lies at $\zeta_{\bf
  k}=0$, indicating that the quasi-electron distribution evolves around
$|{\bf k}|=k_F$. This regime
with $\eta<2\Delta$ is referred to as the weak-pump regime.  
Whereas when
$\eta>2\Delta$, the minimum values of $F({\bf k})$ are realized when $\zeta_{\bf
  k}=-\eta/2\pm\sqrt{\eta^2/4-\Delta^2}$, which is smaller than zero. 
This indicates that during the pump process, the quasi-electron population mainly arises at
$|{\bf k}|<k_{F}$ and hence the hole-like quasi-electrons are mainly pumped. This regime with
$\eta>2\Delta$ is referred to as the strong-pump regime. Actually, in the
experiments, with $\Delta=2.6$~meV for the metal NbN and $\omega=2\Delta$, $\eta\sim
  17.6$~meV when the peak electric field is $50$~kV/cm, indicating that the experiments
  lie in the strong-pump regime.\cite{Matsunaga_1,Matsunaga_2,Matsunaga_3,Matsunaga_4}

\paragraph{Weak-pump regime} We first focus on the weak-pump regime. 
In Figs.~\ref{figyw3}(a), (b) and (c), the temporal evolutions of the Higgs mode $|\delta\Delta|$
 are plotted in the clean (blue solid curves)
   and dirty (red chain and green dashed curves) samples
 with different pump frequencies of the optical field
 $\omega=\Delta$, $2\Delta$ and $4\Delta$, respectively
 ($\Delta=0.8$~meV$\approx 1.15$~THz). The electric field strength
 $E_0=0.2$~kV/cm. Thus, for $\omega=\Delta$, $\eta=0.18$~meV is much smaller
 than $2\Delta$, indicating that the system lies in the weak-pump regime. 
 With this
   electric field strength, the temporal evolutions of the superconducting
     momentum ${\bf p}_s$, which are 
     driven by the optical field [Eq.~(\ref{superfluid_momentum})], are
    presented in Fig.~\ref{figyw3}(d) with $\omega=\Delta$ (the red chain curve) and $2\Delta$ (the
    blue solid curve), respectively. It can be seen in Fig.~\ref{figyw3}(d) that when
$\omega>\Delta$, the induced
supercurrents by the THz pulse is small in magnitude with
\begin{widetext}
  \begin{center}
\begin{figure}[ht]
  {\includegraphics[width=14.2cm]{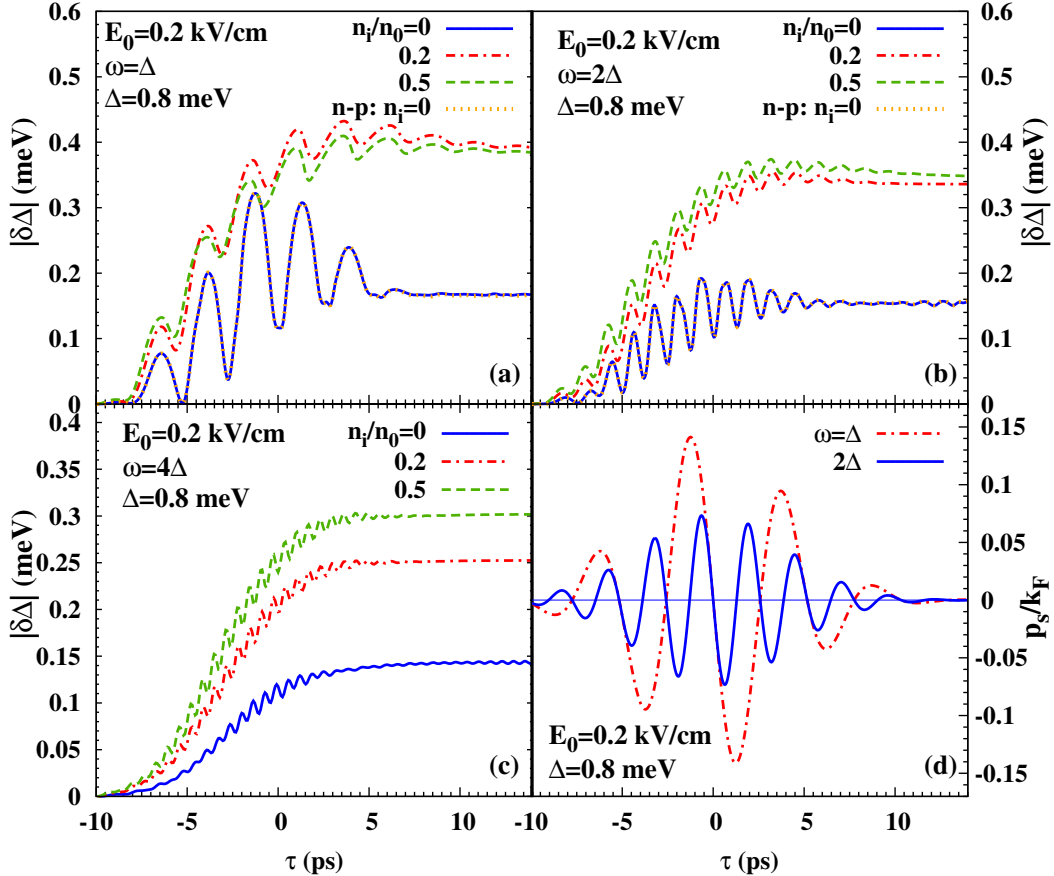}}
  \begin{minipage}[]{18cm}
\begin{center}
  \caption{(Color online) Temporal evolutions of the Higgs mode $|\delta
    \Delta|$ with different pump
    frequencies of the THz pulse $\omega=\Delta$ [(a)], $2\Delta$ [(b)] and
    $4\Delta$ [(c)], respectively. Here, $\Delta=0.8$~meV and the electric field
    strength  
    $E_0=0.2$~kV/cm. With this electric field, the superconducting momentum ${\bf p}_s$ is
    presented in (d) when $\omega=\Delta$ and $2\Delta$. It can be seen that $|{\bf p}_s|<0.15 k_F$ when
    $\omega>\Delta$.
    In (a) and (b), it can be seen that without the
    pump effect, the Higgs
    modes, plotted by the yellow dotted curves, coincide with the ones with both the
    pump and drive effects, represented by the blue solid curves. 
    Moreover, in (a), (b) and (c), it is found that there always exist plateaus after the THz
    pulse, which are suppressed with the increase of the optical-field frequency.
    Finally, it is shown in (a) [or (b), (c)] by the blue solid, red chain and
    green dashed curves that
    with the increase of the impurity density,
    the oscillation amplitude of the Higgs mode is suppressed and the amplitude
    of the plateau of
    the Higgs mode increases.}
  \label{figyw3}
  \end{center}
\end{minipage}
\end{figure}
\end{center}
\end{widetext}
 $|{\bf p}_s|<0.15 k_F$.
By comparing the oscillation frequencies of the Higgs
mode [Figs.~\ref{figyw3}(a), (b) and (c)] with the ones of the supercurrent
[Fig.~\ref{figyw3}(d)], one finds that 
the Higgs mode oscillates with twice the frequency of the THz
field when both the pump and drive effects exist.
Then, the contributions of the pump and drive effects to the Higgs mode
are compared
 in Figs.~\ref{figyw3}(a) and (b) in the impurity-free situation. It can be seen that without the
 pump effect, the Higgs
 modes, plotted by the yellow dotted curves, coincide with the one with both the
pump and drive effects, represented by the blue solid curves.
 This shows that the pump effect
is marginal for the excitation of Higgs mode in the weak-pump regime.
 Moreover, it is found that there always exist plateaus for the Higgs mode after the THz
pulse, which are suppressed with the increase of the optical-field frequency,  
as shown in Figs.~\ref{figyw3}(a), (b) and (c).
Finally, the role of the electron-impurity scattering is addressed. It is shown
in Fig.~\ref{figyw3}(a) [or (b), (c)]
by the blue solid, red chain, and green dashed curves
 that with the increase of the impurity density,
 the oscillation amplitude of the Higgs mode is suppressed and the plateau value of
the Higgs mode increases. These rich features can be understood as follows. 

\begin{widetext}
\begin{center}
\begin{figure}[htp]
\begin{minipage}[]{18cm}
    \hspace{-0cm}\parbox[t]{6cm}{
      \includegraphics[width=5.73cm,height=9.7cm]{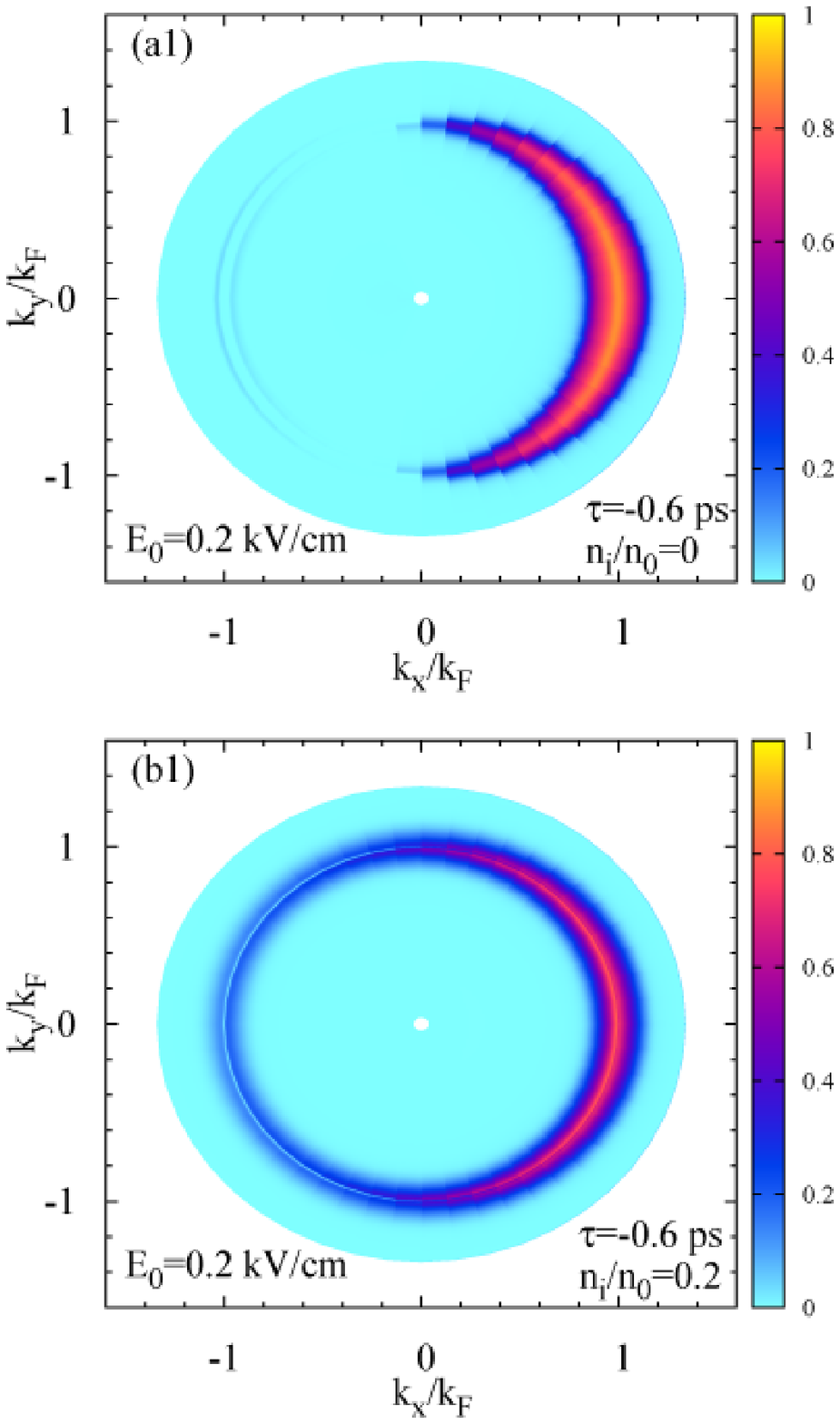}}
    \hspace{-0.4cm}\parbox[t]{6cm}{
      \includegraphics[width=5.73cm,height=9.7cm]{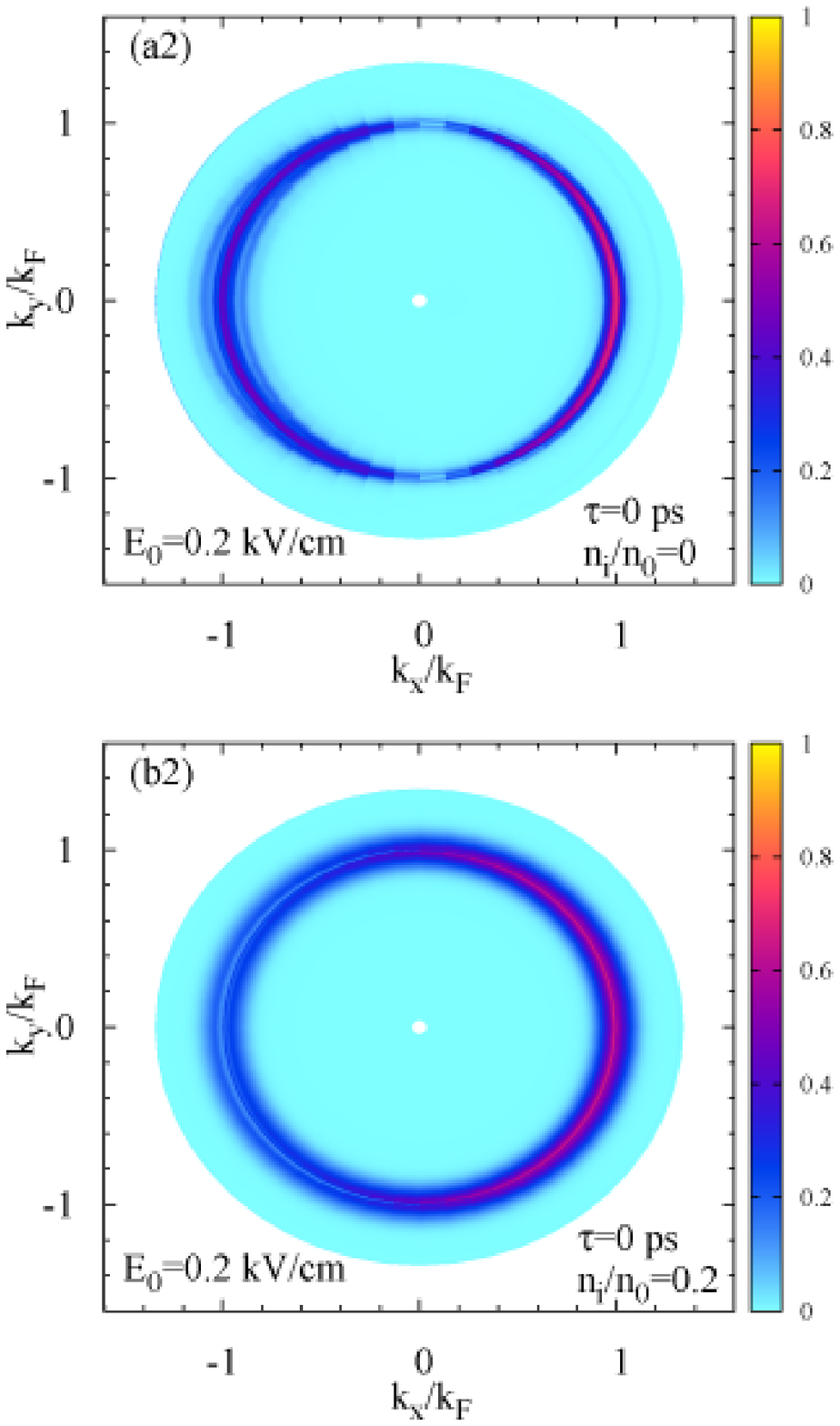}}
\hspace{-0.4cm}\parbox[t]{6cm}{
      \includegraphics[width=5.73cm,height=9.7cm]{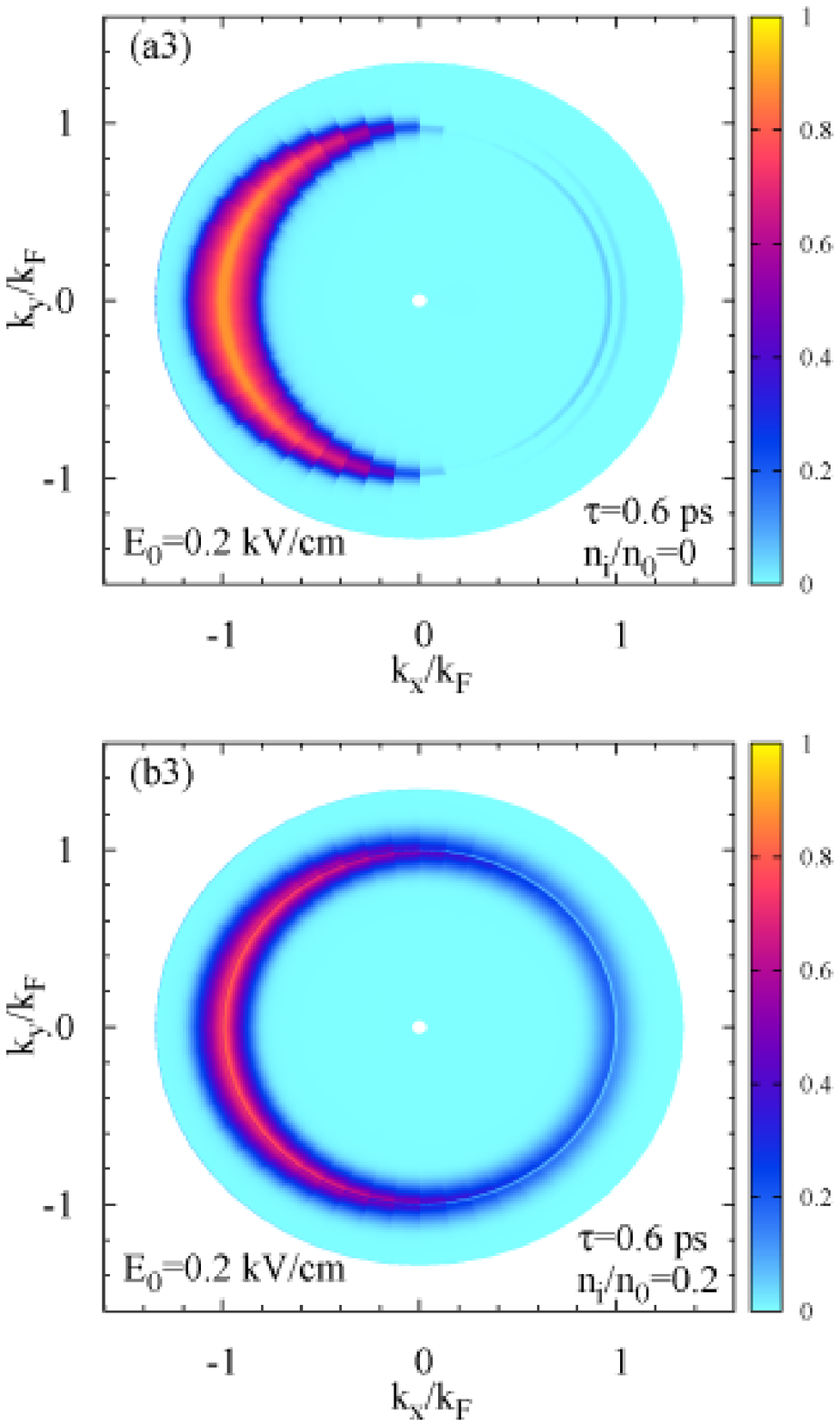}}
  \end{minipage}
  %{\includegraphics[width=17.cm]{figyw4.eps}}
\begin{minipage}[]{18cm}
\begin{center}
  \caption{(Color online) Quasi-electron distributions $\rho^h_{{\bf
          k},11}$ in the momentum space at $\tau=-0.6$, 0, and $0.6$~ps in
    the clean [(${\rm a1}$), (${\rm a2}$) and (${\rm a3}$) with $n_i=0$] and
    dirty [(${\rm b1}$), (${\rm b2}$) and (${\rm b3}$) with $n_i=0.2n_0$]
    samples. $\omega=2\Delta$ with $\Delta=0.8$~meV. The electric field strength 
    $E_0=0.2$~kV/cm, with which ${\bf p}_s\approx 0.13k_F$, 0 and $-0.13k_F$ at
    $\tau=-0.6$, 0 and $0.6$~ps, respectively.}
  \label{figyw4}
\end{center}
\end{minipage}
\end{figure}
\end{center}
\end{widetext}

We first address the role of the drive effect on the anomalous correlation. It
has been well investigated that in the {\em static} case when the center-of-mass
momentum ${\bf q}$ of the Cooper pairs emerges, which can originate from the
spontaneous symmetry-breaking, e.g., in the 
FFLO
state\cite{FF,LO,FFLO_Takada,Yang_FFLO} or with a 
supercurrent,\cite{supercurrent,Tao_2,Yang} a blocking region occupied by the
quasiparticles can appear, 
 in which the anomalous correlation for the Cooper pair can be significantly
 suppressed.\cite{FF,LO,FFLO_Takada,supercurrent,Tao_2,Yang} Then
 it is expected that when the {\em time-dependent} supercurrent emerges with the
 excitation of the center-of-mass momentum of Cooper pairs,
 the blocking region can be dynamically excited, in which the Cooper-pair
 anomalous correlation is also suppressed. Specifically, in
   Fig.~\ref{figyw1},
   a comprehensive physical picture
   has been presented, in which one finds that the driven blocking region, shown
   by the blue
   region in crescent form, directly suppresses the anomalous correlation
   between two electrons (labeled by ``M'' and ``N''). In our
 calculation, with the drive of the electron and hole (particle space) in the
 opposite directions [refer to $\tau_3$ in the drive term in
 Eq.~(\ref{homogeneous})],
 the blocking region for the quasiparticles surely appears,
 with typical examples presented in Fig.~\ref{figyw4} with $E_0=0.2$~kV/cm at
 different times $\tau=-0.6$, $0$ and $0.6$~ps, respectively.

In Figs.~\ref{figyw4}(${\rm a1}$), (${\rm a2}$) and (${\rm a3}$)
when $n_i=0$, one sees that when $\tau=-0.6$~ps [Fig.~\ref{figyw4}(${\rm
      a1}$)] and 0.6~ps [Fig.~\ref{figyw4}(${\rm a3}$)] with finite ${\bf
    p}_s\approx 0.13k_F\hat{\bf x}$ and $-0.13k_F\hat{\bf x}$
  [refer to Fig.~\ref{figyw3}(d)], the blocking regions in the
crescent shape appear, whose positions are consistent with the sign of
  the center-of-mass momentum 
  ${\bf p}_s$ of the Cooper pairs; whereas when $\tau=0$~ps [Fig.~\ref{figyw4}(${\rm a2}$)], with zero
center-of-mass momentum, the blocking region tends to disappear, but there still exists significant
quasiparticle population. Furthermore, it is observed in
Figs.~\ref{figyw4}(${\rm a1}$) and (${\rm a3}$) that inside the blocking
region, the quasi-electron population is close to one.
In the blocking region, the anomalous correlation
\begin{eqnarray}
  \nonumber
  C({\bf k})&=&u_{\bf k}v_{\bf
    k}(\rho^h_{{\bf k},11}-\rho^h_{{\bf k},22})+u_{\bf k}^2\rho^h_{{\bf k},12}-v_{\bf
    k}^2\rho^h_{{\bf k},21}\\
  &\approx& u_{\bf k}v_{\bf
    k}(\rho^h_{{\bf k},11}-\rho^h_{{\bf k},22})
\label{anomalous_correlation}
\end{eqnarray}
is significantly suppressed with $\rho^h_{{\bf k},11}\lesssim 1$ and
$\rho^h_{{\bf k},22}=1-\rho^h_{-{\bf k},11}\lesssim 1$.\cite{Tao_2,FFLO_Takada} Then due to the
suppression of the anomalous correlation, from Eq.~(\ref{Higgs_formula}), the
Higgs mode is significantly excited. 
Furthermore,
the suppression of the anomalous correlation does not
depend on the sign of the center-of-mass momentum of Cooper pairs. Accordingly, although the
center-of-mass momentum of Cooper pairs oscillates with the frequency of the optical field, the Higgs
 mode originating from the suppression of the anomalous correlation oscillates
 with twice the frequency of the optical field. It is noted that in the weak-pump
 regime, the quasi-electrons are mainly pumped around the Fermi surface in the
 absence of the drive effect; whereas 
 the blocking region also arises around the Fermi surface but due to the drive
 effect.
 Thus, thanks to the Pauli blocking
 effect, the emergence of the
 blocking region can efficiently suppress the pump effect.
 Consequently, in the weak-pump regime, the pump effect plays a marginal role and
 the drive effect is dominant in the excitation of the Higgs mode [refer to the
 blue solid and yellow dotted curves in Figs.~\ref{figyw3}(a) and (b)].

 We then focus on the influence of the electron-impurity scattering on the Higgs mode
 dynamics. In Figs.~\ref{figyw4}(${\rm b1}$), (${\rm b2}$) and (${\rm b3}$) with $n_i=0.2n_0$, by comparing with the
 impurity-free situation in Figs.~\ref{figyw4}(${\rm a1}$), (${\rm a2}$) and (${\rm a3}$), it is observed that the
 electron-impurity scattering has significant influence on the
 formation of the blocking region.\cite{FFLO_Takada2} Specifically, 
 on one hand, the electron-impurity
 scattering can suppress the range of the blocking region and hence its
 oscillation. This is because the
 drift effect of the electron and hole, which contributes to the formation of
 the blocking region, can be suppressed
 by the electron-impurity scattering.\cite{Tao_Ge,Yang_hot,Lei_hot}
Thus, the suppression of the oscillation of the blocking region
 tends to suppress the oscillation amplitude of the Higgs mode.
 On the other hand, the electron-impurity scattering tends to
 destroy the blocking region by averaging the quasi-electron distribution.
 Accordingly, from Eq.~(\ref{anomalous_correlation}), the
 emergence of the significant quasiparticle population in the unblocking region
 further suppresses the anomalous
 correlation.  This tends to enhance the magnitude of the Higgs mode.

\begin{figure}[ht]
\begin{minipage}[]{8.6cm}
    \hspace{0cm}\parbox[t]{4.3cm}{
      \includegraphics[width=4.45cm,height=7.57cm]{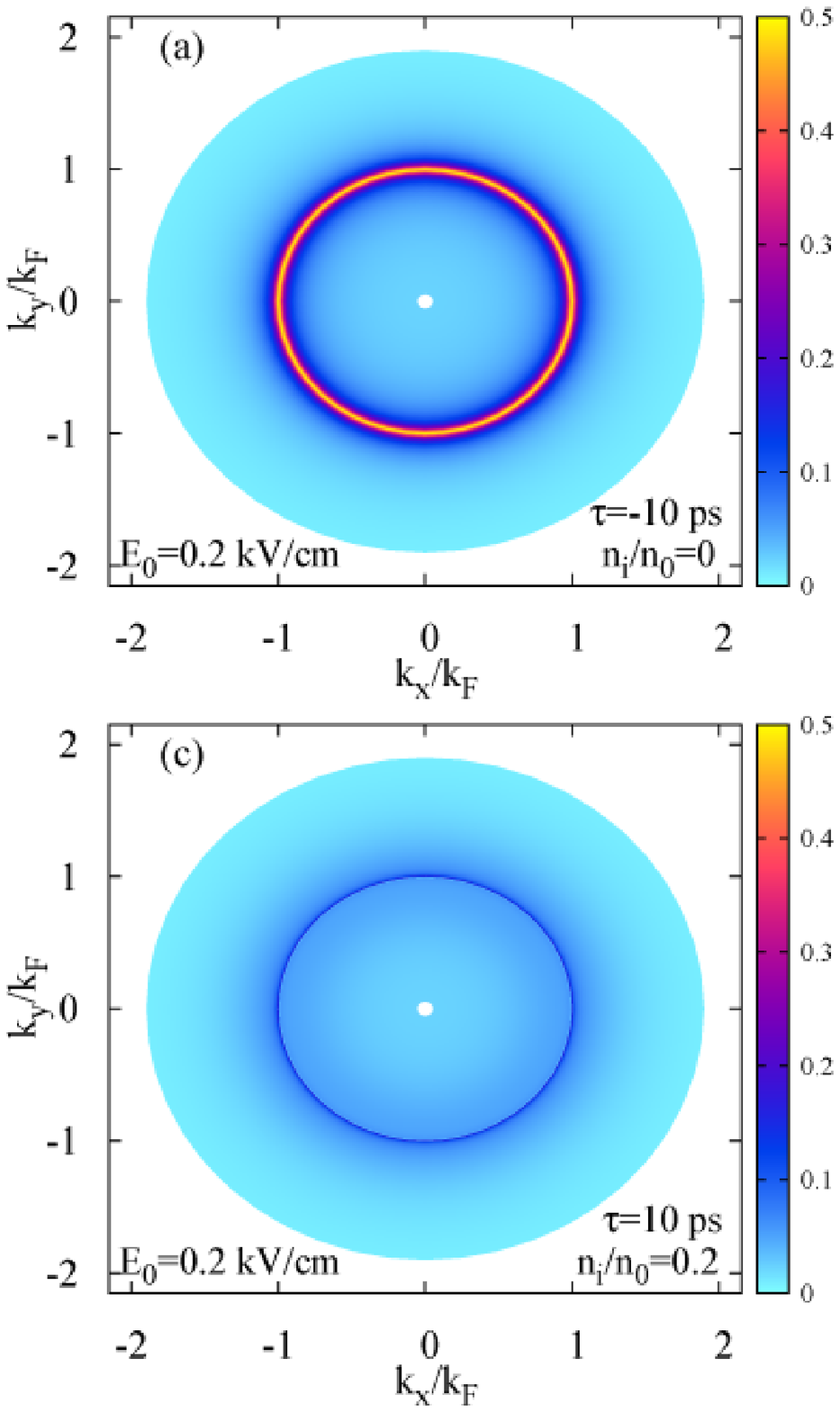}}
    \hspace{-0.1cm}\parbox[t]{4.3cm}{
      \includegraphics[width=4.45cm,height=7.57cm]{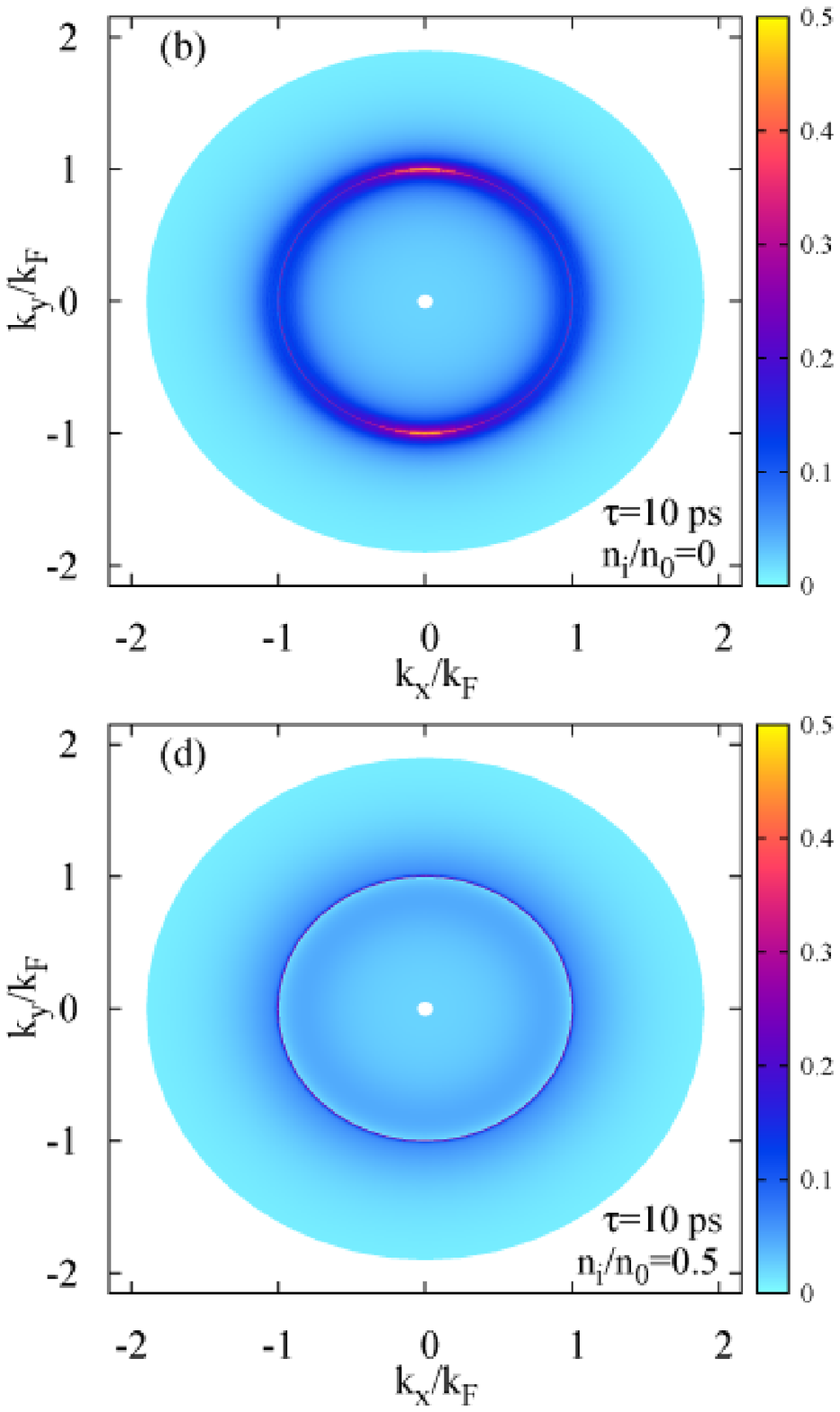}}
  \end{minipage}
  \begin{minipage}[]{8.6cm}
  \caption{(Color online) Anomalous correlations in the momentum space before
    [$\tau=-10$~ps, (a)] and after
    [$\tau=10$~ps, (b), (c) and (d)] the THz pulses with $E_0=0.2$~kV/cm and
    $\omega=2\Delta\approx 2.3$~THz. In (b), (c) and (d),
    the impurity densities
    $n_i=0$, $0.2n_0$ and $0.5n_0$.}
  \label{figyw5}
\end{minipage}
\end{figure}

 To make the above physical picture clearer, in Fig.~\ref{figyw5},
 we further plot the anomalous correlations before [(a), $\tau=-10$~ps] and
 after [(b), (c) and (d), $\tau=10$~ps]
 the THz pulses with $E_0=0.2$~kV/cm and $\omega=2\Delta\approx 2.3$~THz.
 In Figs.~\ref{figyw5}(b), (c) and (d), the impurity densities are set to be 
 $n_i=0$, $0.2n_0$ and $0.5n_0$, respectively.
 In these figures, it can be seen that the anomalous correlation
 is significant only around the Fermi surface.\cite{FFLO_Takada,Tao_2,Yang_FFLO}
 We first address the influence of the THz pulse on the anomalous correlation in
 the impurity-free situation.  
 By comparing the anomalous correlation in Figs.~\ref{figyw5}(a) and (b),
 it can be seen that in the impurity-free situation, the anomalous correlation is suppressed by the
 THz pulse only in the blocking region and the anomalous
 correlation becomes anisotropic in the momentum space.
 This is consistent with the previous works in the static
 situation, in which the anomalous correlation is suppressed only in the
 blocking region.\cite{Tao_2,FF,FFLO_Takada}
 Then the influence of the impurity can be
 seen by comparing Fig.~\ref{figyw5}(c) [or (d)] with (b). It is shown
 in Fig.~\ref{figyw5}(c) [(d)] that the anomalous
 correlation becomes isotropic due to the momentum scattering with $n_i=0.2n_0$
 ($0.5n_0$). This confirms the conclusion from Eq.~(\ref{anomalous_correlation})
 that the existence of the impurity tends to average the quasiparticle
 population and hence 
 the anomalous correlation around the Fermi surface. Furthermore,
 one observes that in Figs.~\ref{figyw5}(c) [or (d)],
the anomalous correlation is further suppressed compared to the free situation
in (b), which shows that the electron-impurity scattering can further suppress
the superconductivity after the THz pulse. Thus, with the increase of the impurity
density, the plateau of
the Higgs mode increases [refer to the red and
blue solid curves in Figs.~\ref{figyw3}(a), (b) and (c)].

The further suppression of the superconductivity due to the impurity after the
THz pulse can be 
understood from another point of view. We find that with the increase of
  the impurity density,
the quasiparticle
density increases during the temporal
evolutions,  
shown in Fig.~\ref{figyw14} in Appendix~\ref{BB}.  
This can be understood from the fact that in the presence of impurities, the optical
absorption is significantly enhanced because the driven electrical current is no
longer in phase to the driven field.\cite{Tao_Ge,Yang_hot,Lei_hot}
The enhancement of the optical absorption by the impurities further suppresses the anomalous
correlation [refer to Eq.~(\ref{anomalous_correlation})].
With the increase of the quasiparticle density, the normal-fluid and
super-fluid densities are expected to deviate from their equilibrium values.
Thus, to further understand the non-equilibrium superconducting state after
the pulse, the normal-fluid and
super-fluid densities are also estimated in Appendix~\ref{BB}, which are often estimated in the pump-probe
experiments.\cite{MgB2_1,MgB2_2,YBCO,BSCCO} It is emphasized that this
estimation is performed by assuming that the system is {\em
  in the Fermi-distribution with an effective temperature},
and hence the two-fluid description
is expected to be effective.\cite{Tinkham_book,two_fluid,MgB2_1,MgB2_2,YBCO,BSCCO}

\begin{figure}[ht]
  {\includegraphics[width=7.8cm]{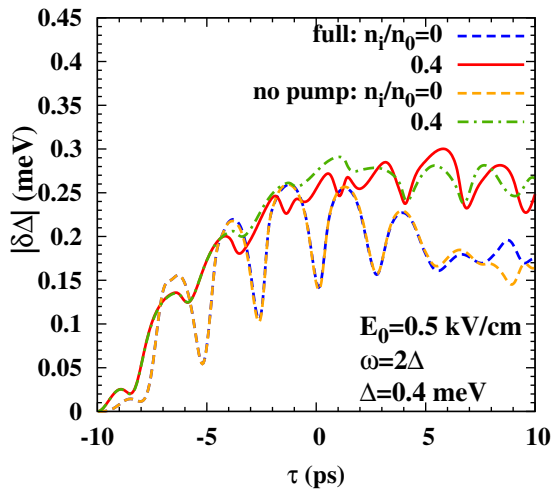}}
  \caption{(Color online) Temporal evolutions of the Higgs mode in the
    strong-pump regime. With $E_0=0.5$~kV/cm and $\omega=2\Delta=0.8$~meV, one
    obtains that 
    $\eta\approx 1.1$~meV, which is larger than $2\Delta$. It can be seen that in the clean (dirty)
    sample, the Higgs mode calculated with both the pump and drive effects,
    represented by the blue dashed (red solid) curve, almost
    coincides with the one calculated without the pump effect, denoted by the yellow
    dashed (green chain) curve.}
  \label{figyw6}
\end{figure}

\paragraph{Strong-pump regime}
\label{condition_pump}
We then extend our calculation to the strong-pump regime. It is noted that a strong
 electrical field in the intense THz pulse can destroy the superconductivity
 (refer to Fig.~\ref{figyw15}).
 Here, we take $E_0=0.5$~kV/cm and $\Delta=0.4$~meV. Then with
$\omega=2\Delta$, it is obtained that $\eta\approx 1.1$~meV, which is larger than $2\Delta$. With
these parameters, we
show that in the superconducting GaAs QWs,
even in the strong-pump regime, the pump effect still plays a marginal role in the excitation of
the Higgs mode. This can be seen in Fig.~\ref{figyw6} that in the clean (dirty)
sample, the Higgs mode calculated with both the pump and drive effects,
represented by the blue dashed (red solid) curve, almost
coincides with the one calculated without the pump effect, denoted by the yellow
dashed (green chain) curve. 

Above we have shown that in both the weak- and strong-pump regimes, with 
  relatively small superconducting momenta $|{\bf p}_s|\ll k_{F}$, the pump effect always plays a 
marginal role in the excitation of the quasiparticle due to the effect of Pauli
blocking. Actually, it can be estimated that as long as $|{\bf p}_s|\lesssim
k_F$, the pump effect cannot be efficient (shown below). This is exactly the situation in the conventional
superconducting metals with large Fermi surfaces, although intense THz fields
are applied.\cite{Matsunaga_1,Matsunaga_2,Matsunaga_3,Matsunaga_4}  
Previously, the explanation of the Higgs-mode oscillation is
based on the pump effect.\cite{Earliest,Matsunaga_1,Matsunaga_2,Matsunaga_3,
  Matsunaga_4,Axt1,Axt2,Anderson1,path_integral}
Our results suggest that it is the drive effect that is really responsible.

Finally, we remark that only when $|{\bf p}_s|\gtrsim k_F$, the pump
effect can contribute to the excitation of the Higgs mode, as estimated as
follows.   
In the strong-pump regime ($\eta\gtrsim 2\Delta$),
 the hole-like quasiparticle is
dominantly pumped around some special momenta labeled by ${\bf k}_0$
[refer to Eqs.~(\ref{condition_1}) and (\ref{condition_2})], which are
determined by 
\begin{equation}
{\bf k}_0^2/(2m^*)-\mu\approx -\eta.
\label{natural}
\end{equation} 
Actually, Eq.~(\ref{natural}) is established only when $|{\bf p}_s|\lesssim
  \sqrt{2}k_F$ with ${\bf
  k}_0^2/(2m^*)\approx \mu-\eta>0$ satisfied. When $|{\bf p}_s|\lesssim
  \sqrt{2}k_F$,  ${\bf k}_0$ is away from
  the Fermi surface by $\Delta k\equiv k_F-|{\bf k}_0|$.  It is noted that the
  boundary of the blocking
region {\em in the clean limit} is away from the Fermi surface by about 
$|{\bf p}_s|$. Thus, when $2\Delta k\gtrsim |{\bf p}_s|$, the pumped hole-like quasiparticles lie out of the
blocking region, which cannot be efficiently blocked. This
requires that $|{\bf
  p}_s|\gtrsim k_F$. Whereas when $|{\bf p}_s|\gtrsim
  \sqrt{2}k_F$, Eq.~(\ref{natural}) is no longer established. In this situation,
  ${\bf p}_s^2/(4m^*)\gtrsim \mu$, i.e., the effective chemical potential
  contributed by the AC Stark effect can be even larger than the one of the
  system. In this situation, the pump effect becomes extremely strong and the
  quasiparticles can be efficiently pumped in the whole momentum space.
  From above analysis, it is estimated that when $|{\bf
  p}_s|\gtrsim k_F$, the pump effect can have contribution to the excitation of
  the Higgs mode. 
Moreover, one sees that one way to realize the significant pump effect is to efficiently suppress the drive
effect and hence the range of the blocking region.

\subsubsection{Charge Imbalance: Creation and Relaxation}
\label{Numerical_s_charge}
The charge imbalance created by the electrical method and its
  relaxation has been intensively
  studied.\cite{Tinkham1,Tinkham2,Takahashi_SHE2,
Tinkham_book,Pethick_review,Larkin,Gray,Tinkham_condensate}
 It is believed that for the isotropic
$s$-wave superconductor, the {\em elastic} scattering due to the impurity cannot cause the relaxation of
the charge
imbalance.\cite{Tinkham1,Tinkham2,Takahashi_SHE2,Tinkham_book,Pethick_review,Larkin,Gray}
 This is because the elastic scattering cannot exchange the electron-like and
hole-like
quasiparticles due to coherence
factor $(u_{\bf k}u_{{\bf k}'}-v_{\bf k}v_{{\bf k}'})$ in the electron-impurity scattering
potential [refer to Eq.~(\ref{s_diagonal})].\cite{Tinkham1,Tinkham2,
  Takahashi_SHE2,Tinkham_book,Pethick_review,Larkin,Gray}
 Nevertheless, in the previous
 studies,\cite{Tinkham1,Tinkham2,Takahashi_SHE2,Tinkham_book,Pethick_review,Larkin,Gray}
 the charge neutrality condition 
is not explicitly considered in the relaxation process of the charge imbalance. 
In other words, the
studies\cite{Tinkham1,Tinkham2,Takahashi_SHE2,Tinkham_book,Pethick_review,Larkin,Gray}
 are actually performed 
in the framework of quasiparticle-number conservation.\cite{Tao_2} Actually, to maintain the charge
neutrality, the Cooper pair condensate has to 
respond to the dynamics of the
quasiparticles.\cite{Takahashi_SHE1,Takahashi_SHE2,Hirashima_SHE,
Takahashi_SHE3,Takahashi_SHE_exp,Hershfield,Spivak,kinetic_book} 
In this part, we investigate the creation of the charge imbalance
by the optical pulse and its relaxation via the optical Bloch equations [Eqs.~(\ref{homogeneous}),
(\ref{self_consistent_s}-\ref{rate_of_change})] in the framework of charge
neutrality. The physical picture for the charge neutrality condition has been
  addressed explicitly in Sec.~\ref{charge_picure}.

\begin{figure}[ht]
  {\includegraphics[width=7.9cm]{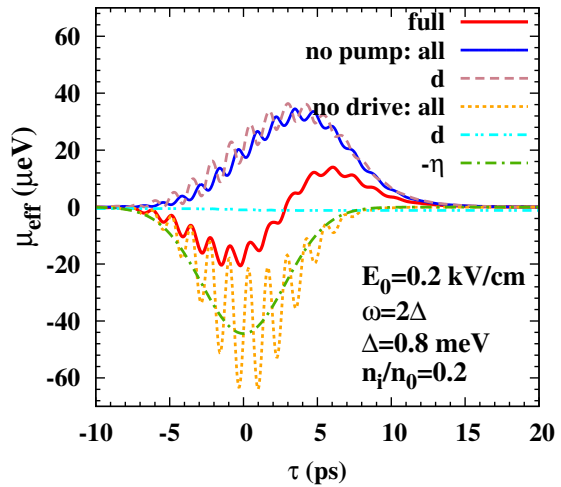}}
  \caption{(Color online) Temporal evolution of the effective chemical potential in the
  condensate in the presence of the optical pulse with the typical 
impurity density $n_i=0.2n_0$. $E_0=0.2$~kV/cm and
$\omega=2\Delta$ with $\Delta=0.8$~meV. The red solid curve shows that
during the evolution,  
  the effective chemical
potential is first negative when $\tau<3$~ps, then becomes positive when
$\tau>3$~ps and finally decays to zero after the pulse. The blue solid
 (yellow dotted)
 curve represents the calculated effective chemical
 potential when only the drive (pump) effect exists. The cyan double-dot--dashed
(purple dashed) 
curve is calculated with only the
diagonal elements in the quasiparticle density matrix retained when
only the pump (drive) effect exists. Finally, the chemical potential
induced by the AC Stark effect, i.e., $\eta$, is presented by the green chain curve, which
depicts the envelope of the yellow dotted curve.}
  \label{figyw7}
\end{figure}

\paragraph{Optical creation of charge imbalance}
 Although in the excitation of the Higgs mode, the pump effect is shown to play a marginal
role (Sec.~\ref{Higgs_s}), it is found that both the pump and drive effects can
be important in the creation of the charge
imbalance. Their contributions can be even distinguished in the time domain.
 This is presented in Fig.~\ref{figyw7}, in which the temporal evolution
of the effective chemical
potential $\mu_{\rm eff}$ is plotted
by the red solid curve with the typical 
impurity density $n_i=0.2n_0$ when $E_0=0.2$~kV/cm and
$\omega=2\Delta=1.6$~meV. 
It can be seen that during the evolution, the effective chemical
potential, represented by the red solid curve, is first negative when
$\tau<3$~ps, then becomes positive when $\tau>3$~ps and {\em finally decays to zero
after the pulse}. From
Eq.~(\ref{self_consistent_s}) with $\Phi=\mu-\mu_{\rm eff}$, one observes that the
negative effective chemical potential means the increase of the total chemical potential
and hence the condensate density; at the same time, the hole-like
 quasiparticle charge becomes larger than the electron-like one.  It is noted
 that the total density of quasiparticles increases during the pulse (refer to
 Fig.~\ref{figyw14}). Thus, with the
induction of the negative effective chemical potential, both the condensate and
quasiparticle densities are increased to maintain the charge
neutrality. This is in contrast to the common belief that the quasiparticle
  densities increase through the breaking of the Cooper pairs. Whereas with the
positive effective chemical potential, the electron-like quasiparticle charge becomes larger than the
hole-like one in accompany with the decrease of the condensate density.

Furthermore, in Fig.~\ref{figyw7}, when only the drive (pump) effect exists, as shown by
 the blue solid (yellow dotted) curve, the effective chemical
potential is positive (negative). Moreover, one observes that 
the red solid curve can be treated as the simple summation of the blue
solid and yellow dotted  
ones. This indicates that the positive and negative
parts of the effective chemical
potential mainly come from the drive and pump effects,
respectively. It is noticed that in the physical situation with both the pump
and drive effects,
for the pump effect, the excitation of quasiparticle population is efficiently
suppressed by the drive effect (Sec.~\ref{Higgs_s}). Nevertheless, as addressed
 in Eq.~(\ref{self_consistent_s}), both
the quasiparticle population and the correlation between the quasi-electron and
quasi-hole can contribute to
the charge imbalance. Then it is speculated that the charge imbalance due to
the pump effect mainly comes from the induction of the correlation between the quasi-electron and
quasi-hole, which cannot be suppressed by the Pauli blocking. Moreover, the fact that the charge imbalance is the simple
  superposition 
of the ones due to the pump and drive effects indicates that
the charge imbalance due to the drive effect is contributed by a different
channel from the pump effect. Thus, it is further speculated that the
charge imbalance contributed by the drive effect comes from the induction of the
quasiparticle population. Both speculations are directly confirmed by the
numerical calculation. This
can be seen in Fig.~\ref{figyw7} by the cyan double-dot--dashed (purple dashed) curve that when
only the pump (drive) effect exists, the quasiparticle populations have no
(dominant) contribution to the charge imbalance.  
Thus, the optical excitation of the charge imbalance can be understood 
by separately studying the charge imbalance due to the pump and drive
effects. It is emphasized that the obtained picture can be applied to both the
  weak and strong-pump regimes because in both situations, the induction of the
  quasiparticle due to the pump effect is suppressed (this is confirmed by the
  numerical calculations directly).

We first analyze the charge imbalance due to the pump effect
by analytically calculating its contribution to the effective chemical potential. 
From Eqs.~(\ref{division}) and (\ref{condition_1}), one obtains 
\begin{equation}
\rho_{{\bf k},12}^{h}+\rho_{{\bf k},21}^{h}=\frac{E_{\bf k}^2+\zeta_{\bf
    k}\eta}{\Delta\eta}
\Big(\frac{\Delta \eta}{E_{\bf k}\mathcal{E}_{\bf
    k}}\Big)^2(1-2f^0_{\bf
  k})\Big(1-\cos2\mathcal{E}_{\bf k}T\Big).
\end{equation}
Then the net charge contributed by the correlation between the
  quasi-electron and quasi-hole is 
\begin{equation}
\delta Q_{c}=-\sum_{\bf k}\frac{\Delta^2\eta}{E_{\bf k}\mathcal{E}_{\bf
    k}^2}(1-2f_{\bf k}^0)(1-\cos2\mathcal{E}_{\bf k}T)
\approx -\sum_{\bf k}\frac{\Delta^2\eta}{E_{\bf k}^3}.
\end{equation}
By further noticing that $2\delta v_{\bf k}^2=-(\Delta^2/E_{\bf
    k}^3)\delta\mu_{\rm eff}$ in Eq.~(\ref{Tinkham_delta}), the charge
  neutrality condition requires that $\delta\mu_{\rm eff}\approx -\eta$. 
  This relation is directly 
  confirmed by the green chain curve in Fig~\ref{figyw7}, in which $\eta$
  depicts the envelope of the yellow dotted curve. Actually, this
  simple relation provides a simple
  physical picture for the pump-induced charge imbalance, in which the AC Stark effect
  directly modifies the total chemical potential.

For the drive effect, the induced positive effective chemical potential indicates
  that the charge carried by the electron-like quasiparticle is
  larger than the hole-like one. The physics picture is qualitatively analyzed based on the
  optical Bloch equations in the quasiparticle space
  [Eq.~(\ref{quasiparticle_s})] as follows. In
the free situation with only the drive term retained,
Eq.~(\ref{quasiparticle_s}) is written as 
\begin{equation}
\frac{\partial\rho_{\bf k}^h}{\partial
  T}+\frac{1}{2}\Big\{eE_x\tilde{\tau}_3,\frac{\partial \rho_{\bf k}^h}{\partial
k_x}\Big\}+\frac{1}{2}\Big\{eE_x\tilde{\tau}_3,\Big[\rho_{\bf
k}^h,\frac{\partial {\mathcal U}_{\bf k}}{\partial k_x}{\mathcal U}_{\bf k}^{\dagger}\Big]\Big\}=0,
\label{simplified_drive}
\end{equation}
in which 
$\tilde{\tau}_3({\bf k})\equiv {\mathcal U}_{\bf k}\tau_3{\mathcal U}_{\bf k}^{\dagger}=(u_{\bf k}^2-v_{\bf
  k}^2)\tau_3-2u_{\bf k}v_{\bf k}\tau_1$
 with both the diagonal and off-diagonal terms retained.
In Eq.~(\ref{simplified_drive}), the second term
 is the conventional drive term for the quasiparticle in
the Boltzmann equation,\cite{Tinkham_book,Kopnin,Tao_2,Larkin}
 whereas the third term is contributed by the Berry phase.\cite{Berry,Berry_0,helix_Berry}
By defining $q_{\bf k}^*=e(\zeta_{\bf k}/E_{\bf k})(\rho_{{\bf
    k},11}^h+1-\rho_{-{\bf k},22}^h)$, which is the net charge for the
quasiparticle with the 
momentum ${\bf k}$,\cite{Tinkham_book,Pethick_review} and further neglecting the
quasiparticle correlation, it is
obtained from Eq.~(\ref{simplified_drive}) that
\begin{eqnarray}
\nonumber
&&\frac{\partial q^*_{\bf k}}{\partial T}+2eE_x\Big(\frac{\zeta_{\bf k}}{E_{\bf
    k}}\Big)^2\frac{\partial q^*_{\bf k}}{\partial k_x}-2eE_x\frac{\zeta_{\bf
    k}}{E_{\bf k}}\frac{k_x}{m^*}\frac{\Delta^2}{E_{\bf k}^3}q_{\bf k}^*\\
&&\mbox{}+eE_x\frac{k_x}{m^*}\frac{\Delta^2}{E_{\bf
    k}^3}\frac{q^*_{\bf k}+q^*_{-{\bf k}}}{2}=e^2E_x\frac{k_x}{m^*}\frac{\zeta_{\bf
  k}}{E_{\bf k}}\frac{\Delta^2}{E_{\bf k}^3}.
\label{q_k}
\end{eqnarray}

Although Eq.~(\ref{q_k}) is complex, one important feature is that there
exists a source term for $q_{\bf k}^*$ on the right-hand side of the equation.
 This source term, which originates from the Berry-phase effect, is
proportional to $\Delta^2$. This indicates that the
 charge-conservation of the {\em quasiparticle} is absent due to the existence
 of the superconducting order
 parameter. This is consistent with
the conclusion in the Blonder-Tinkham-Klapwijk model when studying the Andreev
reflection, which reveals that the order parameter itself directly breaks the charge conservation
of quasiparticles.\cite{BTK} One notices that
 in the situation with relatively small impurity density, only the blocking region should be
considered.
Actually, this source term directly contributes to the formation of the blocking
region. From the source term, it can be seen that
 with $E_x>0$ ($E_x<0$), the quasiparticle charges increase when $k_x<0$
($k_x>0$).  It is further noted that the source term is proportional to $k_x$, which 
 is larger for the electron-like quasiparticle than the hole-like one.
 Then {\em in the blocking region}, the electron-like quasiparticle charge can be
created faster than the hole-like one, which directly contributes to the charge imbalance
with more electron-like quasiparticles.

We emphasize that the optical excitation of the charge imbalance 
is a unique feature for the superconductor with nonzero order parameter,
which cannot be realized in the normal state. When
the order parameter is close to zero, on one hand, the pump term tends to zero and hence
there cannot exist significant correlation between the quasi-electron and quasi-hole
states; on the other hand, the source term in Eq.~(\ref{q_k}) becomes close to
zero and hence no significant quasiparticles can be created from the
condensate. 
Experimentally, the effective
chemical potential induced by the optical field in the charge imbalance can be directly 
measured either through the voltage between the quasiparticle and
condensate measured in the setup of Clarke's
works,\cite{Clarke_first,Clarke_thermal} or through the effective
chemical potential measured in the Josephson
effect.\cite{voltage}

\paragraph {Charge-imbalance relaxation due to the electron-impurity scattering}
In Fig.~\ref{figyw7}, it is anomalous to observe that after
the pulse at $\tau\approx 8$~ps, the induced effective chemical
potential relaxes to zero. This indicates that there exist relaxation channels
for the charge imbalance even in
the presence of the
elastic scattering in the isotropic
$s$-wave superconductivity, which is in
contrast to the previous
studies.\cite{Tinkham1,Tinkham2,Pethick_review,Tinkham_book}
To reveal the mechanism for the charge-imbalance relaxation, a simplified
model in the $s$-wave superconducting QWs is set up with a {\em small} initially-given
charge imbalance, in which ${\bf p}_s$ is set to be zero and the HF self-energy is neglected.
Accordingly, Eq.~(\ref{quasiparticle_s}) is simplified into 
\begin{equation}
\partial_T\rho_{\bf k}^h+i\big[E_{\bf
  k}\tau_3,\rho_{\bf k}^h\big]+i\big[\mu_{\rm eff}\tilde{\tau}_3,\rho_{\bf k}^h\big]
=\partial_t\rho_{\bf k}|_{\rm scat}^{\rm d}+\partial_t\rho_{\bf k}|_{\rm scat}^{\rm off}.
\label{quasi_simplified}
\end{equation}
Specifically, in Eq.~(\ref{quasi_simplified}), the off-diagonal terms in 
$\mu_{\rm eff}\tilde{\tau}_3$ induce the precession between the
  quasi-electron and quasi-hole states and hence the quasiparticle correlation;
  $\partial_t\rho_{\bf k}|_{\rm
    scat}^{\rm off}$ directly breaks the conservation of the quasiparticle
  number\cite{Tao_2} (more discussions are referred to Appendix~\ref{AA}).
The initial state in the quasiparticle space with a small quasiparticle charge
 imbalance is set to be  
\begin{equation}
\rho_{\bf k}^{h,c}=\left(
\begin{array}{cc}
f_0(E_{\bf k}^c) & 0\\
0 & 1-f_0(E_{\bf k}^c) 
\end{array}
\right).
\label{initial_a}
\end{equation}
In Eq.~(\ref{initial_a}), $E_{\bf k}^c=\sqrt{(\varepsilon_{\bf
    k}-\mu-\delta\mu_c)^2+|\Delta|^2}$ with $\delta\mu_c=0.01\mu$ and
$f_0(E_{\bf k}^c)=\{\exp[E_{\bf k}^c/(k_BT_e)]+1\}^{-1}$. With $|\delta\mu_c|\ll
|\mu|$, 
\begin{equation}
\rho_{\bf k}^{h,c}\approx \left(
\begin{array}{cc}
f_0(E_{\bf k}) & 0\\
0 & 1-f_0(E_{\bf k}) 
\end{array}
\right)-\frac{\partial f_0}{\partial E_{\bf k}}\frac{\zeta_{\bf k}}{E_{\bf
    k}}\delta\mu_c\tau_3.
\label{initial_c}
\end{equation} 
With this initial state, the effective chemical potential for the condensate can
be induced due to the charge neutrality condition [Eq.~(\ref{self_consistent_s})].  
Thus, Eqs.~(\ref{quasi_simplified}), (\ref{self_consistent_s}) and Eq.~(\ref{initial_c}) 
provide the consistent equations to study the charge-imbalance relaxation, which
are solved first numerically and then analytically below.

\begin{figure}[ht]
  {\includegraphics[width=7.9cm]{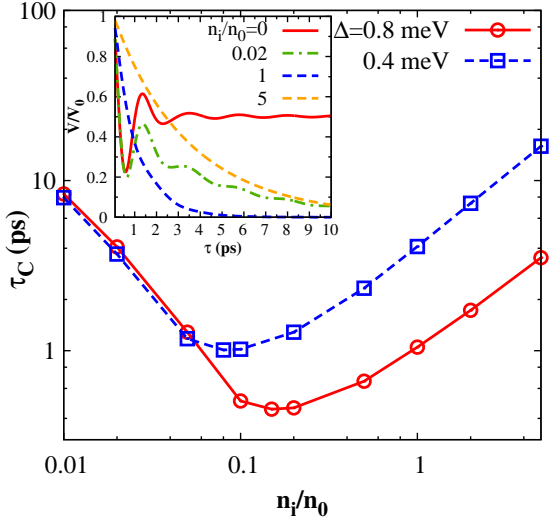}}
  \caption{(Color online) Impurity-density dependencies of the CIRT
    with $\Delta=0.8$~meV (red solid curve with circles) and 0.4~meV (blue
    dashed curve with squares), respectively. The finite CIRT shows that the electron-impurity
    scattering surely can cause the
    charge-imbalance relaxation. 
    In the inset, the temporal evolutions of the normalized effective chemical
    potential $V/V_0$ are shown
    with different impurity densities $n_i=0$ (red solid curve), 0.02$n_0$ (green
    chain curve), $n_0$ (blue dashed curve), and 5$n_0$ (yellow dashed
    curve). Specially, when $n_i=0$, the effective chemical potential does not
    relax to zero but to half of its initial value, indicating infinite
    charge-imbalance lifetime.}
  \label{figyw8}
\end{figure}

 In Fig.~\ref{figyw8}, the impurity-density dependencies of the charge-imbalance
 relaxation time (CIRT) $\tau_C$
 with $\Delta=0.8$~meV and 0.4~meV are plotted by the red solid curve with
 circles and blue dashed curve 
 with squares. It is shown that the CIRT is finite with finite impurity density,
 indicating that the electron-impurity scattering surely can cause the
 charge-imbalance relaxation. Specifically, one sees in Fig.~\ref{figyw8} that with
 the increase of the impurity density, the CIRT
 first decreases and then increases, showing similar features in the spin
 relaxation time (SRT) in the D'yakanov-Perel' (DP)\cite{DP} 
 mechanism.\cite{OptOri,Awschalom,Zutic,Fabian,Dyakonov_book,wureview,Korn,notebook}
 Furthermore, in the inset of Fig.~\ref{figyw8},
 the temporal evolutions of the normalized effective chemical potential $V/V_0$ are shown
 with different impurity densities $n_i=0$ (red solid curve), 0.02$n_0$ (green
 chain curve), $n_0$ (blue dashed curve) and 5$n_0$ (yellow dashed
 curve). Specifically, when $n_i=0$, the effective chemical potential does not
 relax to zero but to half of its initial value, indicating infinite
 charge-imbalance lifetime.

 Although there exist similarities in the momentum-scattering dependence of the
 relaxation rates,
 the DP
 mechanism\cite{OptOri,Awschalom,Zutic,Fabian,Dyakonov_book,wureview,Korn,notebook}
 cannot simply explain the features revealed in the charge-imbalance
 relaxation. In the DP mechanism, the SOC acts as a momentum-dependent
 effective magnetic field ${\bgreek \Omega}({\bf k})$, around
 which the electron spins with different momenta process with different frequencies, i.e.,
 the inhomogeneous broadening.\cite{wureview,inhomogeneous} Without the momentum
 scattering, this inhomogeneous
 broadening can cause a free-induction decay due to the destructive
 interference. Whereas when there
 exists momentum scattering, the system can be divided into the weak and
 strong scattering regimes. In the weak scattering regime with $|{\bgreek
   \Omega}({\bf k})| \tau_{\bf k} \gtrsim 1$, the momentum scattering opens
 a spin relaxation channel and the electron SRT $\tau_s$ is proportional to 
 $\tau_{\bf k}$. Here, $\tau_{\bf k}$ is the momentum relaxation time.
 In the strong scattering regime with $|{\bgreek
   \Omega}({\bf k})| \tau_{\bf k} \ll 1$, the momentum scattering suppresses the
 inhomogeneous broadening and $\tau_s$ is inversely proportional to $\tau_{\bf
   k}$.\cite{OptOri,Awschalom,Zutic,Fabian,Dyakonov_book,wureview,Korn,notebook}
Nevertheless, when the SOC does not depend on the angle of momentum, the elastic
scattering cannot provide the spin relaxation channel,\cite{Lin,wureview}
as long as the SOC is so weak that it can be neglected in the energy
spectrum of the electron.\cite{Tao_valley}

It is interesting to see that although
the effective chemical potential and quasiparticle excitation energy in the
coherent term of Eq.~(\ref{quasi_simplified}) act as
the effective SOC in the DP mechanism, they actually cannot provide the
  inhomogeneous broadening in the presence of
 the elastic scattering because of their momentum-angle independence.\cite{Lin,wureview,inhomogeneous}
Hence, the DP mechanism cannot explain the calculated charge-imbalance relaxation due to
the electron-impurity scattering.\cite{Tao_valley,Lin,wureview}
Moreover, one notes that even in the free situation,
the CIRT is infinite, which is in contrast to the finite SRT in the
DP mechanism.\cite{OptOri,Awschalom,Zutic,Fabian,Dyakonov_book,wureview,Korn,notebook} 
Actually, a new mechanism is expected to be responsible for the charge-imbalance
 relaxation here. 
The concrete physical picture for the charge-imbalance relaxation 
can be obtained from the analytical analysis, which is presented as follows.

Due to the absence of the momentum angle in the coherent terms of 
  Eq.~(\ref{quasi_simplified}),
 the calculation of the charge-imbalance relaxation can be markedly simplified. The density
  matrix can be expanded by its Fourier components, i.e., $\rho^h_{\bf
    k}=\rho^h_k+\sum_{l=1}^{\infty}\rho_k^{h,l} e^{il\theta_{\bf k}}$. With the
  initial state Eq.~(\ref{initial_c}), only the homogeneous
  component $\rho^h_{k}$ involves in the relaxation of the charge imbalance,
  whose kinetic equations are written as
 \begin{eqnarray}
\nonumber
&&\partial_T \rho^h_k+i\big[\tilde{E}_k\tau_3,\rho^h_k\big]+i\big[\tilde{\mu}_{\rm
  eff}\tau_1,\rho^h_k\big]+(\rho^h_k-\tau_3\rho^h_k\tau_3)/\tau^{\rm I}_{\bf
    k}\\
\mbox{}&&-(\tau_1\tau_2\rho^h_k-\tau_1\rho^h_k\tau_3+{\rm h.c.})/\tau^{\rm II}_{\bf
  k}=0.
\label{homogeneous_component}
\end{eqnarray}
In Eq.~(\ref{homogeneous_component}), $\tilde{E}_{k}=E_{\bf k}+\mu_{\rm eff}\zeta_{\bf k}/E_{\bf k}$,
$\tilde{\mu}_{\rm eff}=-\mu_{\rm eff}\Delta/E_{\bf k}$,
\begin{eqnarray}
\label{tau1}
\hspace{-0.7cm}&&\frac{1}{\tau_{\bf k}^{\rm I}}=\frac{n_im^*}{2\pi}\int d\theta_{{\bf k}'-{\bf
    k}}|V_{{\bf k}-{\bf k}'}|^2(u_{\bf k}^2-v_{\bf k}^2)^2\Big|\frac{E_{\bf
    k}}{\zeta_{\bf k}}\Big|,\\
\hspace{-0.7cm}&&\frac{1}{\tau_{\bf k}^{\rm II}}=\frac{n_im^*}{2\pi}\int d\theta_{{\bf k}'-{\bf
    k}}|V_{{\bf k}-{\bf k}'}|^2(u_{\bf k}^2-v_{\bf k}^2)u_{\bf k}v_{\bf
  k}\Big|\frac{E_{\bf k}}{\zeta_{\bf k}}\Big|, 
\label{tau2}
\end{eqnarray}
with $\theta_{\bf k}$ being the angle of momentum ${\bf k}$.
It is noted that $\tau_{\bf k}^{\rm I}$ and $\tau_{\bf k}^{\rm II}$ in Eqs.~(\ref{tau1})
and (\ref{tau2}) come from $\partial_t\rho_{\bf k}|_{\rm scat}^{\rm d}$ and
  $\partial_t\rho_{\bf k}|_{\rm scat}^{\rm off}$ in
  Eq.~(\ref{quasi_simplified}), respectively. Accordingly, $\tau_{\bf k}^{\rm II}$ directly
  breaks the quasiparticle-number conservation.\cite{Tao_2} Furthermore,
 $\tau_{\bf k}^{\rm I}$ and $\tau_{\bf k}^{\rm II}$ are different from the
conventional 
momentum scattering time $\tau_{\bf k}$,\cite{wureview,Tao_valley} although the
former being in the same order as $\tau_{\bf
  k}$. Actually, from Eq.~(\ref{tau2}), $\tau_{\bf k}^{\rm II}>0$ ($\tau_{\bf
  k}^{\rm II}<0$) for the electron-like (hole-like) quasi-electron with
$|{\bf k}|>k_F$ ($|{\bf k}|<k_F$).

By further expanding $\rho^h_k$ by the Pauli matrices in the Nambu space, i.e.,
$\rho^h_k=\rho^h_{k,0}\tau_0+\sum_{i=1}^3\rho^h_{k,i}\tau_i$ with
$\tau_0={\rm diag}\{1,1\}$, from Eq.~(\ref{homogeneous_component}),
 the kinetic equations for the
components $\rho^h_{k,i}$ ($i=1,2,3$) read 
\begin{equation}
\frac{\partial}{\partial T}\left(
\begin{array}{c}
\rho^h_{k,1}\\
\rho^h_{k,2}\\
\rho^h_{k,3}
\end{array}
\right)+
\left(
\begin{array}{ccc}
2/\tau_k^{\rm I}&2\tilde{E}_k&0\\
-2\tilde{E}_k&2/\tau_k^{\rm I}&2\tilde{\mu}_{\rm eff}\\
4/\tau_k^{\rm II}&-2\tilde{\mu}_{\rm eff}&0
\end{array}
\right)\left(
\begin{array}{c}
\rho^h_{k,1}\\
\rho^h_{k,2}\\
\rho^h_{k,3}
\end{array}
\right)=0.
\label{components}
\end{equation}
By using the components $\rho^h_{k,i}$ of $\rho^h_k$,
 the charge neutrality condition [Eq.~(\ref{self_consistent_s})]
becomes 
\begin{equation}
n_0=\sum_{\bf k}\Big[1-\frac{\zeta_{\bf k}+\mu_{\rm eff}}{\sqrt{(\zeta_{\bf
      k}+\mu_{\rm eff})^2+\Delta^2}}+\frac{\zeta_{\bf k}}{E_{\bf
    k}}(1+2\rho^h_{k,3})-\frac{2\Delta}{E_{\bf k}}\rho^h_{k,1}\Big].
\label{charge_components}
\end{equation}
Eq.~(\ref{components}) can be further analyzed in the near-equilibrium situation, in
which the density matrix is composed of its equilibrium and deviation
parts. By writing $\rho^h_{k,i}=\bar{\rho}^h_{k,i}+\delta\rho^h_{k,i}$ with
$\bar{\rho}^h_{k,i}$ and $\delta\rho^h_{k,i}$ being the equilibrium and deviation
parts, Eq.~(\ref{components}) are linearized to be  
\begin{eqnarray}
\label{deviations1}
\hspace{-0.4cm}&&\partial_T\delta\rho^h_{k,1}+2\delta\rho^h_{k,1}/\tau_{\bf k}^{\rm
  I}+2E_{\bf k}\delta\rho^h_{k,2}=0,\\
\label{deviations2}
\hspace{-0.4cm}&&\partial_T\delta\rho^h_{k,2}-2E_{\bf k}\delta \rho^h_{k,1}+2\delta\rho^h_{k,2}/\tau_{\bf
  k}^{\rm I}+\mu_{\rm eff}\Delta/E_{\bf k}=0,\\
\hspace{-0.4cm}&&\partial_T\delta\rho^h_{k,3}+4\delta\rho^h_{k,1}/\tau_{\bf k}^{\rm II}=0.
\label{deviations}
\end{eqnarray}

The features of the charge-imbalance relaxation without and with impurities 
can be understood based on Eqs.~(\ref{charge_components}) and (\ref{deviations1}-\ref{deviations}).
 We first analyze the impurity-free limit with $1/\tau_k^{\rm I}=1/\tau_k^{\rm
   II}=0$
 in Eqs.~(\ref{deviations1}-\ref{deviations}). 
From Eq.~(\ref{deviations}), one observes that in the
  impurity-free limit, 
  $\delta\rho^h_{k,3}$ does not evolve with time, which contributes the charge
  imbalance due to the non-equilibrium quasiparticle population. Furthermore, in
the steady state with the effective chemical potential denoted by $\mu_{\rm
  eff}^{\infty}$, from Eqs.~(\ref{deviations1}) and (\ref{deviations2}),
one obtains $\delta\rho^h_{k,2}=0$ and $\delta\rho^h_{k,1}=\mu_{\rm
  eff}^{\infty}\Delta/(2E_{\bf k}^2)$. Then from the charge neutrality condition
[Eq.~(\ref{charge_components})], in 
the steady state, $\sum_{\bf k}[-\frac{\Delta^2}{E_{\bf
    k}^3}(\mu^0_{\rm eff}-2\mu^{\infty}_{\rm eff})]=0$ with $\mu_{\rm eff}^0$
being the initial effective chemical potential. Hence, the steady-state
effective chemical potential $\mu^{\infty}_{\rm eff}=\mu^0_{\rm eff}/2$, which
explains the steady state found in the numerical calculation (shown by the red solid
curve in the inset of Fig.~\ref{figyw7}).

When there exists the momentum scattering, we first address the role of
$\tau_{\bf k}^{\rm I}$ in the charge-imbalance relaxation. One notes that in
Eq.~(\ref{deviations2}), $\delta\rho^h_{k,3}$ does not directly influence the
evolutions of $\delta\rho^h_{k,1}$ and $\delta\rho^h_{k,2}$, but rather
influences them through the influence on $\mu_{\rm eff}$. By neglecting 
$1/\tau_{\bf k}^{\rm II}$ in Eq.~(\ref{deviations}),
$\delta\rho^h_{k,3}$ still does not evolve with the time. Then from
Eqs.~(\ref{deviations1}) and (\ref{deviations2}), one obtains  
that in the steady
state, $\delta\rho^h_{k,1}=\frac{\Delta}{2E_{\bf k}^2}\mu_{\rm
  eff}^{\infty}/\big[1+\frac{1}{(E_{\bf k}\tau_{k}^{\rm
    I})^2}\big]$. Furthermore, from the charge neutrality condition
[Eq.~(\ref{charge_components})], one obtains
$\sum_{\bf k}\big\{-\frac{\Delta^2}{E_{\bf
    k}^3}\big[\mu^0_{\rm eff}-\mu^{\infty}_{\rm eff}(1+\frac{(E_{\bf k}\tau_{\bf
    k}^{\rm I})^2}{1+(E_{\bf k}\tau_{\bf
    k}^{\rm I})^2})\big]\big\}=0$, which indicates that $\mu_{\rm
  eff}^{0}/2<\mu^{\infty}_{\rm eff}<\mu_{\rm eff}^{0}$. Specifically, this
further indicates that when
$\langle E_{\bf k}\tau_{\bf
    k}^{\rm I}\rangle\ll 1$, the charge-imbalance relaxation can be suppressed by
  $\tau_{\bf k}^{\rm I}$ by suppressing the induction of $\rho^h_{k,1}$, i.e.,
  the correlation between the quasi-electron and quasi-hole.  
Moreover, from Eq.~(\ref{deviations}), one finds that in the presence of
$\tau_{\bf k}^{\rm II}$, the induction of the quasiparticle correlation
$\delta\rho^h_{k,1}$ directly leads to the fluctuation of the quasiparticle number
$\delta\rho^h_{k,3}$. Actually, this directly induces the annihilation of the
extra quasiparticles in the quasi-electron and quasi-hole bands into the
 Cooper pairs.\cite{Bardeen,Tinkham2,Josephson_B,Tao_1}

Therefore, $\tau_{\bf k}^{\rm II}$ can directly open a charge-imbalance relaxation
channel by relaxing the charge imbalance due to the quasiparticle
population, whose rate of change also depends on the value of the correlation between the
quasi-electron and quasi-hole. Accordingly, there exists the competition between the scattering
terms Eqs.~(\ref{tau1}) and (\ref{tau2}), leading to the non-monotonic
dependence on the momentum scattering for the CIRT.
 Specifically, in the weak scattering limit with $\langle E_{\bf k}\tau_{\bf
    k}^{\rm I}\rangle\gg 1$, one expects that the momentum scattering due to $\tau_{\bf
    k}^{\rm II}$ can directly
  open a charge-imbalance relaxation channel with the CIRT proportional to
  the momentum scattering strength. In the strong scattering regime with $\langle E_{\bf k}\tau_{\bf
    k}^{\rm I}\rangle\ll 1$, the induction of the quasi-electron and quasi-hole correlation
can be directly suppressed by the impurity scattering, which can further suppress the
charge-imbalance relaxation through the quasiparticle population. In this
situation, the CIRT is enhanced with the increase of the momentum scattering
strength. From this physical picture, $\langle E_{\bf k}\tau_{\bf
    k}^{\rm I}\rangle\approx \Delta \langle\tau_{\bf
    k}^{\rm I}\rangle=1$ labels the boundaries between the weak and strong
  scattering regimes. Thus, with $\langle\tau_{\bf
    k}^{\rm I}\rangle$ less influenced by the order parameter, the position of
  the boundaries between the weak and strong scattering regimes scales according
  to $1/\Delta$ (refer to the blue dashed and red solid curves in Fig.~\ref{figyw8}).

  Finally, we summarize the physical picture for the charge-imbalance relaxation channels
  provided by the elastic scattering. It is emphasized that the quasiparticle
    correlation between the quasi-electron and quasi-hole states,
    i.e., $\langle \alpha_{{\bf k}\uparrow}\beta_{{\bf
        k}\downarrow}\rangle$,
    is responsible for the charge-imbalance
    relaxation, which is often overlooked in the previous
    studies.\cite{Tinkham1,Tinkham2,Pethick_review}
  Here, the existence of the non-equilibrium effective
  chemical potential itself can cause the precession between the quasi-electron
  and quasi-hole states, directly inducing the quasiparticle 
  correlation. Once the
  quasiparticle correlation is induced, in the presence of the electron-impurity
  scattering, the process involving the annihilation of the quasi-electron and
  quasi-hole into the Cooper pairs, i.e., $\alpha_{{\bf k}\uparrow}\beta_{{\bf
      k}\downarrow}S^{\dagger}$, is inevitably
  triggered [refer to Eq.~(\ref{deviations})],\cite{Bardeen,Tinkham2,Josephson_B,Tao_2}
  whose rate of change is
  directly determined by $|\tau_{{\bf k}}^{\rm II}|$ defined in
  Eq.~(\ref{tau2}). This process has been schematically presented in
  Fig.~\ref{figyw2}.
  Consequently, the annihilation of the extra quasiparticles in the quasi-electron and
    quasi-hole bands directly causes the relaxation of charge imbalance for the
    quasiparticles and contributes to the fluctuation of the effective chemical
    potential for the condensate. Nevertheless, although the presence of the impurity
    scattering directly opens a charge-imbalance relaxation channel due to the quasiparticle
  population, it also suppresses the
  induction of the quasiparticle correlation.  
  This competition
  between the relaxation channels due to the
  quasiparticle correlation and population leads to the non-monotonic dependence
  on the momentum scattering for the charge-imbalance relaxation. Accordingly, 
  although there exist the similarities in the
  momentum-scattering dependence between the CIRT and SRT in the DP
  mechanism,
  their relaxation mechanisms are totally different.

\section{($s$+$p$)-wave superconducting (100) QWs}
\label{p_wave}
 
In this section, we study the optical response to the THz pulses in the ($s$+$p$)-wave
superconducting QWs,
which can be realized in the strong spin-orbit coupled InSb (100) QWs in proximity to an $s$-wave
superconductor.\cite{Tao_1,Gorkov_Rashba} In our
  previous work,\cite{Tao_1} we have shown that in this configuration, due to the
  Rashba-like SOC, not only the $p$-wave triplet Cooper pairing but also the
  corresponding triplet order parameter can be induced, which are in $(p_x\pm
  ip_y)$-type.
  Moreover, we find that the ${\bf l}$-vector for the triplet
  anomalous correlation and ${\bf d}$-vector for the triplet order parameter
  are parallel to the effective magnetic field due to the SOC. Similar
configuration can also be realized in CePt$_3$Si superconducting film,
 which is a heavy-Fermion material.\cite{CePtSi_1,CePtSi_2,Nagaosa} 
Here, based on the understanding on the equilibrium properties of 
  ($s$+$p$)-wave superconductor,\cite{Tao_1,Gorkov_Rashba,CePtSi_1,CePtSi_2,Nagaosa} 
it is intriguing to explore their non-equilibrium properties, especially those of the triplet Cooper pairs.
Below, we first present the Hamiltonian and then extend the gauge-invariant 
optical Bloch equations in the $s$-wave case to the ($s$+$p$)-wave one.
 Finally, we numerically calculate the
optical response by solving the optical Bloch equations, focusing on the
properties related to the spin dynamics of triplet Cooper pairs
(Sec.~\ref{p_Cooper}). 
The features in the Higgs-mode excitation and charge-imbalance dynamics are also
  addressed (Sec.~\ref{p_Higgs}).

\subsection{Hamiltonian}
\label{Hamiltonian_p}
In the ($s$+$p$)-wave superconducting (100) QWs, the Hamiltonian is composed by
the free BdG Hamiltonian $\tilde{H}_0$ and the interaction
Hamiltonian including the electron-electron Coulomb and
electron-impurity
 interactions $\tilde{H}_{ee}$ and $\tilde{H}_{ei}$. Here, the electron-phonon interaction is neglected
  due to its weak contribution at the low temperature (refer to Sec.\ref{Numerical_s}).
Specifically, the free BdG Hamiltonian in the
  presence of an optical field propagating along the $\hat{\bf
      z}$-direction,
 in which the vector potential is assumed to be along the
  $\hat{\bf x}$-direction, is written as\cite{Tao_2,Nagaosa} 
\begin{widetext}
\begin{equation}
\hspace{-0.45cm}\tilde{H}_0=\int\frac{d{\bf r}}{2}\tilde{\Psi}^{\dagger}\left(
\begin{array}{cccc}
\frac{({\bf p}-\frac{e}{c}{\bf A})^2}{2m^*}-\mu+e\phi(x) & -\alpha(k_x-\frac{eA_x}{c})-i\alpha k_y &
\frac{\Delta_p}{2}\{e^{i\zeta(x)},e^{i\theta_{\bf k}}\}&\Delta_se^{i\zeta(x)}\\
-\alpha(k_x-\frac{e A_x}{c})+i\alpha k_y&\frac{({\bf p}-\frac{e}{c}{\bf
    A})^2}{2m^*}-\mu+e\phi(x)&-\Delta_se^{i\zeta(x)}
&-\frac{\Delta_p}{2}\{e^{i\zeta(x)},e^{-i\theta_{\bf k}}\}\\
\frac{\Delta_p}{2}\{e^{-i\zeta(x)},e^{-i\theta_{\bf
    k}}\}&-\Delta_se^{-i\zeta(x)}&-\frac{({\bf p}+\frac{e}{c}{\bf
    A})^2}{2m^*}+\mu-e\phi(x)&-\alpha (k_x+\frac{eA_x}{c})+i\alpha k_y\\
\Delta_se^{-i\zeta(x)}&-\frac{\Delta_p}{2}\{e^{-i\zeta(x)},e^{i\theta_{\bf
    k}}\}&-\alpha (k_x+\frac{eA_x}{c})-i\alpha k_y&-\frac{({\bf
    p}+\frac{e}{c}{\bf A})^2}{2m^*}+\mu-e\phi(x)
\end{array}
\right)\tilde{\Psi}.
\label{p_Hamiltonion}
\end{equation}
\end{widetext}
Here, $\tilde{\Psi}(x)=(\psi_{\uparrow}(x),\psi_{\downarrow}(x),\psi^{\dagger}_{\uparrow}(x),
\psi^{\dagger}_{\downarrow}(x))$ represents the field operator in the Nambu$\otimes$spin
space; $\alpha=\gamma_D(\pi/a)^2$  
  for the infinitely deep well 
  with $\gamma_D$ being the Dresselhaus coefficient;
 and $\Delta_p$ ($\Delta_s$) is the magnitude of the $p$-wave triplet ($s$-wave singlet) order
  parameter.  
The electron-electron Coulomb and electron-impurity interactions are written as 
\begin{eqnarray}
  \nonumber
  \hspace{-0.5cm}&&\tilde{H}_{\rm ee}=\int \frac{d{\bf r}d{\bf
      r}'}{8}U({\bf r}-{\bf
  r}')\big[\tilde{\Psi}^{\dagger}({\bf r}){\mathcal T}_3\tilde{\Psi}({\bf r})\big]
\big[\tilde{\Psi}^{\dagger}({\bf r}'){\mathcal T}_3\tilde{\Psi}({\bf
    r}')\big],\\
\label{p_Coulomb}
\hspace{-0.5cm}&&\\
\hspace{-0.5cm}&&\tilde{H}_{\rm ei}=\frac{1}{2}\int d{\bf r}\tilde{\Psi}^{\dagger}({\bf
  r})V({\bf r}){\mathcal T}_3\tilde{\Psi}({\bf r}),
\label{p_impurity}
\end{eqnarray}
respectively, in which ${\mathcal T}_3={\rm
  diag}\{1,1,-1,-1\}$. Finally, it is addressed 
that in this Hamiltonian [Eqs.~(\ref{p_Hamiltonion}), (\ref{p_Coulomb}) and (\ref{p_impurity})],
 there exits similar gauge
structure as the one in the $s$-wave superconductor [Eqs.~(\ref{vector}-\ref{phase})].

\subsection{Optical Bloch Equations}
\label{KSBE_p}
In this part, we generalize the optical Bloch equations in the $s$-wave
superconducting QWs to the ones in the 
($s$+$p$)-wave situation. Here, in the ${\bf p}_s$-gauge, the optical Bloch equations in the
  homogeneous situation read 
\begin{eqnarray}
\nonumber
\hspace{-0.5cm}&&\frac{\partial \rho_{\bf k}}{\partial T}+
i\Big[\Big(\frac{{\bf k}^2}{2m^*}-\Phi\Big){\mathcal T}_3+h_{\bf k}^{\rm soc},\rho_{\bf k}\Big]
+\frac{1}{2}
\Big\{\frac{\partial {\bf p}_s}{\partial T}{\mathcal T}_3,\frac{\partial \rho_{\bf k}}{\partial {\bf
    k}}\Big\}\\
\nonumber
\hspace{-0.5cm}&&\mbox{}+i\Big[\left(
\begin{array}{cccc}
0&-\alpha {\bf p}^x_{s} &0&0 \\
-\alpha {\bf p}^x_{s}&0&0&0\\
0&0&0&\alpha {\bf p}^x_{s}\\
0&0&\alpha {\bf p}^x_{s}&0
\end{array}
\right),\rho^{\rm off}_{{\bf k}}\Big]+i\Big[\frac{{\bf p}_s^2}{2m^*}{\mathcal T}_3,\rho_{\bf
  k}\Big]\\
\nonumber
\hspace{-0.5cm}&&\mbox{}+i\Big[\left(
\begin{array}{cccc}
0&0&\Delta_pe^{i\phi_{\bf k}}&\Delta_s \\
0&0&-\Delta_s&-\Delta_p e^{-i\phi_{\bf k}}\\
\Delta_pe^{-i\phi_{\bf k}}&-\Delta_s&0&0\\
\Delta_s&-\Delta_pe^{i\phi_{\bf k}}&0&0
\end{array}
\right),\rho_{{\bf k}}\Big]\\
\hspace{-0.5cm}&&\mbox{}=\frac{\partial\rho_{\bf k}}{\partial t}\Big|_{\rm HF}+\frac{\partial\rho_{\bf
  k}}{\partial t}\Big|_{\rm scat}.
\label{homogeneous_p}
\end{eqnarray}
The details of the derivation have been outlined in Sec.~\ref{KSBE}.
In Eq.~(\ref{homogeneous_p}), $\rho_{\bf k}$ is the $4\times 4$ density matrix in the
Nambu$\otimes$spin space;\cite{Tao_2} $\Phi=\mu-\mu_{\rm eff}$; $h_{\bf k}^{\rm soc}=-\alpha k_x\tau_0\otimes
\sigma_x+\alpha k_y\tau_3\otimes \sigma_y$ represents the SOC Hamiltonian in the
Nambu$\otimes$spin space;
 $\rho^{\rm off}_{{\bf k}}=\frac{1}{2}(\rho_{\bf k}-{\mathcal T}_3\rho_{\bf
   k}{\mathcal T}_3)$ only includes the off-diagonal blocks of the
 density matrix. Furthermore, in Eq.~(\ref{homogeneous_p}), in the left-hand
 side of the equation,
 the fourth term comes from the supercurrent-induced effective SOC, which can
 directly induce the dynamics of the Cooper-pair anomalous
 correlation;
 in the right-hand side of the equation,     
 the HF and scattering terms are written as 
\begin{eqnarray}
 \label{HF_p}
\hspace{-0.8cm}&&\partial_t\rho_{\bf k}|_{\rm HF}=i\sum_{{\bf k}'}U_{{\bf k}-{\bf
    k}'}\Big[{\mathcal T}_3(\rho_{{\bf k}'}-\bar{\rho}_{{\bf
      k}'}){\mathcal T}_3,\rho_{\bf k}\Big],\\
\nonumber
\hspace{-0.8cm}&&\partial_t\rho_{\bf k}|_{\rm scat}=-\pi n_i\sum_{{\bf k}'}\sum_{\eta_1,\eta_2=1}^4|V_{{\bf
    k}-{\bf k}'}|^2\delta({\mathcal E}_{{\bf k}'\eta_1}-{\mathcal E}_{{\bf
    k}\eta_2})\\
\hspace{-0.8cm}&&\times\Big[\mathcal{T}_3\tilde{\Gamma}_{{\bf k}'\eta_1}\mathcal{T}_3\tilde{\Gamma}_{{\bf
    k}\eta_2}\rho_{\bf k}-\mathcal{T}_3\rho_{{\bf k}'}\tilde{\Gamma}_{{\bf
    k}'\eta_1}\mathcal{T}_3\tilde{\Gamma}_{{\bf k}\eta_2}+{\rm H.c.}\Big].
\label{scat_p}
\end{eqnarray}

Specifically, in Eq.~(\ref{HF_p}), $\bar{\rho}_{{\bf
      k}}$ is the density matrix in the equilibrium state.\cite{Tao_1} 
In Eq.~(\ref{scat_p}), ${\mathcal E}_{{\bf k}1}=E_{\bf k}^{+}$, ${\mathcal
    E}_{{\bf k}2}=E_{\bf k}^{-}$, ${\mathcal E}_{{\bf k}3}=-E_{\bf k}^{+}$ and
  ${\mathcal E}_{{\bf k}4}=-E_{\bf k}^{-}$, where $E_{\bf k}^{\pm}=\sqrt{(\varepsilon_{\bf
    k}^{\pm}-\mu)^2+\Delta_{\pm}^2}$ with $\varepsilon_{{\bf k}}^{\pm}={\bf k}^2/(2m^*)\pm \alpha
k$ and $\Delta_{\pm}=|\Delta_s\pm \Delta_p|$. The projection operators
$\tilde{\Gamma}_{{\bf k}\eta}=\tilde{{\mathcal U}}_{\bf k}\tilde{Q}_{{\bf
    k}\eta}\tilde{{\mathcal U}}_{{\bf k}}^{\dagger}$ with $\tilde{Q}_{{\bf k}1}={\rm
  diag}(1,0,0,0)$, $\tilde{Q}_{{\bf k}2}={\rm
  diag}(0,1,0,0)$, $\tilde{Q}_{{\bf k}3}={\rm
  diag}(0,0,1,0)$ and $\tilde{Q}_{{\bf k}4}={\rm
  diag}(0,0,0,1)$ being the projection operators in the quasiparticle space.
Here, $\tilde{{\mathcal U}}_{\bf k}$ is the unitary transformation matrix from the particle
space to the quasiparticle one, which is 
 written as 
\begin{equation}
\tilde{{\mathcal U}}_{\bf k}=\frac{\sqrt{2}}{2}
\left(
\begin{array}{cccc}
u_{{\bf k}}^+e^{-i\phi_{\bf k}}&-u_{{\bf k}}^+&v_{{\bf k}}^+&v_{{\bf k}}^+e^{-i\phi_{\bf k}} \\
-u_{{\bf k}}^-&-u_{{\bf k}}^-e^{i\phi_{\bf k}}&v_{{\bf k}}^-e^{i\phi_{\bf k}}&-v_{{\bf k}}^-\\
v_{{\bf k}}^+&-v_{{\bf k}}^+e^{i\phi_{\bf k}}&-u_{{\bf k}}^+e^{i\phi_{\bf k}}&-u_{{\bf k}}^+\\
-v_{{\bf k}}^-e^{-i\phi_{\bf k}}&-v_{{\bf k}}^-&-u_{{\bf k}}^-&u_{{\bf k}}^-e^{-i\phi_{\bf k}}
\end{array}
\right).
\label{unitary_p}
\end{equation}
In Eq.~(\ref{unitary_p}), $u_{{\bf k}}^{\pm}=\sqrt{1/2+\zeta_{{\bf k}}^{\pm}/(2E_{{\bf
        k}}^{\pm})}$ and $v_{{\bf k}}^{\pm}=\sqrt{1/2-\zeta_{{\bf k}}^{\pm}/(2E_{{\bf
        k}}^{\pm})}$ with $\zeta_{{\bf k}}^{\pm}=\varepsilon_{\bf
  k}^{\pm}-\mu$.

Finally, it is addressed that in Eq.~(\ref{homogeneous_p}), 
 $\mu_{\rm eff}$ in $\Phi$ is unspecified, which can
also be 
determined by the charge neutrality condition from the
 self-consistent equation\cite{Takahashi_SHE1,Takahashi_SHE2,Hirashima_SHE,
Takahashi_SHE3,Takahashi_SHE_exp,Hershfield,Spivak,kinetic_book}    
\begin{eqnarray}
\nonumber
\hspace{-0.5cm}&&N_0=\frac{1}{2}\sum_{\bf k}\Big\{2-\frac{\displaystyle \varepsilon_{{\bf
        k}}^+-\Phi}{\displaystyle \sqrt{(\varepsilon_{{\bf k}}^+-\Phi)^2+\Delta_+^2}}-\frac{\displaystyle \varepsilon_{{\bf
        k}}^--\Phi}{\displaystyle \sqrt{(\varepsilon_{{\bf k}}^--\Phi)^2+\Delta_-^2}}\\
\hspace{-0.5cm}&&\mbox{}+\mbox{Tr}\big[\tilde{{\mathcal U}}_{\bf k}(\rho_{\bf
  k}^h-\frac{1-{\mathcal T}_3}{2})\tilde{{\mathcal U}}_{\bf k}^{\dagger}{\mathcal T}_3\big]\Big\}.
\label{charge_p}
\end{eqnarray}
Here, $N_0$ is the electron density in the QWs, and $\rho_{\bf k}^{h}=\tilde{{\mathcal U}}_{\bf k}^{\dagger}\rho_{\bf
  k}\tilde{{\mathcal U}}_{\bf k}$ is the density matrix in the quasiparticle space. In the
numerical calculation below, 
${\bf p}_s$ takes
the same form as the one in Eq.~(\ref{superfluid_momentum}).

\subsection{Numerical Results}
\label{Numerical_p}
In this subsection, we present the numerical results
by solving the optical Bloch equations [Eqs. (\ref{homogeneous_p}),
(\ref{superfluid_momentum}) and (\ref{charge_p})] in the
specific material InSb (100) QWs in proximity to an $s$-wave
superconductor.
 All parameters used in our computation are
listed in Table~\ref{material_p}.\cite{material_book,SOC_p}

\begin{table}[h]
\caption{Parameters used in the computation for InSb (100) QWs
in proximity to an $s$-wave superconductor.\cite{material_book,SOC_p}}
 \label{material_p} 
\begin{tabular}{ll|ll}
    \hline
    \hline    
    $m^*/m_0$&\;\;\;$0.015$&\;$N_0$~(cm$^{-2}$)&\;\;\;$8\times10^{14}$\\[4pt]
    $\kappa_0$&\;\;\;$16.8$ &\;$\gamma_D~({\rm eV}\cdot{\rm \mathring{A}}^3)$&\;\;\;$389$\\[4pt]
    $T_e~({\rm K})$&\;\;\;$5$&\;$a~({\rm nm})$&\;\;\;$3$\\[4pt]
    $\Delta_p~({\rm meV})$&\;\;\;$0.05$&\;$\sigma_t~({\rm ps})$&\;\;\;$4$\\[4pt]
    \hline
    \hline
\end{tabular}
\end{table}

\begin{figure}[ht]
  {\includegraphics[width=7.9cm]{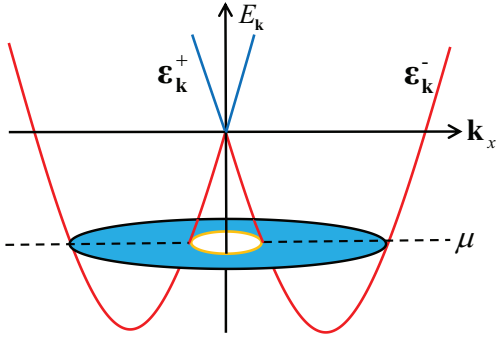}}
  \caption{(Color online) Schematic of the band structures of
 $\varepsilon_{\bf k}^+$- and $\varepsilon_{\bf k}^-$-bands,
shown by the blue and red solid curves, respectively. The dashed line
labeled by $\mu$ corresponds to the chemical potential, with which only the lower band
is efficiently populated. In this situation,
 one sees that the Fermi ``sphere'' (the blue region) is in
the shape of an annulus with the inner and outer Fermi surfaces represented by
the yellow and black circles.}
\label{figyw9}
\end{figure}

In our computation, the electron density is chosen to be $n_e\le 3N_0$. 
With this density, electrons mainly populate at the lower branch
of the energy band ($\varepsilon_{\bf k}^-$-band), as shown in Fig.~\ref{figyw9} by the red solid
curve.\cite{Tao_2} 
In this situation, one sees in Fig.~\ref{figyw9} that the Fermi ``sphere'' (the blue
region) is in
the shape of an annulus with the inner and outer Fermi surfaces represented by
the yellow and black circles.  
Then some approximations can be
made for the scattering term [Eq.~(\ref{scat_p})] to reduce the computation
complication.\cite{Tao_valley}
 On one hand, with $\mu\lesssim 0$ at low temperature, the
 scattering between $\varepsilon_{\bf k}^+$- and $\varepsilon_{\bf k}^-$-bands 
contributes marginally to the scattering process; on the other hand,  
the scattering between the inner and outer Fermi surfaces can be neglected, 
because in this process, large momentum magnitude needs to be changed.\cite{Tao_valley}

\subsubsection{Optical Excitation of Spin Polarization of Cooper Pairs}
\label{p_Cooper}
In our previous work,\cite{Tao_1} it has been revealed that in the InSb (100) QWs in proximity to an $s$-wave
superconductor, due to the Rashba-like SOC,  
the spin polarization of electrons in the momentum space is parallel to the effective magnetic
field ${\bgreek \Omega}({\bf k})$ due to the SOC in the
equilibrium state. This feature is all the same as the
  one in the normal state.\cite{Tao_valley,Tao_2} In the {\em normal} state,
 it has been well understood that when
there exists electrical field, with the drift
of electron states due to the applied field, this
parallel relation is broken and hence the effective magnetic field
${\bgreek \Omega}({\bf k})$ can induce the momentum-dependent out-of-plane spin
polarization, which accounts for the spin current of
electrons.\cite{Tao_valley} Moreover, the center-of-mass momentum driven by the
electrical field contributes to the effective magnetic field, which tends to
polarize the electron states along this effective magnetic
 field.\cite{Jinluo,Jianhua_Liouville,Ivchenko,topological_0}  
 In the superconducting state in our configuration, we show that both the spin
polarization and spin current can be induced by the optical field, which
oscillate with the same frequency of the optical field. Nevertheless, with our
computation 
parameters, we find that the order parameters have little influence on the
optical generations of the spin
polarization and spin current. The details are presented in Appendix~\ref{CC}.

Furthermore, in the superconducting InSb (100) QWs,
we have revealed that there exists $p$-wave triplet Cooper
pairing in $(p_x\pm ip_y)$-type.\cite{Tao_1}  With the frequency and
momentum-dependent\cite{Leggett,Sigrist,Tkachov,non_unitary} Cooper pairing written as
$[f_s({\bf k},\omega)+{\bf f}({\bf k},\omega)\cdot {\bgreek
  \sigma}]i\sigma_y$, 
it can be further shown that the
${\bf f}$-vector of the triplet Cooper pairing is also parallel to the
effective magnetic field ${\bgreek \Omega}({\bf k})$ due to the SOC.\cite{Tao_1}
By analogy with the optical generation of the spin polarization addressed above,
one expects that the ${\bf
  f}$-vector can also be controlled by the optical field. Actually, very recently,
Tkachov indeed showed that in noncentrosymmetric superconductors, the driven
center-of-mass momentum ${\bf q}$ of the Cooper pairs can induce the nonunitary
triplet pairing, which contributes to the spin polarization of the triplet
Cooper pairs.\cite{Tkachov} Specifically, with the 
spin polarization of Cooper pairs ${\bf S}^{\rm CP}_{\bf k}$ described by $i{\bf f}({\bf k},\omega=0)\times {\bf
  f}^*({\bf k},\omega=0)$, Tkachov showed that  
with small superfluid velocity, the Cooper-pair spin
polarization\cite{Tkachov}
\begin{equation}
{\bf S}^{\rm CP}_{\bf k}\propto {\bgreek \Omega}_{\bf k}\times[{\bgreek
    \Omega}_{\bf k}\times({\bf q}\cdot\partial_{\bf k}){\bgreek \Omega}_{\bf
    k}].
\label{Tkachov}
\end{equation}

It is noted that Tkachov's calculation is performed in the static situation,
in which the supercurrent is induced from the proximity effect.\cite{Tkachov} Nevertheless, with
the super-fluid velocity dynamically generated by the optical field, we 
expect that the
Cooper-pair spin polarization can also be induced, which has not yet been
  reported in the literature. Particularly, the optical method can avoid
  the complexity when introducing the supercurrent by the proximity effect.
Furthermore, the study of the dynamics of the Cooper-pair spin polarization can
provide more understanding from the microscopic point of view.
Actually, before the concrete calculation, the feature of Cooper-pair
spin polarization in (100) InSb QWs can be
roughly conjectured based on Eq.~(\ref{Tkachov}).
Here, for the Dresselhaus SOC in the Rashba-like type, i.e., ${\bgreek \Omega}_{\bf
  k}=-\alpha k_x\hat{\bf x}+\alpha k_y\hat{\bf
  y}$, with the center-of-mass momentum ${\bf q}=q_x\hat{\bf x}$, we
conjecture from Eq.~(\ref{Tkachov}) 
\begin{equation}
{\bf S}^{\rm CP}_{\bf k}\propto{\bgreek \Omega}_{\bf k}\times[{\bgreek
    \Omega}_{\bf k}\times({\bf q}\cdot\partial_{\bf k}){\bgreek \Omega}_{\bf
    k}]=q_xk_y^2\hat{\bf x}+q_xk_xk_y\hat{\bf y}.
\label{Rashba}
\end{equation}
This indicates that the $\hat{\bf x}$-component ($\hat{\bf y}$-component) of the Cooper-pair spin
polarization is even (odd) in momentum, which is proportional to $k_y^2$
($k_xk_y$). Then the {\em total} Cooper-pair spin polarization, which is in sum of the momentum,
is along the $\hat{\bf x}$-direction.

 In our
  framework, in the density matrix, the information about the frequency has been integrated
  [Eq.~(\ref{integration})], in which the Cooper pairing is integrated to be the
  anomalous correlation.\cite{Tao_1,Tao_2,Leggett_book} Actually, it is
  reasonable to define the Cooper-pair spin polarization by using the anomalous
  correlation but not Cooper pairing at zero frequency.\cite{Tkachov}
Microscopically, the Cooper-pair spin
polarization is calculated by the wavefunction of the triplet Cooper pairs
[Eq.~(\ref{pair_function})],
which does not depend on the relative temporal coordinate (refer to
Ref.~\onlinecite{Leggett_book}). Furthermore, the Fourier
components of the Cooper-pair wavefunction are exactly the  {\em anomalous
  correlation},\cite{Leggett_book} which can be directly calculated by the optical Bloch equations. 
 Here, 
 the anomalous correlation of the Cooper pairs is 
 convenietly expressed as 
\begin{equation}
\left(
\begin{array}{cccc}
\rho_{13}({\bf k})&\rho_{14}({\bf k}) \\
\rho_{23}({\bf k})&\rho_{24}({\bf k})
\end{array}
\right)=\big[l_0({\bf k})\sigma_0+{\bf l}({\bf k})\cdot{\bgreek \sigma}\big]i\sigma_y,
\end{equation}
in which the ${\bf l}$-vector ($l_0$) describes the anomalous correlation of the triplet
(singlet) Cooper pairs. Accordingly, from definition of the 
  Cooper-pair spin polarization in Eq.~(\ref{polarization_definition}),
by using the ${\bf l}$-vector, the total spin
polarization of the Cooper pair reads 
\begin{equation}
{\bf P}_{\rm C}=\frac{\sum_{\bf k}i{{\bf l}({\bf k})\times {\bf l}^*({\bf
    k})}}{\sum_{\bf k}\big[|{\bf l}_x({\bf k})|^2+|{\bf l}_y({\bf k})|^2+|{\bf
  l}_z({\bf k})|^2\big]},
\label{polarization_definition2}
\end{equation}
in which the denominator is introduced to act as the normalization factor.

\begin{figure}[ht]
  {\includegraphics[width=7.9cm]{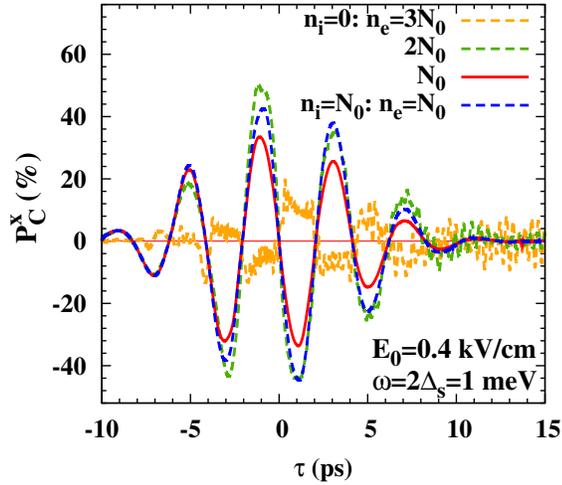}}
  \caption{(Color online) Temporal evolutions of the $\hat{\bf x}$-component
    of the total Cooper-pair spin polarization ${\bf P}_{\rm C}^x$ 
    with different electron densities
    $n_e=3N_0$ (yellow dashed curve), $2N_0$ (green chain curve) and $N_0$ [red
    solid (blue dotted) curve in the absence (presence) of the
    impurity]. $E_0=0.4$~kV/cm.
    The blue
    dotted curve denotes the total Cooper-pair spin polarization in the presence
  of the impurity with $n_i=N_0$.}
  \label{figyw10}
\end{figure}

Then the temporal evolution of ${\bf P}_{\rm C}$ is calculated by the optical Bloch
equations, in which it is found that only the $\hat{\bf x}$-component ${\bf
  P}_{\rm C}^x$ exists, in agreement
with the analysis in Eq.~(\ref{Rashba}). 
Specifically, in Fig.~\ref{figyw10}, the temporal evolutions of ${\bf P}_{\rm C}^x$
 are plotted with different
 electron densities
$n_e=3N_0$ (yellow dashed curve), $2N_0$ (green chain curve) and $N_0$ [red
solid (blue dotted) curve in the absence (presence) of the impurity].
 It is noted that for these three electron densities, the chemical
potentials are about $4.2$, $-9.9$ and $-15.6$~meV, respectively, which are larger than the
minimal value of the energy spectra $-m^*\alpha^2/2\approx -17.9$~meV.\cite{Tao_1} 
In Fig.~\ref{figyw10}, it can be seen that in the impurity-free situation, when $n_e=2N_0$ and $N_0$, the total 
spin polarizations of the Cooper
pairs are significant, which oscillate with the frequency of the optical
field. Whereas when $n_e=3N_0$, corresponding to a positive chemical
potential, the total Cooper-pair spin polarization is efficiently suppressed,
 whose oscillation does not show a particular pattern. This reveals that the low
 electron density with {\em single-band population} is in favor of the realization of 
 significant Cooper-pair total spin polarization. This is in contrast to the
 optical excitation of the electron spin polarization, which is less influenced
 by the electron density (refer to Appendix~\ref{CC}). 
Finally, we also calculate the total Cooper-pair spin polarization with the
impurity density $n_i=N_0$, which is presented by the blue dotted curve in
Fig.~\ref{figyw10}. One finds that the impurity can even enhance the 
optical excitation of the total Cooper-pair spin polarization.

\begin{figure}[ht]
   {\includegraphics[width=7.6cm]{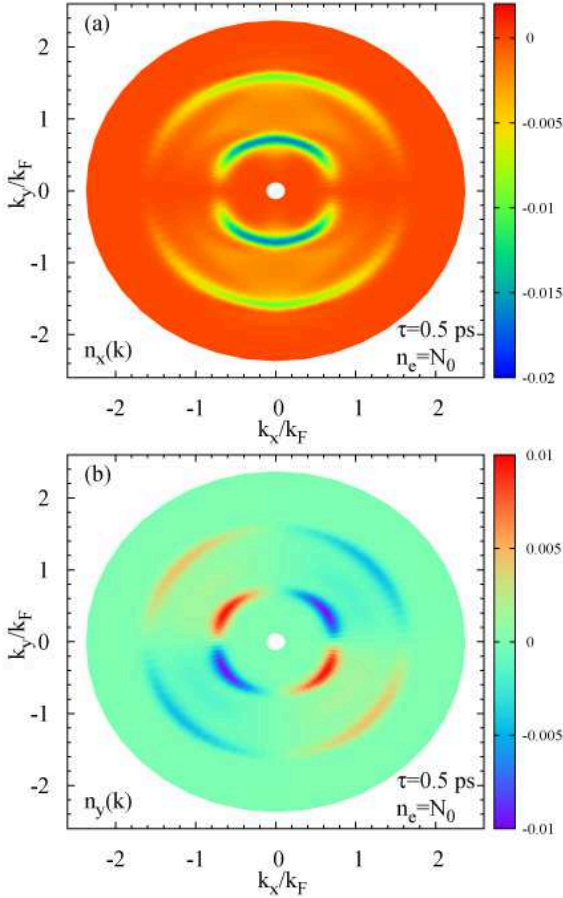}}
   \caption{(Color online) Momentum distributions of the Cooper-pair spin vectors ${\bf
       n}_x({\bf k})$ [(a)] and ${\bf n}_y({\bf k})$ [(b)] at
     $\tau=0.5$~ps when $n_e=N_0$ and $n_i=0$. $E_0=0.4$~kV/cm. At this particular time $\tau=0.5$~ps, ${\bf
       p}_s^x\approx -0.4k_F<0$ with $k_F=\sqrt{2\pi N_0}$. In the figures,
     it can be seen that ${\bf n}_x({\bf k})\propto -k_y^2$ and ${\bf n}_y({\bf
       k})\propto -k_xk_y$ in both the inner and outer Fermi surfaces, in agreement with Eq.~(\ref{Rashba}).
     One further observes that both ${\bf n}_x({\bf k})$ and ${\bf
       n}_y({\bf k})$ are larger in the inner Fermi surface than the those in the
     outer Fermi one.}
   \label{figyw11}
 \end{figure}

To further reveal the dynamical features of ${\bf P}_{\rm C}$ from the microscopic
viewpoint, we calculate the momentum distribution of the Cooper-pair spin
vector ${\bf n}({\bf k})\equiv i{{\bf l}({\bf k})\times {\bf l}^*({\bf
    k})}$ in the dynamical evolution. When
$n_e=N_0$ in the impurity-free situation, at a particular time $\tau=0.5$~ps with ${\bf
  p}_s^x\approx -0.4k_F<0$ ($k_F=\sqrt{2\pi N_0}$),
 the $\hat{\bf x}$- and $\hat{\bf y}$-components of ${\bf n}({\bf
  k})$ in the momentum space are plotted in Figs.~\ref{figyw11}(a) and (b),
 respectively. With this electron density $n_e=N_0$, the Fermi ``sphere'' is in
the shape of an annulus with the inner and outer Fermi surfaces (refer to
Fig.~\ref{figyw9}).
It can be seen
that ${\bf n}({\bf k})$ is significant only around the inner and outer Fermi
surfaces.\cite{Tao_1,Tao_2}
Moreover, one finds that ${\bf n}_x({\bf k})\propto -k_y^2$ and ${\bf n}_y({\bf
  k})\propto -k_xk_y$ 
in both the inner and outer Fermi surfaces, in agreement with
Eq.~(\ref{Rashba}).
Thus, only ${\bf n}_x({\bf k})$
contributes to the total
Cooper-pair spin polarization after the sum of momentum.
Furthermore, a specific feature is found by comparing the Cooper-pair spin
vectors in the inner and outer Fermi surfaces. 
It is intriguing to observe that both ${\bf n}_x({\bf k})$ and ${\bf
  n}_y({\bf k})$ are larger in the inner Fermi surface than those in the
outer one. These numerical results can be understood from the analytical
analysis in the following.  

It is convenient to set up simplified kinetic equations for $l_0$ and ${\bf
    l}$-vector from the optical Bloch equations [Eq.~(\ref{homogeneous_p})],
  which can be used to analyze the dynamics of Cooper pairing
  directly. Moreover, from the simplified kinetic equations, 
  the kinetic equations for the Cooper-pair spin vectors can
  be obtained. Specifically, with the third, fifth and sixth
  terms in the left-hand side and the HF-term in the right-hand side of
  Eq.~(\ref{homogeneous_p}) neglected,
  (which have been checked to be unimportant in the excitation of Cooper-pair spin
  polarization by the numerical
  calculations), the kinetic equations for $l_0$ and ${\bf
    l}$-vector become 
\begin{eqnarray}
\nonumber
&&\partial_T(l_0+{\bf l}\cdot{\bgreek
    \sigma})+i\{\zeta_{\bf k}-\alpha k_x\sigma_x+\alpha k_y\sigma_y,l_0+{\bf
    l}\cdot{\bgreek \sigma}\}\\
&&\mbox{}+i\big[-\alpha {\bf p}_s^x\sigma_x,l_0+{\bf
    l}\cdot{\bgreek \sigma}\big]=0.
\end{eqnarray} In the matrix form,
\begin{eqnarray}
  \nonumber
\hspace{-0.5cm}&&\partial_T\left(
\begin{array}{c}
l_{\bf k}^0\\
{\bf l}_{\bf k}^x\\
{\bf l}_{\bf k}^y\\
{\bf l}_{\bf k}^z
\end{array}
\right)+2i\left(
\begin{array}{cccc}
\zeta_{\bf k}&-\alpha k_x&\alpha k_y&0\\
-\alpha k_x&\zeta_{\bf k}&0&0\\
\alpha k_y&0&\zeta_{\bf k}&i\alpha {\bf p}_s^x\\
0&0&-i\alpha {\bf p}_s^x&\zeta_{\bf k}
\end{array}
\right)
\left(
\begin{array}{c}
l_{\bf k}^0\\
{\bf l}_{\bf k}^x\\
{\bf l}_{\bf k}^y\\
{\bf l}_{\bf k}^z
\end{array}
\right)\\
\hspace{-0.5cm}&&\mbox{}=0.
\label{l_vectors}
\end{eqnarray}

In Eq.~(\ref{l_vectors}), one finds that $l_0$ is coupled to the ${\bf l}$-vector due
to the SOC. It is noted that in the equilibrium state, the
${\bf l}$-vector of the triplet Cooper correlation is parallel to the effective
magnetic field ${\bgreek \Omega}({\bf k})$ due to the SOC in the momentum space.
 When the optical
field with in-plane vector potential is applied to the
  superconducting system, the superconducting velocity is induced, which
  directly contributes to the effective SOC, i.e., $\pm i\alpha {\bf p}_s^x$, in Eq.~(\ref{l_vectors}). 
  This effective SOC can cause the
  precession of the ${\bf l}$-vectors, with the component perpendicular to ${\bgreek
    \Omega}({\bf k})$ induced. Thus, from the definition ${\bf n}({\bf k})=i{\bf l}({\bf k})\times {\bf l}^*({\bf
    k})$,\cite{Tkachov} the Cooper-pair spin polarization
  ${\bf n}({\bf k})$ is expected.

To make this point more concrete, from Eq.~(\ref{l_vectors}),
 the kinetic equations for the ${\bf n}$-vector can be
derived, written as
\begin{eqnarray}
\label{n_vector_1}
\hspace{-0.5cm}&&\partial_T^2{\bf n}_x+4\alpha^2k_y(k_x{\bf n}_y+k_y{\bf n}_x)
+4\alpha^2k_y{\bf p}_s^x({\bf l}_y^*l_0+{\bf l}_yl_0^*)=0,\\
\label{n_vector_2}
\hspace{-0.5cm}&&\partial_T{\bf n}_y=(k_x/k_y)\partial_T{\bf n}_x+2\alpha {\bf
  p}_s^x{\bf n}_z,\\
\hspace{-0.5cm}&&\partial_T^2{\bf n}_z+4\alpha^2k^2{\bf n}_z+2\alpha{\bf
  p}_s^x\partial_T{\bf n}_y+4\alpha^2k_x{\bf p}_s^x({\bf l}_z^*l_0+{\bf l}_zl_0^*)=0.
\label{n_vector_3}
\end{eqnarray}
From Eqs.~(\ref{n_vector_1}-\ref{n_vector_3}), it is concluded that ${\bf n}_x\propto {\bf p}_s^x$,
 ${\bf n}_y\propto {\bf p}_s^x$, and ${\bf n}_z\propto ({\bf p}_s^x)^2$. This is
 because in the equilibrium state, $\bar{\bf l}_z=0$ (``bar'' labels the
 equilibrium state), whose first order
 deviation 
 is proportional to ${\bf p}_s^x$. By only keeping the quantities in the
 first order of ${\bf p}_s^x$, one obtains from 
 Eq.~(\ref{n_vector_2}) that 
\begin{equation}
k_y{\bf n}_y=k_x{\bf n}_x,
\label{simple_relation}
\end{equation}
 which satisfies the
 numerical results presented in Figs.~\ref{figyw11}(a) and (b). Thus, from 
 Eq.~(\ref{n_vector_1}), one obtains
\begin{equation}
\partial_T^2{\bf n}_x+4\alpha^2k^2{\bf n}_x+8\alpha^2k_y{\bf p}_s^x\bar{\bf l}_y\bar{l}_0=0,
\end{equation}
whose solution reads
\begin{equation}
{\bf n}_x({\bf k})\approx\frac{8\alpha^2k_y\bar{\bf
    l}_y\bar{l}_0}{\omega^2-4\alpha^2k^2}{\bf p}_s^x.
\label{solution}
\end{equation}
Here, $\bar{l}_0({\bf k})$ and $\bar{\bf l}_y({\bf k})$ in the equilibrium state are\cite{Tao_1,Tao_2}
\begin{eqnarray}
  \label{l_0_equilibrium} 
\hspace{-0.9cm}&&\bar{l}_0=u^{+}_{\bf k}v^{+}_{\bf k}\big[2f_0(E_{\bf k}^+)-1\big]+u^{-}_{\bf
  k}v^{-}_{\bf k}\big[2f_0(E_{\bf k}^-)-1\big],\\
\hspace{-0.9cm}&&\bar{\bf l}_y=\frac{k_y}{k}\big\{u^{+}_{\bf k}v^{+}_{\bf k}[2f_0(E_{\bf k}^+)-1]-u^{-}_{\bf
  k}v^{-}_{\bf k}[2f_0(E_{\bf k}^-)-1]\big\}.
\label{l_y_equilibrium}
\end{eqnarray}
It is noted that when $\omega$ tends to zero, Eq.~(\ref{solution}) directly
recovers the static results in the work of Tkachov.\cite{Tkachov} Furthermore,
Eq.~(\ref{solution}) also describes the dynamical situation especially at the
high frequency. Moreover, it is expected that the
Cooper-pair spin polarization can be resonantly excited for particular momenta
around $k^*\equiv \omega/(2\alpha)$. Nevertheless, this resonance is pronounced only
when $k^*\approx k_F$, at which the anomalous correlation in the equilibrium
state $\bar{l}_0({\bf k})$
and $\bar{\bf l}({\bf k})$ are significant.

The features for the dynamics of Cooper-pair spin polarization revealed in the
numerical calculation can be understood based on
Eq.~(\ref{solution}). Specifically, from Eqs.~(\ref{simple_relation}),
 (\ref{solution}-\ref{l_y_equilibrium}), one obtains that ${\bf n}_x({\bf k})\propto
 k_y^2{\bf p}_s^x$ and ${\bf n}_y({\bf k})\propto
 k_xk_y{\bf p}_s^x$, confirming the conjecture in
 Eq.~(\ref{Rashba}). Furthermore, it is observed that in
 Eq.~(\ref{l_y_equilibrium}), the contributions of the anomalous correlation
 from the lower and upper bands are opposite, which is relatively large in the
 situation with {\em single-band population}\cite{Tao_1} (refer to Fig.~\ref{figyw10}). Thus, the induced
 total Cooper-pair polarization is significant when only the lower-band is
 populated. It is noted when $n_e\geq N_0$, $\omega\ll \alpha
 k$. Thus,
\begin{equation}
 {\bf n}_x({\bf k})\approx -2(k_y/k^2)\bar{\bf l}_y\bar{l}_0{\bf
   p}_s^x,
 \end{equation}
which is inversely proportional to the momentum. This relation naturally
  explains the calculated results that the induced
 Cooper-pair spin vector is larger in the inner Fermi surface than the one in
 the outer Fermi surface (refer to Fig.~\ref{figyw11}).  Finally, it is addressed that from
 Eq.~(\ref{solution}), it is obtained that the Cooper-pair spin polarization is stabilized by the
 superconducting momentum.

\subsubsection{Higgs Mode and Charge Imbalance}
\label{p_Higgs}
In this part, we consider the dynamics of the Higgs mode
and charge imbalance excited by the
THz pulses in the ($s$+$p$)-wave superconducting InSb (100) QWs. In this configuration,
  although there exists strong
  SOC, most features for the dynamics of the Higgs
  mode and charge imbalance are revealed to be similar to those in the
  $s$-wave one (refer to Sec.~\ref{s_wave}).
  Nevertheless, a novel regime with $|{\bf
    p}_s|$ larger than the Fermi momentum can be realized without destroying the superconductivity. 
  One notes that the regime with $|{\bf
    p}_s|$ larger than the Fermi momentum is hard to be realized in the QWs with a single Fermi
  surface, e.g., the $s$-wave superconducting GaAs QWs considered in Sec.~\ref{s_wave}, 
  in which the superconductivity can be destroyed when $|{\bf
    p}_s|\gtrsim k_F$.  
  Nevertheless, with typical electron densities $n_e\lesssim
  3N_0$, in InSb (100) QWs,
  the specific band structure results in two
  Fermi surfaces due to the strong SOC, which are labeled by the yellow and
  black circles  in Fig.~\ref{figyw9}. It is observed that the Fermi momentum of
    the inner Fermi surface 
  (inner Fermi momentum) can
  be much smaller than one of the outer
  Fermi surface (outer Fermi momentum). Thus, in InSb QWs, the superconducting momentum can be tuned to be
    larger than the inner Fermi momentum but smaller than the outer one,
   but without destroying the superconductivity.

Particularly, compared to the situation in the $s$-wave superconducting
  GaAs QWs,
  when $|{\bf
  p}_s|$ is tuned to be larger than the inner Fermi momentum in InSb QWs,
  some new features in the excitation
    of the Higgs mode and creation of the charge imbalance are expected.
 For the excitation of the Higgs mode, in GaAs QWs, we have
 shown that the pump effect plays a marginal role (refer to
 Sec.~\ref{Higgs_s}). Nevertheless, it is estimated that the pump effect
 influences
 the excitation of the Higgs mode as long as $|{\bf
  p}_s|$ is larger than the Fermi momentum. Accordingly, with $|{\bf
  p}_s|$ larger than the inner Fermi momentum in
    InSb (100) QWs, the pump of the quasiparticles around the inner Fermi
    surface is now expected.
    For the creation of the charge imbalance, in GaAs QWs, we have found that the
    pump effect still has significant 
contribution (refer to Sec.~\ref{Numerical_s_charge}).
Furthermore, it is 
revealed that the charge imbalance is contributed by the pump and drive effects
separately,
through influencing the quasiparticle correlation and
  quasiparticle population, respectively (refer to
Fig.~\ref{figyw7}). This fact indicates that in GaAs QWs, the system still lies in the
``linear'' regime without any interplay between the pump and drive effects.
  Nevertheless, when $|{\bf
  p}_s|$ is tuned to be larger than the inner Fermi momentum in InSb QWs,
  the interplay between the pump and drive effects is expected.

\begin{figure}[ht]
  {\includegraphics[width=7.7cm]{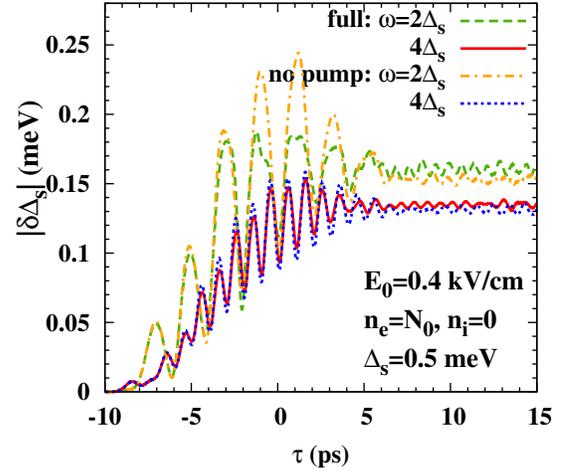}}
  \caption{(Color online) Temporal evolutions of the Higgs mode for the
      singlet order parameter $|\delta\Delta_s|$ with different frequencies
      $\omega=2\Delta_s$ (green dashed curve) and $4\Delta_s$ (red solid curve)
      when $n_e=N_0$ and $n_i=0$. $E_0=0.4$~kV/cm. When the pump effect is neglected,
      $|\delta\Delta_s|$ are plotted for $\omega=2\Delta_s$ (yellow chain curve)
    and $4\Delta_s$ (blue dotted curve). By comparing the green dashed (red
    solid) curve with the yellow chain (blue dotted) curve, one finds
    that when $\omega=2\Delta_s$ ($\omega=4\Delta_s$), the pump effect can have
    significant (marginal) role in the excitation of the Higgs mode. }
  \label{figyw12}
\end{figure}

\paragraph{Higgs mode} We first focus on the
Higgs mode. As with the electron densities we
considered ($n_e\leq 3N_0$), the triplet order parameter and its fluctuation are much smaller than
the singlet one,\cite{Tao_2} here we focus on the Higgs mode contributed
by the singlet order parameter $|\delta\Delta_s|$. 
In Fig.~\ref{figyw12}, the temporal evolutions of the Higgs mode $|\delta\Delta_s|$
are plotted with different optical-field frequencies $\omega=2\Delta_s$ (green dashed curve)
and $4\Delta_s$ (red solid curve) when $n_e=N_0$. The electrical field strength
is taken to be relatively large
$E_0=0.4$~kV/cm, with which the peak value of ${\bf p}_s$ is about $0.6\tilde{k}_F$ ($0.3\tilde{k}_F$) when
$\omega=2\Delta_s$ ($\omega=4\Delta_s$). Here, $\tilde{k}_F=\sqrt{2\pi N_0}$. With this
electric field, the peak
value of $\eta\approx 2$~meV when
 $\omega=2\Delta_s$, is larger than $2\Delta_s=1$~meV, indicating that the
system lies in the strong-pump regime. It can be seen
in Fig.~\ref{figyw12} that the Higgs modes $|\delta\Delta_s|$ oscillate with
twice the frequency of the optical field and plateaus appear after the
optical pulse. These features are similar to the ones in the $s$-wave
superconducting QWs.  

However, it is observed that the pump effect can have significant contribution to
the excitation of the Higgs mode when
$\omega=2\Delta_s$, in contrast to the situation in the $s$-wave
  superconducting QWs (refer to Sec.~\ref{Higgs_s}).
It is shown in Fig.~\ref{figyw12} that with the pump effect
neglected in the calculation, when $\omega=2\Delta_s$ the yellow chain curve is
markedly different 
from the
green dashed curve; nevertheless, when
$\omega=4\Delta_s$, the blue dotted curve
still almost coincides with the red solid one. These can be understood as follows. It is noted
 that when $n_e=N_0$ here, the Fermi momentum of the inner (outer) Fermi surface
 $k_F^{<}\approx 0.5\tilde{k}_F$ ($k_F^{>}\approx 1.4\tilde{k}_F$). Thus, when
 $\omega=2\Delta_s$, $|{\bf p}_s|\lesssim 0.6\tilde{k}_F$ can be comparable to the Fermi momentum of
 the inner Fermi surface, indicating that the system lies in the regime
   with $|{\bf p}_s|\gtrsim k_{F}^<$.
In this regime, it has been estimated that the pump effect can survive from the Pauli blocking
due to the drive effect and hence contribute to the excitation of the Higgs mode
(refer to Sec.~\ref{condition_pump}). Furthermore, by comparing the yellow chain and
green dashed curves, one observes that the pump effect actually suppresses,
rather enhances, the
excitation of the Higgs mode.

\begin{figure}[ht]
  {\includegraphics[width=7.6cm]{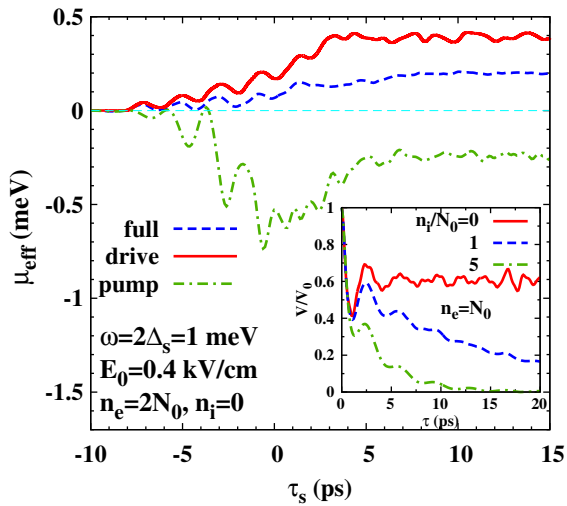}}
  \caption{(Color online) Temporal evolutions of the effective chemical potential in the
    free situation. The electron density $n_e=2N_0$ and
    $\omega=2\Delta_s\approx 1.15$~THz.
    The electric
    field strength is relatively large with $E_0=0.4$~kV/cm, with which the system
    lies in the strong-pump and strong-drive regimes. In
    the figure, the blue dashed curve denotes the full calculation with
    both the drive and pump effects included, whereas the red solid (green chain)
    curve represents the calculated results with only the drive (pump) effect
    retained.  
    In the inset, the temporal evolution of the charge imbalance is presented
    with $n_e=N_0$ with different impurity densities $n_i=0$ (the red solid
    curve), $n_i=N_0$ (the blue
    dashed curve) and
    $n_i=5N_0$ (the green
    chain curve), respectively.}
\label{figyw13}
\end{figure}

\paragraph{Charge imbalance} 
Then we analyze the optical excitation of the charge imbalance.
The temporal evolutions of the effective chemical potential are shown in Fig.~\ref{figyw13} in the
free situation. In the calculation, the electron density $n_e=2N_0$ and 
 $\omega=2\Delta_s=1$~meV. The electric
field strength is relatively large with $E_0=0.4$~kV/cm, with which the system
lies not only in the strong-pump regime but also in the regime with $|{\bf p}_s|\gtrsim k_{F}^<$. In
  Fig.~\ref{figyw13}, the blue dashed curve denotes the full calculation with
  both the drive and pump effects included, whereas the red solid (green chain)
  curve represents the calculated results with only the drive (pump) effect
  retained. It can be seen that the effective chemical potential is always
  positive, and no longer equals to the
  sum of the ones contributed by the drive and pump effects separately. 
 This is in contrast to the features revealed in the $s$-wave superconducting
 GaAs QWs, indicating
 that there exists significant interplay between the
pump and drive effects. Moreover, our calculated results indicate that in the competition between the
pump and drive effects in the creation of the charge imbalance, the drive effect
is dominant.

Finally, it is addressed that the electron-impurity scattering can still provide the
charge-imbalance relaxation channel in the ($s$+$p$)-wave superconducting QWs,
whose features are similar to the situation in the $s$-wave one (refer to
Sec.~\ref{Numerical_s_charge}).
This can be seen in the inset of the
Fig.~\ref{figyw13} that when $n_i=0$, the normalized effective chemical
potential $V/V_0$ does not
relax to zero, represented by the red solid curve; whereas when there exists impurities, the
relaxation channel for the charge imbalance can be opened, as shown by the blue
dashed ($n_i=N_0$) and green
chain ($n_i=5N_0$) curves.

\section{Conclusion and discussion}
\label{summary}
In conclusion, we have investigated
 the optical response to the THz pulses in both the $s$-wave and
($s$+$p$)-wave superconducting semiconductor QWs. We set up
 the gauge-invariant optical Bloch equations via the gauge-invariant
nonequilibrium Green
function approach,\cite{Haug,Levanda,GKB} with the gauge structure revealed by
Nambu explicitly retained.\cite{Nambu_gauge} In the gauge-invariant Green
function approach, the
gauge-invariant Green function with the Wilson line is constructed.\cite{Wilson_line}
By choosing the ${\bf p}_s$-gauge, in the gauge-invariant optical Bloch
equations,
not only can the microscopic description
for the quasiparticle dynamics be realized, but also the dynamics of the
condensate is included, with the superfluid
momentum ${\bf p}_s$ and the effective chemical potential $\mu_{\rm eff}$
naturally incorporated. It is addressed that ${\bf p}_s$ directly contributes to
the center-of-mass momentum and $\mu_{\rm eff}$ corresponds to the collective
excitation revealed by Nambu,\cite{Nambu_gauge,Ambegaokar,Enz,Griffin} evolving
   with time in the homogeneous limit. We show that ${\bf p}_s$ plays an important role in the
   dynamics of quasiparticles. Its nonlinear
term $\propto {\bf p}_s^2$ contributes to the   
 pump of the
quasiparticles (pump effect), and its rate of
change $\partial_t{\bf p}_s$ acts as a drive field to drift the quasiparticles (drive effect).
 Specifically, the drive effect can contribute to the formation of the blocking
 region\cite{Tao_2,FF,LO,FFLO_Takada,supercurrent,Yang,Doppler_1,Doppler_2}
 for the quasiparticle, which directly suppresses the anomalous correlation of Cooper
 pairs (refer to Fig.~\ref{figyw1}). It is found that both the
pump and drive effects contribute to the excitation of the Higgs mode, which
oscillates with
twice the frequency of the optical field. However, it is shown that the contribution from
the drive effect to the excitation of
Higgs mode is dominant as long as the driven superconducting momentum is less 
than the Fermi momentum. This is because in this condition, the pump of the quasiparticle
  population is efficiently suppressed thanks to the Pauli blocking.
This is in sharp contrast to the conclusions obtained from the Liouville\cite{Axt1,Axt2,
   Higgs_e_p} or Bloch\cite{Anderson1,multi_component,Leggett_mode,Xie,Tsuji_e_p,multiband}
equations in the
literature, in which the drive effect is overlooked with only the
pump effect considered. Actually, in these treatments,\cite{Axt1,Axt2,
   Higgs_e_p,Anderson1,multi_component,Leggett_mode,Xie,Tsuji_e_p,multiband} the
 contribution of the Cooper-pair 
  center-of-mass momentum to the suppression of the anomalous
  correlation of Cooper pairs is overlooked. In our framework, the role of the electron-impurity scattering on the
excitation of the superconducting state is also revealed,
 which is found to further suppress the Cooper pairing on the basis of the drive
 effect.

In the gauge-invariant optical Bloch equations, 
the charge neutrality condition is self-consistently considered based on the two-component model for the
charge. In this model, the deviation from the equilibrium state for the
quasiparticle, i.e., the charge imbalance, can cause the fluctuation of the
effective chemical
potential $\mu_{\rm eff}$ for the
condensate.\cite{Tinkham1,Tinkham2,Takahashi_SHE2,
  Tinkham_book,Pethick_review,Larkin,Gray}
This consideration is actually consistent with the one in the determination of the
  collective mode based on the gauge structure and charge conservation for the
superconductivity.\cite{Nambu_gauge,Ambegaokar,Enz,Griffin}
We predict that during the optical
process, the charge imbalance 
  can be created by both the pump and drive effects, with the former arising from the
AC Stark effect and the latter coming from the breaking
of Cooper pairs by the electrical field. Specifically, when $|{\bf p}_s|$
  is much smaller than the Fermi momentum, the charge imbalance is contributed by the pump and drive effects
separately, through influencing the quasiparticle correlation and
  quasiparticle population, respectively.

The induction
of the charge imbalance of quasiparticles directly causes the fluctuation
of the effective chemical potential of the condensate.
 This fluctuation of the effective chemical potential is found to directly provide
a charge-imbalance relaxation channel even with the elastic scattering
due to impurities. This is in contrast to the previous understanding in the
literature that in
the isotropic $s$-wave superconductivity, the impurity scattering cannot cause
any charge-imbalance relaxation.\cite{Tinkham1,Tinkham2,Pethick_review} 
Actually, the previous understanding is based on the framework with
  quasiparticle-number conservation but not the charge conservation, 
in which the charge-imbalance relaxation is induced
by the direct scattering of quasiparticles between
the electron- and hole-like branches in the presence of
the impurities (refer to Fig.~\ref{figyw2}). This inter-branch scattering
  is forbidden for the electron-impurity scattering in the isotropic $s$-wave
  superconductivity thanks to the coherence 
factor $(u_{\bf k}u_{{\bf k}'}-v_{\bf k}v_{{\bf k}'})$ in the scattering
potential.\cite{Tinkham1,Tinkham2,Pethick_review}
Furthermore, the momentum-scattering dependence of the charge-imbalance
relaxation is revealed. When the momentum scattering is weak
(strong), the charge-imbalance relaxation
is enhanced (suppressed) by the momentum scattering.

 Although the above momentum-scattering dependencies of the
 charge-imbalance relaxation seemingly resemble
the ones in the DP mechanism,\cite{DP,
  OptOri,Awschalom,Zutic,Fabian,Dyakonov_book,wureview,Korn,notebook}
 we point out that the DP mechanism cannot explain the charge-imbalance
 relaxation in the presence of the elastic
 scattering.\cite{Tao_valley,Lin,wureview}   
 In fact, a new mechanism is revealed to be responsible for the charge-imbalance
 relaxation here. We demonstrate that the
charge-imbalance relaxation here is caused by the
direct annihilation of the quasiparticles in the quasi-electron
and quasi-hole bands (refer to Fig.~\ref{figyw2}), in which the quasiparticle-number
conservation is {\em broken}. The source of the breaking of quasiparticle-number
conservation is the quasiparticle correlation between the quasi-electron and
quasi-hole states,\cite{Tao_2} which is contributed by the quasiparticle
precession induced by the non-equilibrium
chemical potential of the condensate. 
Then, due to the electron-impurity scattering, the induction of the
quasiparticle correlation further triggers the process of the
condensation with two quasiparticles binding into one
Cooper pair in the condensate, or vice versa.\cite{Bardeen,Tinkham2,Josephson_B}

These processes can directly cause the annihilation of
the extra quasiparticles in the quasi-electron or quasihole
bands, due to which the charge-imbalance relaxation for
the quasiparticles is induced. Meanwhile, with the condensation or breaking
of the Cooper pairs in the condensate, the fluctuation
of the effective chemical potential is also induced. Thus, through the
quasiparticle correlation, the electron-impurity
scattering opens a charge-imbalance relaxation channel due to the fluctuation of the
quasiparticle population. Based on this picture, it is emphasized that the
  creation and relaxation of charge imbalance is a unique feature for the
  superconductivity with non-zero order parameter, in which the particle-number
  or quasiparticle-number 
  fluctuation inherently exists due to the breaking of the global U(1)
  symmetry.
It is further found that the induction of the quasiparticle correlation by $\mu_{\rm
eff}$ is directly suppressed
 by the impurity scattering. 
 Consequently, the competition between the relaxation channels
due to the quasiparticle correlation and population
leads to the non-monotonic dependence on the momentum
scattering for the charge-imbalance relaxation.

By using the optical Bloch equations, the optical creation of the spin
polarization for the Cooper pairs is 
investigated in the ($s$+$p$)-wave 
superconducting InSb (100) QWs. We predict that when the optical field with the
in-plane vector potential applied, the total 
spin polarization of triplet Cooper pairs can be induced, which is shown to be parallel
to the vector potential and oscillates with the frequency of the optical
field. This can be understood as follows. In the equilibrium state,
  in the InSb superconducting (100) QWs with
  the Rashba-like SOC,  the
  ${\bf l}$-vector of the anomalous correlation for the triplet Cooper pairs
  is parallel to the effective
  magnetic field ${\bgreek
    \Omega}({\bf k})$ due to the SOC.\cite{Tao_1} Here, ${\bf l}$-vector is
  defined from
$\big[{\bf l}({\bf k})\cdot {\bgreek \sigma}\big]i\sigma_y=\left(
\begin{array}{cccc}
F_{\uparrow\uparrow}({\bf k}) & F_{\uparrow\downarrow}({\bf k})\\
F^*_{\uparrow\downarrow}({\bf k})& F_{\downarrow\downarrow}({\bf k})
\end{array}
\right)$
with $F({\bf k})$ representing the Fourier components of triplet Cooper-pair
wavefunction in spatial space.\cite{Leggett_book}
When the optical
field with the in-plane vector potential is applied to this 
superconducting system, the superconducting velocity is induced, which is
shown to contribute to an effective SOC. 
This induced effective SOC can cause the
precession of the ${\bf l}$-vectors, with the component perpendicular to ${\bgreek
  \Omega}({\bf k})$ induced. Thus, the Cooper-pair spin vector can be
induced by its definition ${\bf n}({\bf k})=i{\bf l}({\bf k})\times {\bf
  l}^*({\bf k})$,\cite{Tkachov} whose summation by momentum contributes to the total spin polarization
of Cooper pairs ${\bf P}_{\rm C}$ [Eq.~(\ref{polarization_definition})].
It is demonstrated that the induced Cooper-pair spin
vector ${\bf n}({\bf k})$ is inversely proportional to the momentum.
Moreover, it is revealed that the Cooper-pair spin polarization is proportional
to the superconducting velocity, which oscillates with the frequency of the
optical field. This shows that the Cooper-pair spin polarization is
  directly stabilized by the superconducting momentum.

  Although our calculations are
  performed in the two-dimensional superconducting semiconductor QWs in particular materials
  with small and simple Fermi surfaces, the obtained predictions can still
  shed light on the optical response in the film of the superconducting metal, 
  even with complex
  Fermi surfaces. In our set up, the optical field and the correspondingly-induced superconducting velocity
    are treated to be homogeneous in the whole material. This is because with
  our material parameters, the London penetration depth
  $\lambda_L\approx \sqrt{m^*/(\rho_s e^2)}$ for the magnetic
  field\cite{Schrieffer} is in the order of micrometer, much
  larger than the well width of the QWs. In this situation, the Meissner effect
  can be neglected and hence the
  optical field can efficiently penetrate into the material. Actually, even in the 
  superconducting film of metal, the efficient penetration of the optical field is often considered to be
  satisfied,\cite{MgB2_1,MgB2_2,YBCO,BSCCO,voltage,Earliest,Matsunaga_1,Matsunaga_2,Matsunaga_3,Matsunaga_4}
  to which the framework used in this work can be extended.

  From the experimental point of view, we remark the possible experimental
  detections for our predictions, including the
  Higgs mode induced by the drive effect, the induction of the charge
  imbalance by the optical method, the novel relaxation channel for the charge
  imbalance due to the elastic scattering and the induction of the Cooper-pair
  spin polarization by the optical technique. 
  Specifically, for the Higgs mode induced by the drive effect, our calculation
 shows that its oscillation amplitude is suppressed and plateau value after the
 pulse is enhanced by the electron-impurity
 scattering. Particularly, the latter feature is in contrast to the ones in the
influence of the impurity on the pump effect.\cite{footnote}
 Thus, the experimental observation on the 
 impurity-density dependence of the Higgs-mode oscillation can help to distinguish the
 contribution to the Higgs mode from the drive and pump effects. For the charge imbalance induced
 by the optical method, it can be directly 
detected either through the voltage between the quasiparticle and
condensate measured in the setup of Clarke's
works,\cite{Clarke_first,Clarke_thermal} or through the effective
chemical potential measured in the Josephson
effect.\cite{voltage} These techniques with time resolution can also be used to
measure the charge-imbalance relaxation due to the impurity scattering,
which should be performed at low
temperature with significant impurity density. As to the Cooper-pair
 spin polarization induced by the optical method, its direct observation is not
 as easy as the spin polarization of electrons. Nevertheless, it was
 proposed by Tkachov that the Cooper-pair spin
 polarization can be detected by the
 magnetoelectric Andreev effect.\cite{Tkachov}

 Finally, we remark the physical origin of the effective chemical
 potential from another point of view, which has been presented based on
 the consideration of the
 charge conservation in the two-component model for the
 charge.\cite{Tinkham1,Tinkham2,Takahashi_SHE2,Tinkham_book,Pethick_review,Larkin,Gray,Tinkham_condensate}
 From the gauge structure in the superconductivity, the
  effective chemical potential origins from the rate of change of the
  superconducting phase. Actually, based on the work of Ambegaokar and
  Kadanoff,\cite{Ambegaokar} in the long-wave limit, the excited superconducting phase in the optical
  process is
  exactly the collective mode revealed by Nambu with the consideration of the
  gauge invariance in the superconductivity,\cite{Nambu_gauge}
  which is referred to as the Nambu-Goldstone mode in the field
  theory.\cite{Goldstone,Wilson_line} In both the experiment\cite{Goldstone_Nb_Pb,Cea}
  and theory,\cite{Cea_theory} 
   the Nambu-Goldstone mode was reported to directly
  contribute to the optical absorption, especially when the photon energy is
  below the superconducting gap. Based on this understanding,
  we conjecture that the effective chemical potential is contributed by the temporal variations of
  the Nambu-Goldstone mode, which is excited by the optical pulse. Therefore, the
  study on the effective chemical potential not only helps to reveal the
  dynamics of the charge imbalance, but also can shed light on the understanding of the
  optical excitation for the Nambu-Goldstone mode.
 
\begin{acknowledgments}
This work was supported
 by the National Natural Science Foundation of China under Grant
No. 11334014 and  61411136001. One of the authors (T.Y.) would like
to thank M. Q. Weng for helpful discussions. 
\end{acknowledgments}

\begin{appendix}
\section{OPTICAL BLOCH EQUATIONS IN QUASIPARTICLE SPACE}
\label{AA}
It is convenient to perform the analytical analysis for the dynamical process of
the {\em quasiparticle} by the optical Bloch equations in the
  quasiparticle space. Here, we transform the optical Bloch equations in the
  particle space, i.e., Eq.~(\ref{homogeneous}), into the ones in the
  quasiparticle space by the unitary transformation Eq.~(\ref{unitary}), which are written as
\begin{eqnarray}
\nonumber
\hspace{-0.6cm}&&\frac{\partial \rho_{\bf k}^h}{\partial T}+i\Big[E_{\bf
  k}\tau_3,\rho_{\bf k}^h\Big]+i\Big[\mu_{\rm eff}\tilde{\tau}_3,\rho_{\bf
  k}^h\Big]
+i\Big[\frac{{\bf p}_s^2}{2m^*}\tilde{\tau}_3,\rho_{\bf
  k}^h\Big]\\
\nonumber
\hspace{-0.6cm}&&\mbox{}+\frac{1}{2}\Big\{\frac{\partial {\bf p}_s}{\partial
  T}\tilde{\tau}_3,
\frac{\partial \rho_{\bf k}^h}{\partial {\bf
    k}}\Big\}+\frac{1}{2}\Big\{\frac{\partial {\bf p}_s}{\partial
  T}\tilde{\tau}_3,\Big[\rho_{\bf k}^h,
\frac{\partial {\mathcal U}_{\bf k}}{\partial
{\bf k}}{\mathcal U}_{\bf k}^{\dagger}\Big]\Big\}\\
\nonumber
\hspace{-0.6cm}&&=i\sum_{{\bf k}'}U_{{\bf k}-{\bf k}'}\Big[({\mathcal U}_{\bf
  k}\tau_3{\mathcal U}_{{\bf k}'}^{\dagger})(\rho_{{\bf k}'}^h-\rho_{{\bf
      k}'}^{h,0})({\mathcal U}_{{\bf k}'}\tau_3{\mathcal U}_{\bf
  k}^{\dagger}),\rho_{\bf k}^h\Big]\\
\nonumber 
\hspace{-0.6cm}&&\mbox{}
-\pi n_i\sum_{{\bf
  k}'\eta=\pm}|V_{{\bf k}-{\bf k}'}|^2\delta(E_{{\bf k}'\eta}-E_{{\bf
  k}\eta})\big[({\mathcal U}_{\bf k}\tau_3{\mathcal U}_{{\bf
      k}'}^{\dagger})Q_{\eta}({\mathcal U}_{{\bf k}'}\tau_3{\mathcal U}_{\bf k}^{\dagger})\\
\hspace{-0.6cm}&&\mbox{}\times Q_{\eta}\rho_{\bf k}^h
-({\mathcal U}_{\bf k}\tau_3{\mathcal U}_{{\bf k}'}^{\dagger})
\rho_{{\bf k}'}^hQ_{\eta}({\mathcal U}_{{\bf k}'}\tau_3{\mathcal U}_{\bf k}^{\dagger})Q_{\eta}+{\rm H.c.}\big],
\label{quasiparticle_s}
\end{eqnarray}
whose structure is analyzed as follows.

In the second and third terms in Eq.~(\ref{quasiparticle_s}), the
diagonal terms in $\tilde{\tau}_3$ renormalize the quasiparticle excitation
energy; whereas the off-diagonal terms cause the precession between the quasi-electron and
quasi-hole sates, which act as the pump term similar to the inter-band optical excitation in the
semiconductor.\cite{Haug,Axt_optics,Rossi_optics} Specifically, it can be seen that the fluctuation of the
condensate, i.e., $\mu_{\rm eff}$, can also contribute to the pump term, which
definitely influences the dynamics of the quasiparticle. Moreover, in the quasiparticle space, the drive term is
contributed by the fourth and fifth terms, with the latter originating from the Berry
phase effect.\cite{Berry,Berry_0,helix_Berry} Finally, in the scattering term,
only the electron-impurity scattering is presented here with the
  electron-phonon one [Eq.~(\ref{scat_ep})] negligible at the low temperature. $Q_{\pm}=1/2\pm \tau_3/2$ are the
projection operators in the quasiparticle space. 
It is noted that the derived scattering term here is different from the one used in the
Boltzmann equation for the Bogoliubov quasiparticle, in which the
contribution from the off-diagonal terms in ${\mathcal U}_{\bf k}\tau_3{\mathcal
  U}_{{\bf
    k}'}^{\dagger}=(u_{\bf k}u_{{\bf k}'}-v_{\bf k}v_{{\bf
    k}'})\tau_3-(u_{\bf k}v_{{\bf k}'}+v_{\bf k}u_{{\bf k}'})\tau_1$ is
neglected by
 neglecting the correlation between the
quasi-electron and quasi-hole states.\cite{Kopnin,Tinkham2,Pethick_review,Tao_2}

Specifically, in the scattering term, the contributions from the diagonal and
off-diagonal terms in ${\mathcal U}_{\bf k}\tau_3{\mathcal U}_{{\bf
    k}'}^{\dagger}$ can be separated, which are represented by
$\partial_t\rho_{\bf k}|_{\rm scat}^{\rm d}$ and $\partial_t\rho_{\bf k}|_{\rm
  scat}^{\rm off}$, respectively. For the diagonal contribution,   
\begin{eqnarray}
\nonumber
&&\partial_t\rho_{\bf k}|_{\rm scat}^{\rm d}=-2\pi n_i\sum_{{\bf k}'}|V_{{\bf
    k}-{\bf k}'}|^2(u_{\bf k}u_{{\bf k}'}-v_{\bf k}v_{{\bf
      k}'})^2\\
&&\mbox{}\times\delta(E_{{\bf k}'}-E_{\bf k})(\rho_{{\bf k}}^h-\tau_3\rho_{{\bf
    k}'}^h\tau_3),
\label{s_diagonal}
\end{eqnarray}
which recovers to the scattering term used in the Boltzmann equation for the
Bogoliubov quasiparticle when the off-diagonal term in $\rho_{\bf k}^h$ is
neglected.\cite{Kopnin,Tinkham2,Pethick_review,Tao_2}
For the off-diagonal contribution, 
\begin{eqnarray}
\nonumber
&&\hspace{-0.45cm}\partial_t\rho_{\bf k}|_{\rm scat}^{\rm off}=\pi n_i\sum_{{\bf k}'}|V_{{\bf
    k}-{\bf k}'}|^2(u_{\bf k}u_{{\bf k}'}-v_{\bf k}v_{{\bf
      k}'})(u_{\bf k}v_{{\bf k}'}+u_{{\bf k}'}v_{\bf k})\\
&&\hspace{-0.45cm}\mbox{}\times\delta(E_{{\bf k}'}-E_{\bf k})(\tau_1\tau_3\rho_{{\bf
    k}}^h-\tau_1\rho_{{\bf k}'}^h\tau_3+{\rm H.c.}),
\label{s_off}
\end{eqnarray} 
Obviously, for the equilibrium distribution for the quasiparticle $\rho_{{\bf k},0}^{h}=1/2+[f_0(E_{\bf k})-1/2]\tau_3$,
Eqs.~(\ref{s_diagonal}) and (\ref{s_off}) are exactly zero.

\section{QUASIPARTICLE AND SUPER-FLUID DENSITIES}
\label{BB}
\begin{figure}[ht]
  {\includegraphics[width=7.7cm]{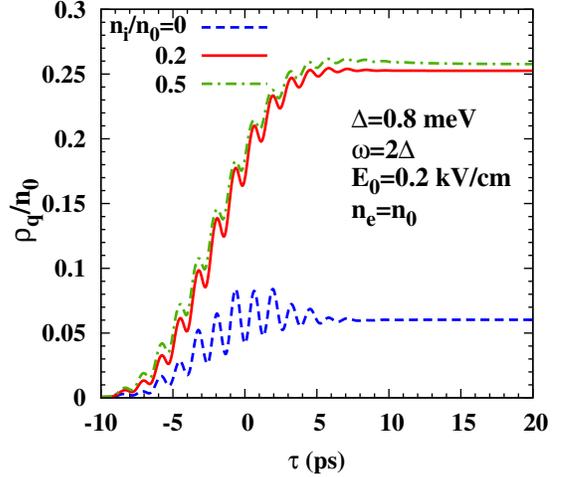}}
  \caption{(Color online) Temporal evolutions of the quasiparticle density
    $\rho_q$ in the $s$-wave superconducting GaAs QWs under the 
optical THz pulse with different impurity
densities $n_i=0$ (blue dashed curve), $0.2n_0$ (red solid curve) and $0.5n_0$
 (green chain curve). $E_0=0.2$~kV/cm and $\omega=2\Delta\approx 2.3$~THz.}
  \label{figyw14}
\end{figure}

In this part, we present the calculated quasiparticle and super-fluid densities under the 
optical THz pulse in the $s$-wave
superconducting GaAs QWs.
In Fig.~\ref{figyw14}, the temporal evolutions of the quasiparticle density
$\rho_q$
 are plotted with different impurity
densities $n_i=0$ (blue dashed curve), $0.2n_0$ (red solid curve) and $0.5n_0$
(green chain curve). It is shown that after the pulse $\tau\gtrsim 5$~ps,
there exist plateaus in the quasiparticle density, whose values increase with the
increase of the impurity density. This is because the existence of the impurity
density can enhance the optical absorption. These
populations of the 
hot quasiparticles can efficiently suppress the Cooper pairing.     
  
Then the normal-fluid and super-fluid densities $\rho_n$ and $\rho_s$ after the pulse are {\em
  estimated}  based on the two-fluid model in the
  equilibrium state.\cite{Kerson_Huang,Schrieffer} Specifically,  
for the order parameter $\Delta=|\Delta|e^{i{\bf q}\cdot{\bf r}}$ with the
center-of-mass momentum ${\bf
  q}=2m^*{\bf v}_s$ along the $\hat{\bf x}$-direction, 
the momentum supercurrent is calculated to be\cite{Schrieffer,Tao_2}
\begin{eqnarray}
\nonumber
\hspace{-0.5cm}&&{\bf J}_s=2m^*{\bf v}_s\sum_{\bf k}\Big[v_{\bf k}^2+(u_{\bf k}^2-v_{\bf
  k}^2)f_0({\bf k}\cdot{\bf v}_s+\sqrt{\Gamma_{\bf
    k}^2+|\Delta|^2})\Big]\\
\hspace{-0.5cm}&&\mbox{}+2\sum_{\bf k}{\bf k}f_0({\bf k}\cdot{\bf v}_s+\sqrt{\Gamma_{\bf
    k}^2+|\Delta|^2}).
\end{eqnarray}
with $\Gamma_{\bf k}=k^2/(2m^*)-\mu+m^*{\bf v}_s^2/2$.
For the linear response, ${\bf q}$ is small, hence, 
\begin{equation}
{\bf J}_s\approx 2{\bf v}_s\sum_{\bf k}\Big[k_x^2\frac{\partial f_0(E_{\bf
    k})}{\partial E_{\bf k}}+m^*v_{\bf k}^2+m^*(u_{\bf k}^2-v_{\bf k}^2)f_0(E_{\bf
k})\Big].
\end{equation}
Thus, with ${\bf J}_s\equiv {\bf v}_sm^*\rho_s$, one obtains
\begin{equation}
\rho_s=2\sum_{\bf k}\Big[\frac{k_x^2}{m^*}\frac{\partial f_0(E_{\bf
    k})}{\partial E_{\bf k}}+v_{\bf k}^2+(u_{\bf k}^2-v_{\bf k}^2)f_0(E_{\bf
k})\Big].
\label{super_density}
\end{equation}

For the normal parts, by assuming the drift distribution $f_0(E_{\bf
    k}-{\bf k}\cdot{\bf v}_n)$ with ${\bf v}_n=v_n\hat{\bf
    x}$,\cite{Kerson_Huang}
 the momentum normal-current reads
\begin{equation}
\nonumber
{\bf J}_n=2\sum_{\bf k}{\bf k}f_0(E_{\bf k}-{\bf k}\cdot{\bf v}_n)\approx
2v_n\hat{\bf x}\Big[-\sum_{\bf
  k}k_x^2\frac{\partial f_0(E_{\bf k})}{\partial E_{\bf k}}\Big].
\end{equation}
Consequently, for the linear response with ${\bf J}_n={\bf v}_{n}m^*\rho_n$, one
has 
\begin{equation}
\rho_n=-2\sum_{\bf
  k}\frac{k_x^2}{m^*}\frac{\partial f_0(E_{\bf k})}{\partial E_{\bf k}}.
\label{normal_density}
\end{equation}
Obviously, from Eqs.~(\ref{super_density}) and (\ref{normal_density}),
 $\rho_s+\rho_n=2\sum_{\bf k}\Big[v_{\bf k}^2+(u_{\bf k}^2-v_{\bf k}^2)f_0(E_{\bf
k})\Big]$, which is exactly the total particle density conserved due to the
charge neutrality [Eq.~(\ref{self_consistent_s})].

It is noticed that Eqs.~(\ref{super_density}) and (\ref{normal_density})
are established for the equilibrium state with $f_0(E_{\bf k})$ representing
the equilibrium quasiparticle distribution.\cite{Kerson_Huang,Schrieffer}
 To estimate the super-fluid and normal-fluid
densities at the non-equilibrium state, Eqs.~(\ref{super_density})
 and (\ref{normal_density}) are extended with
$f_0(E_{\bf k})$ replaced by the non-equilibrium quasiparticle distribution
calculated by optical Bloch equations [Eq.~(\ref{homogeneous})], which is isotropic in the momentum
space after the pulse.\cite{MgB2_1,MgB2_2,YBCO,BSCCO} This extension is based on
the fact
that after the pulse, the quasiparticle distribution can be effectively described
by an effective temperature.\cite{Tao_Ge,Yang_hot,Lei_hot}

\begin{figure}[ht]
  {\includegraphics[width=7.7cm]{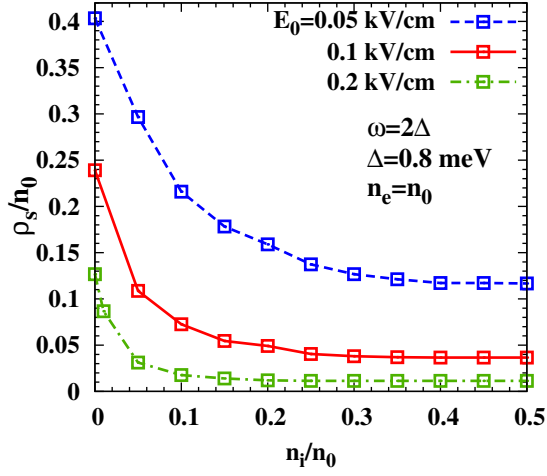}}
  \caption{(Color online) Impurity density dependence of the super-fluid
    density $\rho_s$
    after the pulse, estimated from Eq.~(\ref{super_density}),
    with different electrical fields $E_0=0.05$~kV/cm (blue
    dashed curve with squares), 0.1~kV/cm (red solid curve with squares) and
    0.2~kV/cm (green chain curve with squares). }
  \label{figyw15}
\end{figure}

In Fig.~\ref{figyw15}, the impurity density dependencies of super-fluid density
after the pulse are plotted with different electrical fields $E_0=0.05$~kV/cm (blue
dashed curve with squares), 0.1~kV/cm (red solid curve with squares) and
0.2~kV/cm (green chain curve with squares).
 It is shown that with the
increase of the impurity density, the super-fluid density decreases. This is
consistent with the fact that with the optical pulse, the presence of the
impurity can further suppress the
Cooper pairing [refer to Fig.~\ref{figyw4}(c)]. Specifically, one sees that
although there exists a significant order parameter after the pulse, the super-fluid density can
be extremely small at the non-equilibrium state.

\section{OPTICAL GENERATIONS OF SPIN POLARIZATION AND SPIN CURRENT}
\label{CC}

We first define the spin polarization and spin current in the ($s$+$p$)-wave
superconducting (100) QWs. 
The temporal evolution of the total spin polarization of the system is calculated
by\cite{Shiba,Tao_2}
\begin{equation}
{\bf P}\equiv(P_x,P_y,P_z)=(1/2)\sum_{\bf k}\mbox{Tr}(\rho_{\bf k}{\bgreek
  \alpha})/n_0,
\label{total_spin}
\end{equation}
in which
\begin{equation}
{\bgreek \alpha}=\frac{1+\tau_3}{2}\otimes {\bgreek
  \sigma}+\frac{1-\tau_3}{2}\otimes \sigma_y{\bgreek \sigma}\sigma_y.
\end{equation}
The velocity operator is calculated by the Heisenberg equation $\hat{\bf v}=-i[\hat{\bf
  r},\tilde{H}_0({\bf A}=0)]$. Specifically,
\begin{equation}
v_y=\left(
\begin{array}{cccc}
k_y/m^* & -i \alpha & \Delta_pf({\bf k})& 0\\
i\alpha &k_y/m^*&0&-\Delta_pf^*({\bf k})\\
\Delta_pf^*({\bf k})&0&-k_y/m^*&i\alpha\\
0&-\Delta_pf({\bf k})&-i\alpha &-k_y/m^*
\end{array}
\right),
\end{equation}
with $f({\bf k})=(-k_xk_y+i k_x^2)/k^3$. Then the spin current
 is defined as\cite{Tao_valley,SHE_MacDonald,SHE_KSBE,SHE_Glazov,SHE_Ka} 
\begin{equation}
{\bf J}_{\hat{\bf v}}^{\hat{\bgreek \alpha}}=\sum_{\bf k}\mbox{Tr}\big(\{{\bf
  v},{\bgreek \alpha}\}\rho_{\bf k}\big)/2.
\label{spin_current}
\end{equation}

\begin{figure}[ht]
  {\includegraphics[width=7.7cm]{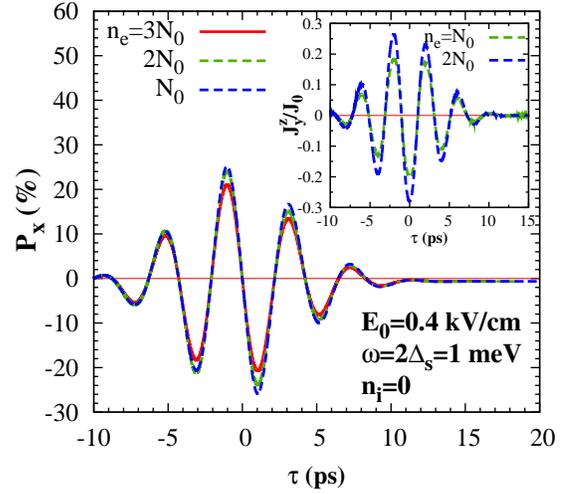}}
  \caption{(Color online) Temporal evolutions of 
optically-generated spin
polarization with different
electron densities $n_e=N_0$ (red solid curve), $2N_0$ (green dashed curve) and
$3N_0$ (blue dashed curve). $E_0=0.4$~kV/cm. The optically-generated spin polarization is along
the $\hat{\bf x}$-direction, i.e., in parallel to the optically-induced
  supercurrent,
and oscillates
with the same frequency of
the optical field. The spin current is presented in the inset, which is calculated   
according to Eq.~(\ref{spin_current}). $J_0=n_ek_F(k_BT_e/E_F)$ with
$k_F=\sqrt{2\pi n_e}$ and $E_F=\pi n_e/m^*$. }
  \label{figyw16}
\end{figure}

According to Eq.~({\ref{total_spin}}), the temporal
evolutions of the
optically-generated spin
polarization are shown in Fig.~\ref{figyw16} with different
electron densities $n_e=N_0$ (red solid curve), $2N_0$ (green dashed curve) and
$3N_0$ (blue dashed curve).
 It is shown that the optically-generated spin polarization is along the
 $\hat{\bf x}$-direction, i.e., in parallel to the optically-induced
  supercurrent,
and oscillates with the same frequency of
the optical field. This is consistent with the previous studies in the 
system with Rashba SOC.\cite{Jinluo,Jianhua_Liouville,Ivchenko} Furthermore, the calculated results with
different electron densities show that the generated spin polarization is less influenced by the
electron density. In the inset of Fig.~\ref{figyw16},  the spin current is
presented, which is calculated 
according to Eq.~(\ref{spin_current}). It is shown that the induced spin current is perpendicular to the total
electrical current with the spin polarization being along the $\hat{\bf
  z}$-direction, oscillating with the frequency of the optical field.
 By noticing that the spin current is divided by $J_0=n_ek_F(k_BT_e/E_F)\propto
 \sqrt{n_e}$ with
$k_F=\sqrt{2\pi n_e}$ and $E_F=\pi n_e/m^*$,
 our calculation further shows that the induced spin current is also insensitive to the
electron density. Finally, it is addressed that with our material parameter, the
numerical calculations indicate that the superconducting order
parameter has little influence on the optical generations of both the spin
polarization and spin current (not shown in Fig.~\ref{figyw16}).

\end{appendix}


\begin{thebibliography}{0}
\bibitem{footnote} Actually, in our framework beyond the Boltzmann equation, even without the drive effect,
  the electron-impurity scattering also
  suppresses the oscillation amplitude of the Higgs mode due to the pump
  effect. This is because in the electron-impurity scattering terms, i.e.,
  Eqs.~(\ref{s_diagonal}) and (\ref{s_off}), both the quasiparticle correlation and
  population are still influenced by the impurity.
  The details are addressed in Sec.~\ref{Numerical_s_charge}.
\end{thebibliography}
\end{document}